\documentclass[12pt,a4paper]{article}

\usepackage{graphicx}

\usepackage{authblk}

\usepackage{cite}

\usepackage{hyperref}

\usepackage{geometry}
\usepackage{amsmath,amssymb}
\usepackage{subfigure}

\newcommand{\be}{\begin{equation}}
\newcommand{\ee}{\end{equation}}

\newcommand{\bea}{\begin{eqnarray}}
\makeatletter
\newcommand*{\rom}[1]{\expandafter\@slowromancap\romannumeral #1@}
\makeatother
\newcommand{\eea}{\end{eqnarray}}

\newcommand{\ba}[1]{\begin{array}{#1}}

\newcommand{\ea}{\end{array}}

\begin{document}
\begin{titlepage}

\vspace{0.5cm}

\begin{center}
\bf{Non-local cosmological models}
\end{center}

\vspace{0.3cm}

\begin{center}
Utkarsh Kumar$^{\dagger}$
\end{center}
\begin{center}
\it{Indian Institute of Science Education and Research, Bhopal}
\end{center}

\vspace{0.2cm}

\begin{center}
Sukanta Panda$^{\ast}$
\end{center}

\begin{center}
\it{Indian Institute of Science Education and Research, Bhopal}
\end{center}

\vspace{0.3cm}

\begin{center}
ABSTRACT
\end{center}
\hspace{0.3cm} Nonlocal cosmological models are studied extensively in recent times because of their interesting cosmological consequences. In this paper, we have analyzed background cosmology on a class of non-local models which are motivated by the perturbative nature of gravity at infrared scale. We show that inflationary solutions are possible in all constructed non-local models. However, exit from inflation to RD era is not possible in most of the models.

\vspace{0.3cm}
\begin{flushleft}
$^{\dagger}$ e-mail: utkarshk@iiserb.ac.in \\
$^{\ast}$ e-mail: sukanta@iiserb.ac.in
\end{flushleft}
\vspace{0.3cm}
\end{titlepage}

\section{Introduction}
Usually, we deal with local theories to study the cosmology of our universe. Generally, local theories contain a classical action which may be Einstien Hilbert(EH) action with matter or beyond. Our universe is accelerating in recent times as it is evident from supernovae observations \cite{Ade:2015xua}.  To explain the acceleration of our current universe, a cosmological constant is usually added to the EH action. However, a physical picture describing the appearance of such a constant in a basic underlying theory is absent. To overcome such issue, other alternative models are floated by modifying the gravity or matter sector. Here in this paper, we work with purely modified gravity models. Our models are inspired by the nonlocal terms in the EH action which appear as IR corrections in gravity theory. In this connection, Infrared effects are calculated for de Sitter space in detail\cite{Belgacem:2017cqo}. There are still issues in computing IR effects in de Sitter space\cite{Antoniadis:1986sb,Antoniadis:1991fa,Tsamis:1994ca,Miao:2011ng,Rajaraman:2016nvv}. Apart from these issues, a different approach is taken, which is phenomenological, to see the effect of nonlocal terms on the background cosmological evolution.

In the year 1998 Wetterich, first proposed a nonlocal gravity model with a nonlocal term as $ \frac{1}{\Box} R$ and a dimensionless parameter  \cite{Wetterich:1997bz}. This model did not produce correct cosmological background evolution. Much later in 2007, Deser and Woodard \cite{Deser:2007jk} modified the action with a nonlocal term as $R f\left(\frac{1}{\Box}R\right),$ for more on this model refer to  \cite{Woodard:2014iga}.In addition to above stated model, similar nonlocal models involving operators such as $ R^{\mu\nu} \Box^{-1} G_{\mu\nu}$ have been presented by Barvinsky \cite{Barvinsky:2003kg,Barvinsky:2011hd,Barvinsky:2011rk}. In these models all couplings are dimensionless. Another class of  nonlocal models  was introduced to study cosmology in IR regime where an additional mass scale corresponding  to cosmological constant,$ \Lambda  $ arises \cite{Dirian:2014xoa,Nersisyan:2016hjh,Amendola:2017qge,Foffa:2013vma,Kehagias:2014sda}.

 The construction of phenomenological non local action is based on the behaviour of the theory in the parameter space where perturbation calculation around de-Sitter space is valid. Our nonlocal action is inspired from gravity correction to the hubble parameter in the perturbative regime in a de-Sitter space-time. Notation in this paper is same as that of ref. \cite{Tsamis:2016boj}. Perturbative IR corrected hubble parameter is given by \cite{NctRpw2,Prokopec:2007ak}
\begin{equation}
3H_{eff}^{2} = \Lambda + 8 \pi G\rho \label{1}
\end{equation}
\begin{equation}
H_{eff}(t) \, = \,
H_{\rm in} \, \Big\{ 1 - G \Lambda \Big( 
c_2 G \Lambda \ln[a(t)]
+ c_3 (G \Lambda)^2 \ln^2[a(t)] 
+ \dots \Big) \Big\} 
\;\; .   \label{2}
\end{equation}
\paragraph{}
The perturbation result fails when $\ln a(t) \sim (G\Lambda)^{-1} $.  To take an account of the IR correction of $O(\ln a(t))$, we need to introduce term $\frac{R}{\Box}$ into the action. In de Sitter space time it is shown that $\frac{R}{\Box}\simeq -4 H_{in}t$ for large observation times \cite{NctRpw3}. However, beyond the perturbative limit, non-perturbative action can be modelled by a function $ f(\frac{R}{\Box}) $, so that we recover our perturbative results when $G\Lambda << 1$.
\paragraph{}
In terms of these corrections, the non-perturbative action takes form,
\begin{equation}
S = \int d^{4}x \sqrt{-g}\big[R + f(X)\big]
\end{equation}
where $X = -G\Lambda \frac{1}{\Box} R $.
\paragraph{}
This was originally motivated to explain inflation in a purely gravitational action. This describes an inflation model followed by a very rapid reheating epoch. Here inflation is ended by a rapid oscillation of hubble parameter after which universe enters into radiation dominated era. However, this model experiences several problems like longer duration for inflation, inflation ends with rapid oscillations in hubble parameters, very fast reheating  etc. For details refer to \cite{NctRpw2,Tsamis:2016boj}.

To overcome these shortcomings in this model, a new model has been proposed in \cite{Tsamis:2016boj}. This model does not suffer from "sign problem" as well as "magnitude problem" as described in \cite{Tsamis:2016boj}.(Sign problem is related to the graceful exit from inflation to radiation dominated era, where as magnitude problem is related to the theoretical value of $ \Lambda $ which is quite large compared to its observed value at late times.) Then this model may describe unification of inflation era at very early times  with dark energy dominated era at late times. 

The paper is organised as follows. In section \ref{sec0} we define the different non local models and derive the expressions required for background cosmological evolution. We have solved background evolution for five non local models and analyse their results in section \ref{sec1}-\ref{sec5}. Finally we summarise all our findings in section \ref{sec6}.  
\section{Non-local Models} \label{sec0}
We study background cosmology in flat FRW metric given by
\begin{equation}
ds^2 \; = \;
- dt^2 \, + \, a^2(t) \, d{\bf x} \cdot d{\bf x}
\;\; . \label{frw}
\end{equation}
where $ a(t)$ is scale factor. Time variation of $a(t)$ gives the Hubble parameter and first slow roll parameter defined as
\begin{equation}
H(t) = \frac{\dot{a}(t)}{a(t)}, 
\end{equation}
\begin{equation}
\epsilon(t) = 1 -\frac{a(t)\ddot{a}(t)}{\dot{a}^2(t)} = -\frac{\dot{H}(t)}{H^{2}(t)}. 
\end{equation}
The required curvature invariants for the our models in flat FRW metric are
\begin{eqnarray}
& \mbox{} &
R \, = \, 6 ( 2 - \epsilon ) H^2
\quad , \quad 
R^2 \, = \, 36 ( 2 - \epsilon )^2 H^4
\;\; , \label{R} \\
& \mbox{} &
R_{\mu\nu} R^{\mu\nu} \, = \,
12 ( 3 - 3\epsilon + \epsilon^2 ) H^4 
\;\; , \label{Rmn} \\
& \mbox{} &
R_{\mu\nu\rho\sigma} R^{\mu\nu\rho\sigma} \, = \,
12 ( 2 - 2\epsilon + \epsilon^2 ) H^4 
\;\; , \label{Rmnrs}
\end{eqnarray}
We also look for  inverse differential  Invariant operators that are responsible for nonlocalities and in FRW geometry they are expressed as
\begin{align}
 &\Box = \frac{1}{\sqrt{-g}}\partial_\mu (\sqrt{-g}g^{\mu\nu}\partial_\nu) = -\partial_t^2 -3H\partial_t {\hskip 1em},\\ & \Box_c = \Box - \frac{1}{6}R = -\partial_t^2 -3H\partial_t-2H^2 -\dot{H} {\hskip 1em},
\end{align}
 while acting them to co-moving time. Their inverses are
 \begin{align}
  & \frac{1}{\Box} = \int_{t_{in}}^{t} dt^{'} \frac{1}{a^{3}(t')}\int_{t_{in}}^{t^{'}} dt^{''} a^{3}(t^{''}) {\hskip 1em } ,\\ & \frac{1}{\Box}_c = \frac{1}{a(t)}\int_{t_{in}}^{t} dt^{'} \frac{1}{a(t')}\int_{t_{in}}^{t^{'}} dt^{''} a^{2}(t^{''}) {\hskip 1em } ,
 \end{align}
these  operators are defined with retarded boundary conditions to prevent the extra degree of freedom \cite{deser}. \paragraph{}
The model given in \cite{Tsamis:2016boj} takes the form 
\begin{equation}
S = \int d^4 x \sqrt{-g}\big[ \frac{1}{16 \pi G} R + f(X) \big], \label{main action}
\end{equation}
where $ X =  G \frac{1}{\Box}R\frac{1}{\Box_{c}}\left(\frac{1}{3}R^2 - R_{\mu\nu} R^{\mu\nu}\right) $ and $\Box_{c} = \Box - \frac{1}{6} R.$
\paragraph{}
We name this model as Model \rom{1} throughout the paper. Here in our work, apart from this form of X[g], we consider other forms of X[g] that may be equally motivated phenomenologically. The proposed forms of X[g] are following:

\begin{equation*}
(\rom{2}) :{\hskip 2em}  X[g] = G \frac{1}{\Box_{c}}R\frac{1}{\Box}\left(\frac{1}{3}R^2 - R_{\mu\nu} R^{\mu\nu}\right), \label{case 1}
\end{equation*}
\begin{equation*}
(\rom{3}) :{\hskip 2em}X[g] = G \frac{1}{\Box_{c}}\left(\frac{1}{3}R^2 - R_{\mu\nu} R^{\mu\nu}\right)\frac{1}{\Box}R,  \label{case 2}
\end{equation*}
\begin{equation*}
(\rom{4}) :{\hskip 2em}  X[g] = GR \frac{1}{\Box}\frac{1}{\Box_{c}}\left(\frac{1}{3}R^2 - R_{\mu\nu} R^{\mu\nu}\right),  \label{case 3}
\end{equation*}
\begin{equation*}
(\rom{5}) :{\hskip 2em} X[g] = G\left(\frac{1}{3}R^2 - R_{\mu\nu} R^{\mu\nu}\right) \frac{1}{\Box}\frac{1}{\Box_{c}}R,  \label{Case 4}
\end{equation*}
\begin{equation*}
(\rom{6}) :{\hskip 2em} X[g] = G \frac{1}{\Box}\frac{1}{\Box_{c}} R \left(\frac{1}{3}R^2 - R_{\mu\nu} R^{\mu\nu}\right), \label{Case 5}
\end{equation*}
\begin{equation*}
(\rom{7}) :{\hskip 2em} X[g] = G \frac{1}{\Box_{c}}\frac{1}{\Box} R \left(\frac{1}{3}R^2 - R_{\mu\nu} R^{\mu\nu}\right), \label{Case 6}
\end{equation*}
from equations (\ref{R}) - (\ref{Rmnrs}) we get 
\begin{equation}
\frac13 R^2 - R_{\mu\nu} R^{\mu\nu} 
\, = \,
12 ( 1 - \epsilon ) H^4 
\;\; . \label{sign}
\end{equation}
Equation (\ref{sign}) changes sign as $\epsilon $ passes through 1.\footnote{There are other possible combination of curvature invariants that can also  change their sign as we cross $ \epsilon = 1$. Some of them are $R^{2} - 3R_{\mu\nu\rho\sigma}R^{\mu\nu\rho\sigma}$ and $ R_{\mu\nu} R^{\mu\nu}-R_{\mu\nu\rho\sigma}R^{\mu\nu\rho\sigma}$} 

 In this Paper we will not work with (\rom{6}) and (\rom{7}) forms of $ X[g] $ since they give rise to unavoidable higher derivative  of $ \epsilon (> 2)$  in the background equation of motion. 
\section{Model \rom{1}} \label{sec1}
For completeness, here we repeat the whole analysis done by \cite{Tsamis:2016boj}. Form of $ X[g]$ considered in \cite{Tsamis:2016boj} is
\begin{equation}
X[g] = G \frac{1}{\Box}R\frac{1}{\Box_{c}}\left(\frac{1}{3}R^2 - R_{\mu\nu} R^{\mu\nu}\right). 
\end{equation}
By introducing two auxilary scalar A,C and two lagrange multiplier B, D, we write equivalent scalar tensor lagrangian \cite{Nojiri:2007uq,Tsamis:2014hra} as 
\begin{equation}
\begin{split}
\mathcal{L} = & \Lambda^{2}h(GC)\sqrt{-g} + B\left[\Box_{c}A -\left(\frac{1}{3}R^2 - R_{\mu\nu} R^{\mu\nu}\right)\right]\sqrt{-g} \\&+ D[\Box C -RA]\sqrt{-g} {\hskip 1em}.
\end{split}
\end{equation}
Varying equivalent lagrangian w.r.t auxilary and lagrange multiplier fields, we get 
\begin{align}
&\frac{1}{\sqrt{-g}}\frac{\delta(\mathcal{L})}{\delta B} = \Box_{c}A -\left(\frac{1}{3}R^2 - R_{\mu\nu} R^{\mu\nu}\right) = 0{\hskip 1em},\\& \frac{1}{\sqrt{-g}}\frac{\delta( \mathcal{L} )}{\delta D} = \Box C - RA = 0 {\hskip 1em}, 
\end{align}
\begin{align}
&\frac{1}{\sqrt{-g}}\frac{\delta(\mathcal{L})}{\delta A} = \Box_{c}B - RD = 0 {\hskip 1em}\Rightarrow B{\hskip 1em} = \frac{1}{\Box_c} RD {\hskip 1em}, \\& \frac{1}{\sqrt{-g}}\frac{\delta(\mathcal{L})}{\delta C} = \Box D + G \Lambda^{2}h^{'}(GC) = 0 {\hskip 1em} \Rightarrow {\hskip 1em} D  = - \frac{1}{\Box}G \Lambda^{2}h^{'}(GC) {\hskip 1em}. 
\end{align} 
 Equation of motion for fields for FRW metric canbe written in following forms 
\begin{eqnarray}
{\ddot A} &\!\! = \!\!&
- 3H{\dot A} - (2 - \epsilon) H^2 A
- 12 (1 - \epsilon) H^4
\;\; , \label{eomAM1} \\
{\ddot B} &\!\! = \!\!&
- 3H{\dot B} - (2 - \epsilon) H^2 B
- 6 (2 - \epsilon) H^2 D
\;\; , \\ \label{eomBM1}
{\ddot C} &\!\! = \!\!&
- 3H{\dot C} - 6 (2 - \epsilon) H^2 A
\;\; , \\ \label{eomCM1}
{\ddot D} &\!\! = \!\!&
- 3H{\dot D} + G \Lambda^2 h'(GC)
\;\; .  \label{eomDM1}
\end{eqnarray}

The $(00)$-component of Einstein field equations is 
\begin{equation}
\begin{split}
\frac{3H^2}{16\pi G} + \frac{1}{2}\Lambda^{2}h(GC)  & -\frac{1}{2}(\dot{A}\dot{B} + \dot{C}\dot{D}) -6H^{3}\dot{B}   \\- &3(H\partial_{t} + H^{2})(\frac{1}{6}AB + AD) = \frac{\Lambda}{16 \pi G}{\hskip 1em},
\end{split} \label{fe00m1}
\end{equation}
and $(11) $-component of Einstein field equation is
\begin{equation}
\begin{split}
&-(3-2\epsilon) \frac{H^{2}}{16 \pi G} - \frac{1}{2}\Lambda^{2}h(GC) + G\Lambda^{2}Ah'(GC) -\frac{1}{6}\dot{A}\dot{B} -\frac{1}{2}\dot{C}\dot{D} + 2\dot{A}\dot{D} \\& {\hskip 1em}-2(1+2\epsilon)H^{3}\dot{B} -2(3- 2\epsilon)H^{4}(B+6D) -(H\partial_{t} + H^{2})(\frac{1}{6}AB + AD)  \\&{\hskip 1em}= \frac{\Lambda}{16 \pi G}{\hskip 1em}.
\end{split}.  \label{fe11m1}
\end{equation}
 From equations (\ref{fe00m1}) and (\ref{fe11m1}), we find constraint equation for first slow roll parameter as
\begin{equation}
\begin{split}
 & \frac{2\epsilon H^{2}}{16 \pi G} + G\Lambda^{2}Ah'(GC)- \frac{2}{3}\dot{A}\dot{B} -\dot{C}\dot{D} + 2\dot{A}\dot{D} -4(2 + \epsilon)H^{3}\dot{B} \\& {\hskip 1em} -2(3-2\epsilon)H^{4}(B+ 6D) -4(H\partial_{t} + H^{2})(\frac{1}{6}AB + AD) = 0.
\end{split} \label{00+11m1}
\end{equation}
 In order to see the background evolution, we convert equations (\ref{eomAM1}) -(\ref{00+11m1}) into dimensionless form.  We define dimensionless time as
\begin{equation}
\tau \equiv  H_{in}t {\hskip 2em} \Rightarrow {\hskip 2em} \partial_{t} = H_{in}\partial_{\tau}. \label{dmlm1}
\end{equation}
 Other dimensionless quantities involving Hubble parameter and scalar fields can also be defined as 
\begin{align}
 & H^{2} \equiv \frac{\chi^{2}}{G}{\hskip 2em} \Rightarrow{\hskip 2em} H^{2}_{in} \equiv \frac{\chi^{2}_{in}}{G}{\hskip 2em} , {\hskip 2em} \Lambda \equiv 3H_{in}^{2} \equiv \frac{3\chi^{2}_{in}}{G}{\hskip 1em} , \\& A \equiv -\frac{3\alpha}{G}{\hskip 2em},{\hskip 2em} B  \equiv -3\beta {\hskip 2em},{\hskip 2em} C \equiv \frac{9\gamma}{G}{\hskip 2em}, {\hskip 2em} D \equiv \delta {\hskip 1em},\\& h(GC) = h(9\gamma)\equiv f(\gamma){\hskip 1em}.
\end{align}
 We wish to solve for $\left\lbrace\alpha ,\beta,\gamma,\delta,\chi,\epsilon\right\rbrace $ subjected to following initial conditions  at $ \tau = \tau_{in}$,
\begin{align}
 & \alpha = \alpha' = \beta = \gamma = \gamma^{'} = \delta^{'} = \delta = 0 ,\beta^{'} = 10 ,\\ & \chi = \chi_{in} , \epsilon = 0 .
\end{align} 
where $' = \frac{d}{d\tau} $ \paragraph{}
 Using these dimensionless variables, equation of motion for $\left\lbrace\alpha ,\beta,\gamma,\delta \right\rbrace $  can be cast as 
\begin{align}
 & \alpha^{''} + 3\frac{\chi}{\chi_{in}}\alpha^{'} + (2 -\epsilon)\frac{\chi^{2}}{\chi^{2}_{in}}\alpha = 4(1-\epsilon)\frac{\chi^{4}}{\chi^{2}_{in}} {\hskip 1em},\\&  \beta^{''} + 3\frac{\chi}{\chi_{in}}\beta^{'} + (2 -\epsilon)\frac{\chi^{2}}{\chi^{2}_{in}}\beta = 2(2-\epsilon)\frac{\chi^{2}}{\chi^{2}_{in}}\delta {\hskip 1em}, \\& \gamma^{''} + 3\frac{\chi}{\chi_{in}}\gamma{'} = 2(2-\epsilon)\frac{\chi^{2}}{\chi^{2}_{in}}\alpha {\hskip 1em},\\& \delta^{''} + 3\frac{\chi}{\chi_in} \delta^{'} = \chi_{in}^{2}f'(\gamma) {\hskip 1em},
\end{align}
Furthermore, the variable $ \chi $ is obtained  from equation (\ref{fe00m1})
\begin{equation}
\begin{split}
 \left[2\chi_{in}\beta^{'}\right]\chi^{3} & + \left[\frac{1}{3} -\frac{1}{2}\alpha \beta + \alpha \delta \right]\chi^{2} + \chi_{in}\partial_{t}\left[-- \frac{1}{2}\alpha \beta + \alpha \delta \right] \chi \\& -\chi^{2}_{in}\left[\frac{1}{3} -\frac{1}{2}\chi_{in}^{2}f(\gamma) + \frac{1}{2}(\alpha^{'}\beta^{'} + \gamma^{'}\delta^{'})\right] = 0{\hskip 1em},
\end{split}  \label{chim1}
\end{equation}
Moreover, we get equation for $\epsilon $ in terms of dimensionless variables from the  equation(\ref{00+11m1}) as
\begin{equation}
\begin{split}
 & \left[2\chi^{2}  + 12 \chi_{in}\chi^{3}\beta^{'} + 
12\chi^{4}(-\beta + 2\delta)\right]\epsilon   =
 3 \Bigg\{ \chi_{in}^{4}\alpha f'(\gamma) + \chi_{in}^{2}(2\alpha' \beta'\\& + 3\gamma'\delta' + 2\alpha'\delta') -8\chi_{in}\chi^{3}\beta' -6\chi^{4}(\beta -2\delta) + 4(\chi_{in}\chi\partial_{t} + \chi^{2})\left(\frac{1}{2}\alpha \beta -\alpha \delta \right) \Bigg\}
\end{split} \label{epsm1}
\end{equation}
Finally, we can further simplify equation (\ref{epsm1}) and we arrive at
\begin{equation}
\begin{split}
\epsilon\chi^{2} & = \frac{3}{2 + \beta'\chi_{in}\chi + 12(-\beta + 2\delta)\chi^{2}}\times \\& \Bigg\{\Big[ \alpha f'(\gamma) + 2f(\gamma)\Big]\chi_{in}^{4} -\frac{4}{3}\chi_{in}^{2} +\Big(2\alpha' +\gamma'\Big)\delta'\chi_{in}^{2} + \frac{4}{3}\chi^{2} + 6(-\beta + 2\delta)\chi^{4} \Bigg\}
\end{split}   \label{eps2m1}
\end{equation}
\subsection{Results}
\paragraph{}
 Assuming scale factor to take  its de Sitter  value $ a(\tau) = e^{{\tau}}$ during inflationary phase, all scalar fields decay as powers of $ e^{\tau} $ which can be seen from following equations \label{infm1}
\begin{eqnarray}
\alpha(\tau) &\!\! = \!\!& 
2 \chi_{\rm in}^2 \Bigl( 1 \!-\! e^{-\tau}\Bigr)^2 
\; \longrightarrow \; 
2 \chi_{\rm in}^2 
\; , \label{earlyalphaom} \\
\beta(\tau) &\!\! = \!\!& 
\frac23 \chi_{\rm in}^2 \Bigl( \tau \!-\! \frac{11}{6}
\!+\! 3 e^{-\tau} \!-\! \frac32 e^{-2\tau} 
\!+\! \frac13 e^{-3\tau} \Bigr) 
\; \longrightarrow \;
\frac23 \chi_{\rm in}^2 \Bigl( \tau \!-\! \frac{11}{6} \Bigr) 
\; , \label{earlybetaom} \\
\gamma(\tau) &\!\! = \!\!& 
\frac83 \chi_{\rm in}^2 \Bigl( \tau \!-\! \frac{11}{6}
\!+\! 3 e^{-\tau} \!-\! \frac32 e^{-2\tau} 
\!+\! \frac13 e^{-3\tau} \Bigr) 
\; \longrightarrow \;
\frac83 \chi_{\rm in}^2 \Bigl( \tau \!-\! \frac{11}{6}
\Bigr) \; , \label{earlygammaom} \\
\delta(\tau) &\!\! = \!\!& 
\frac13 \chi_{\rm in}^2 \Bigl( \tau \!-\! \frac{1}{3}
\!+\! \frac{1}{3} e^{-3\tau} \Bigr) 
\; \longrightarrow \; 
\frac13 \chi_{\rm in}^2 
\Bigl( \tau \!-\! \frac{1}{3} \Bigr) 
\; . \label{earlydeltaom}
\end{eqnarray}
\begin{figure}[!ht]
  \centering
   \includegraphics[scale =0.4]{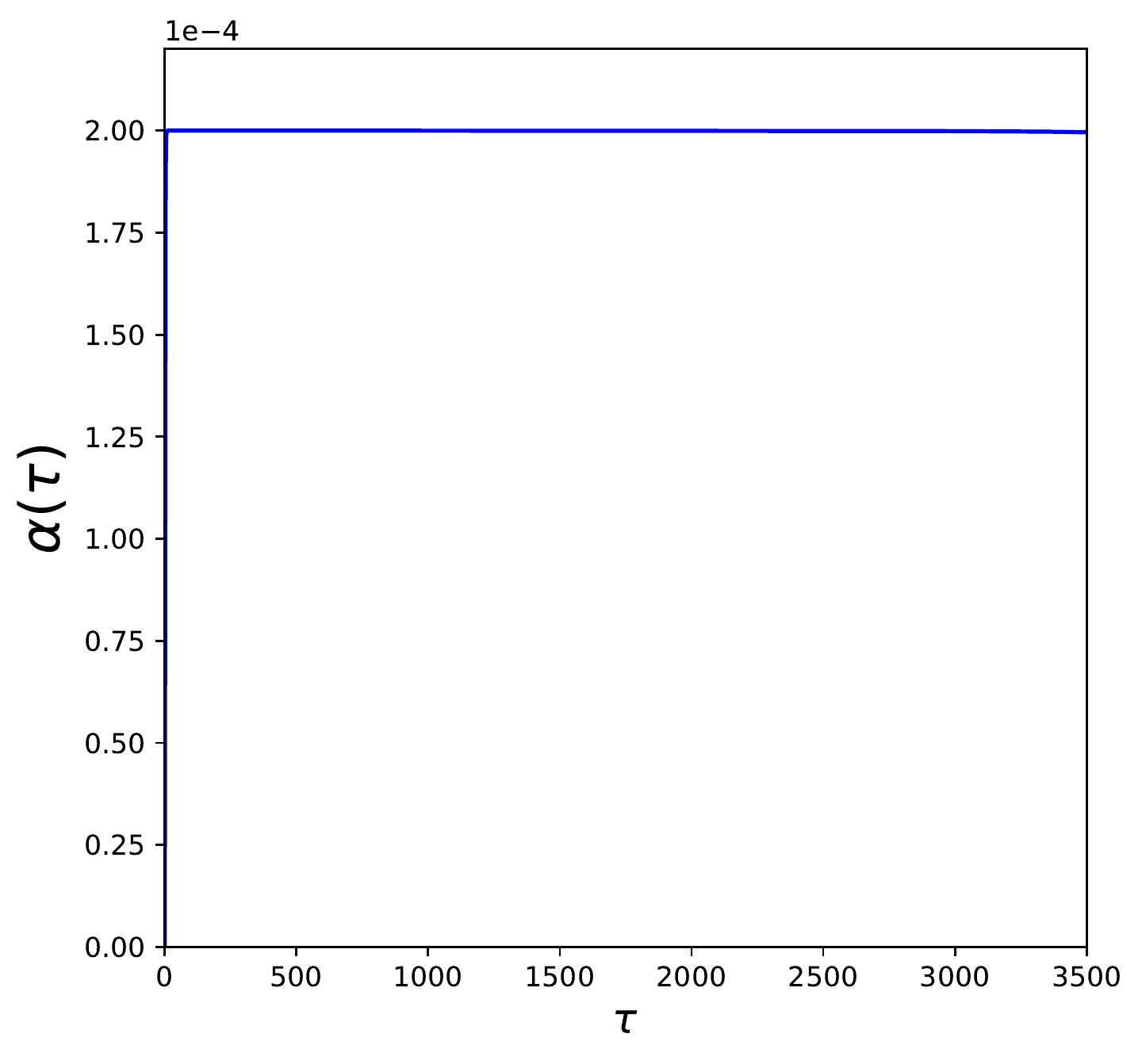}  {\hskip 2em}
  \includegraphics[scale =0.4]{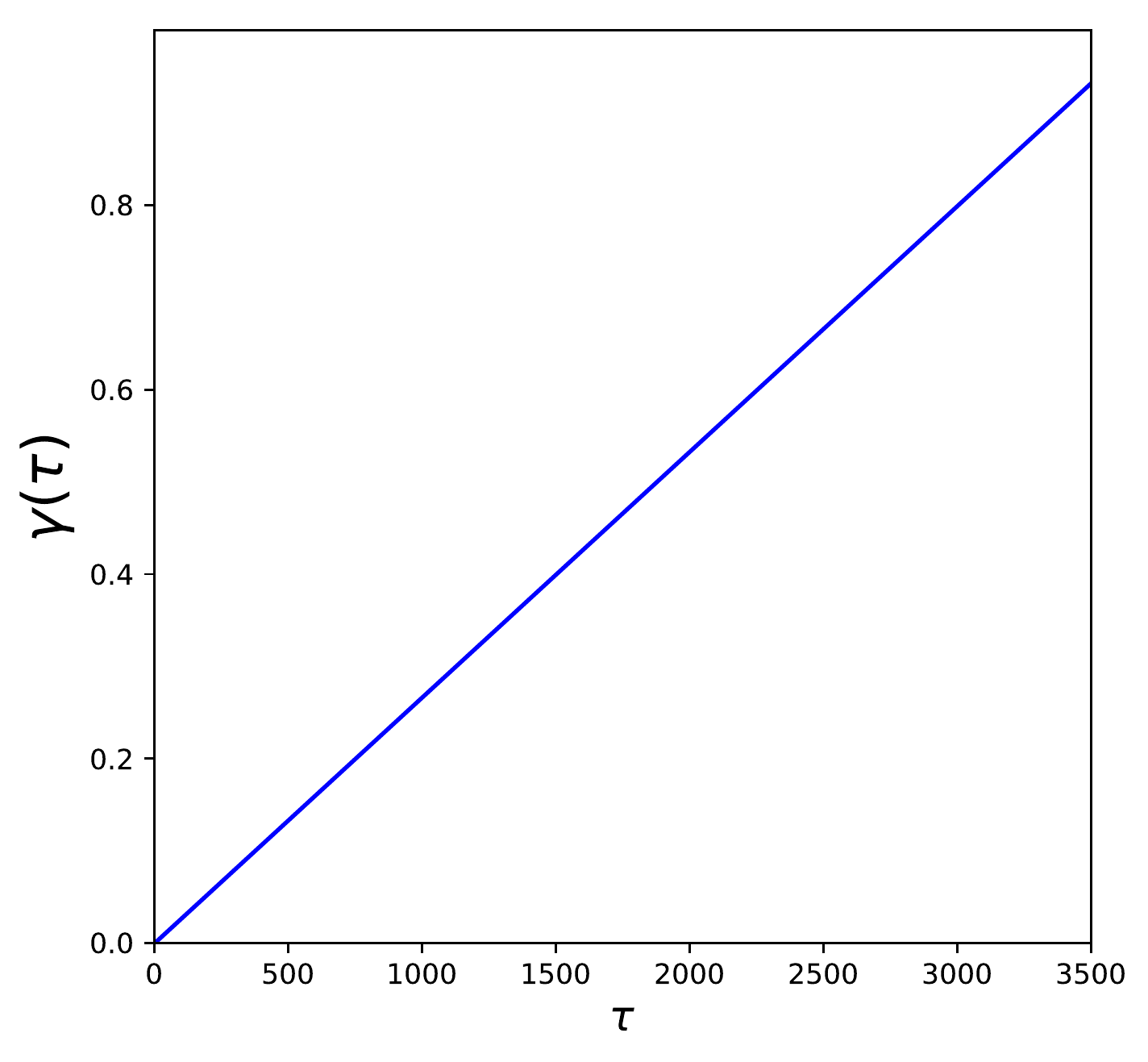}  
\caption[Plots for scalar field $\alpha(\tau)$ and $\gamma(\tau)$ for case 1]{Plots for scalar field $\alpha(\tau)$ and $\gamma(\tau)$ for $\chi_{in} = \frac{1}{100}$ and $ f(X) = \frac{X}{1-X} $.}  \label{earlyalpha}
\end{figure}

\begin{figure}[!ht]
  \centering
     \includegraphics[scale =0.4]{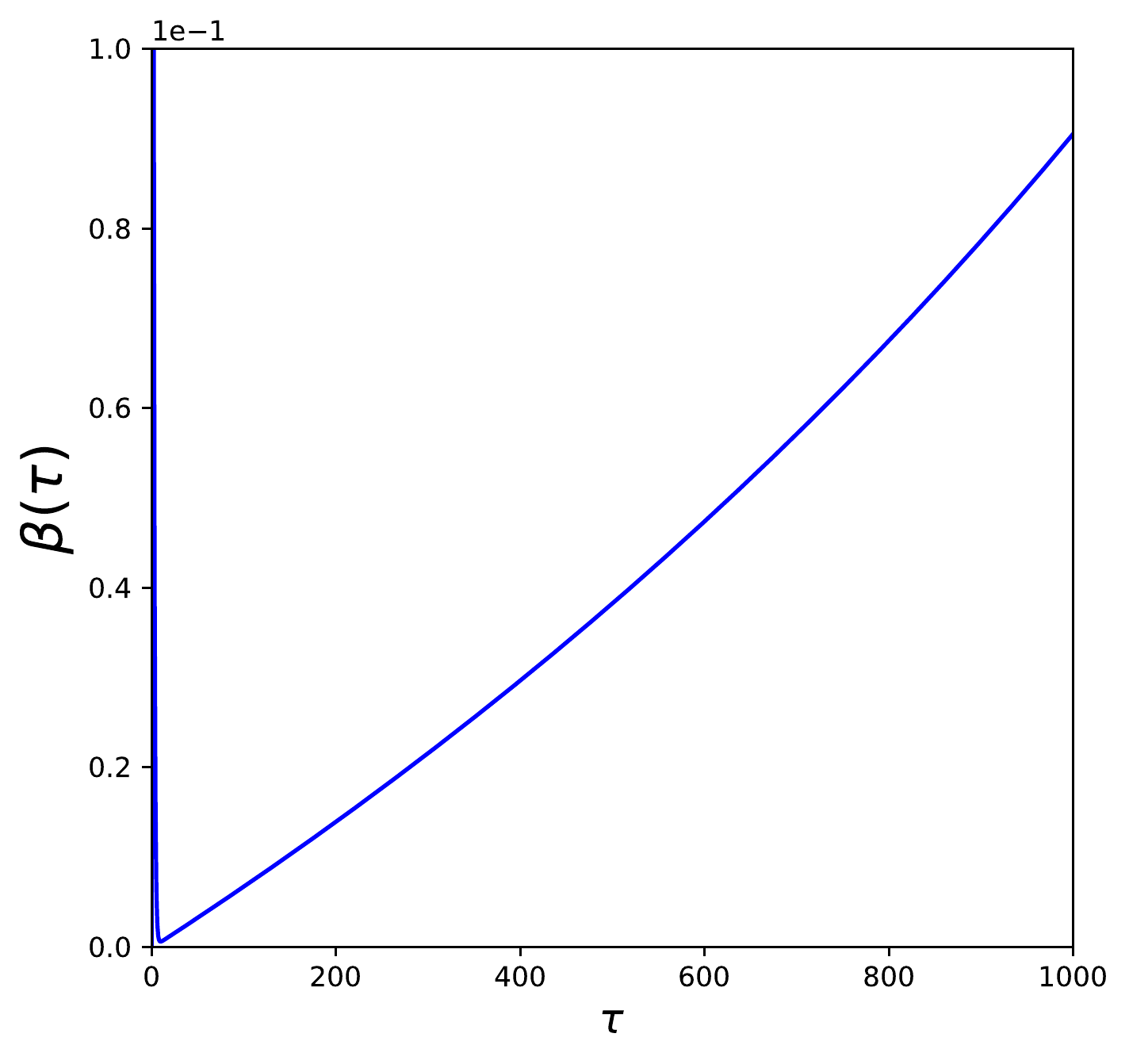}  {\hskip 2em}
  \includegraphics[scale =0.4]{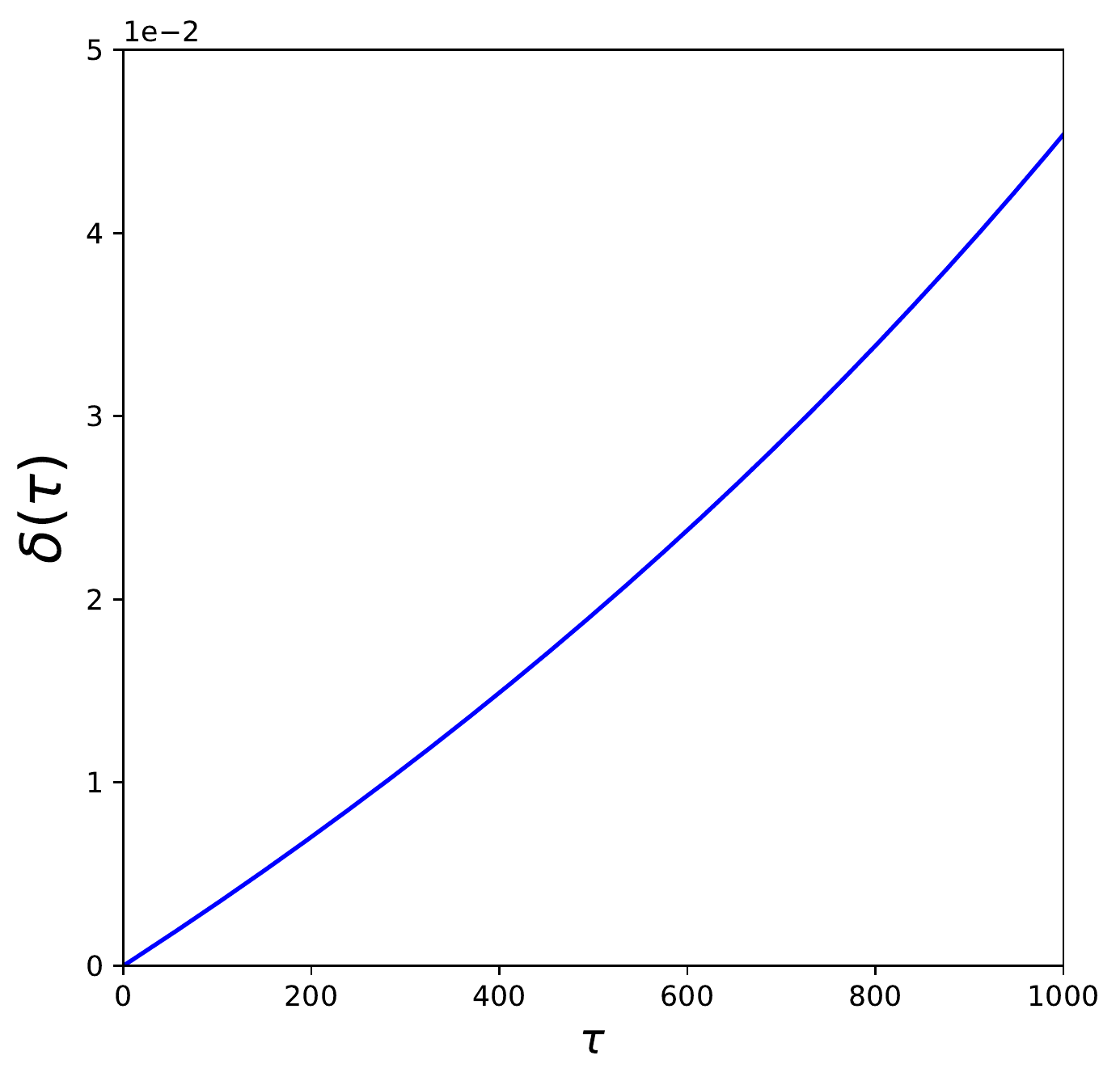} 
\caption[Plots for scalar field $\beta(\tau)$ and $\delta(\tau)$ for case 1]{Plots for scalar field $\beta(\tau)$ and $\delta(\tau)$ for $\chi_{in} = \frac{1}{100}$ and $ f(X) = \frac{X}{1-X} $.}   \label{earlybeta}
\end{figure}

\paragraph{}
Furthermore, using equations (\ref{earlyalphaom})-(\ref{earlydeltaom}) derivatives of scalar fields during this epoch take the form 
 \begin{equation}
 \alpha'(\tau) \rightarrow 0 {\hskip 1em},{\hskip 1em} \beta'(\tau) \rightarrow \frac{2}{3}\chi_{in}^{2}{\hskip 1em}, {\hskip 1em}\gamma'(\tau) \rightarrow \frac{8}{3}\chi_{in}^{2}{\hskip 1em},{\hskip 1em} \delta'(\tau)\rightarrow \frac{1}{3}\chi_{in}^{2}{\hskip 1em},
 \end{equation}
  It is easy to obtain the values for the Hubble parameter and the slow roll parameter \cite{Tsamis:2016boj} as 
  \begin{equation}
  \chi(\tau) \rightarrow \chi_{in}(1- 2\chi_{in}^{4}\tau){\hskip 2em},{\hskip 2em} \epsilon(\tau) \rightarrow 2\chi_{in}^{4} {\hskip 1em}. \label{hm1}
\end{equation}  

\begin{figure}[!ht]
  \centering
    \includegraphics[scale =0.4]{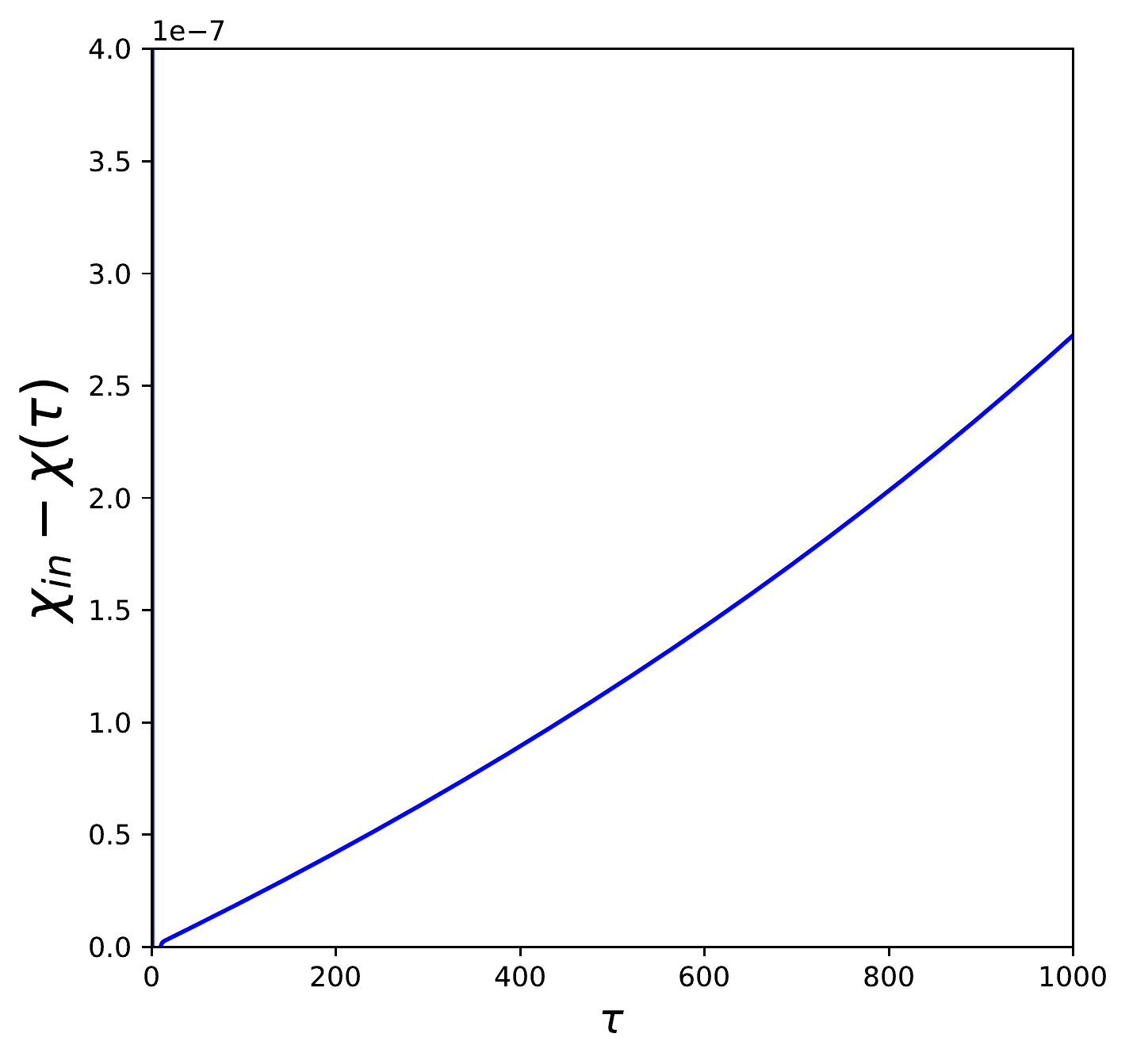}  {\hskip 2em}
   \includegraphics[scale =0.4]{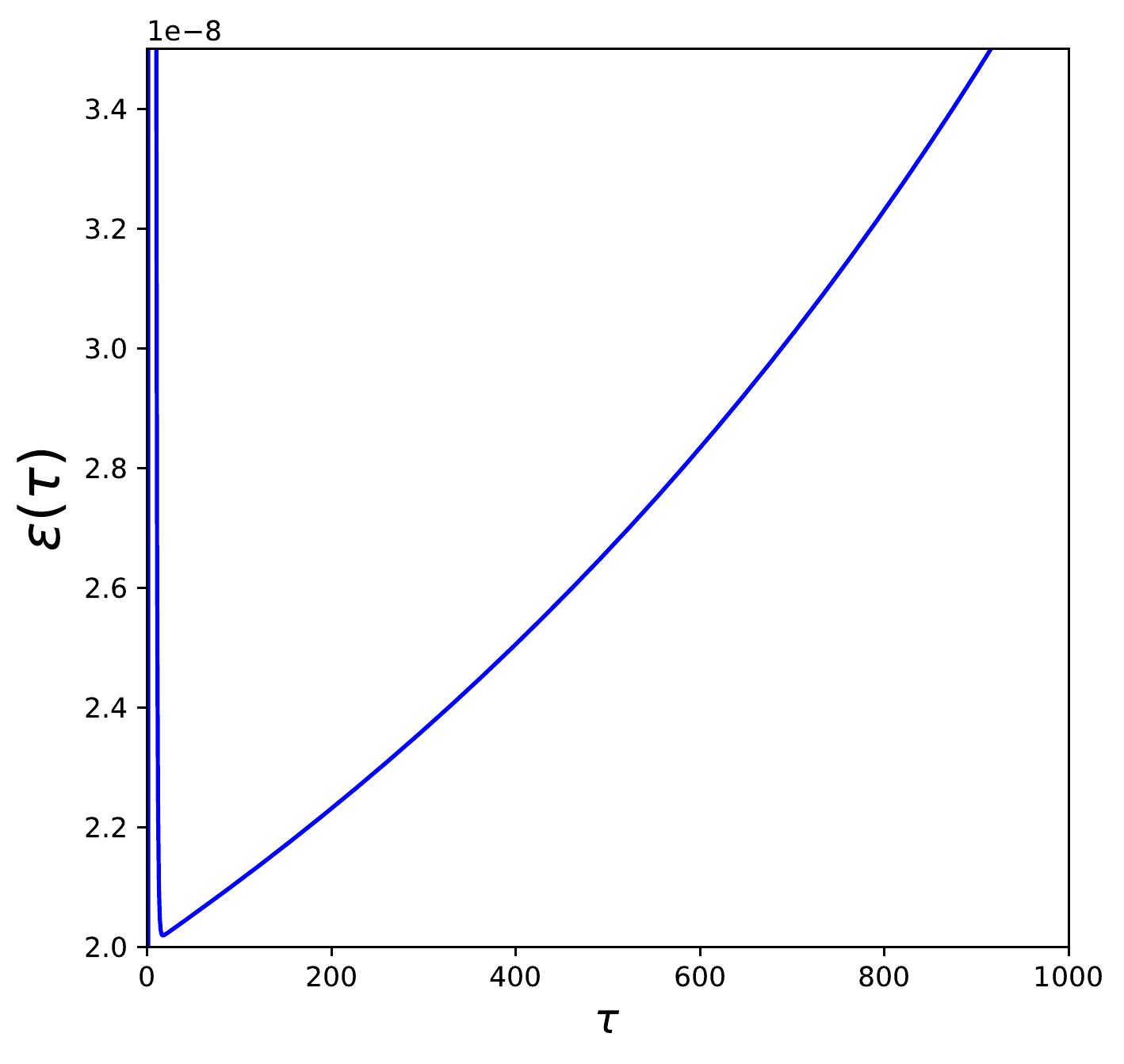} 
\caption[Plots for $\chi_{in}- \chi(\tau)$ and $\epsilon(\tau)$ for case 1]{Plots for geometrical quantities $\chi_{in}- \chi(\tau)$ and $\epsilon(\tau)$ for $\chi_{in} = \frac{1}{100}$ and $ f(X) = \frac{X}{1-X} $.}   \label{chiom}
\end{figure}
\begin{figure}[!ht]
     \includegraphics[scale =0.4]{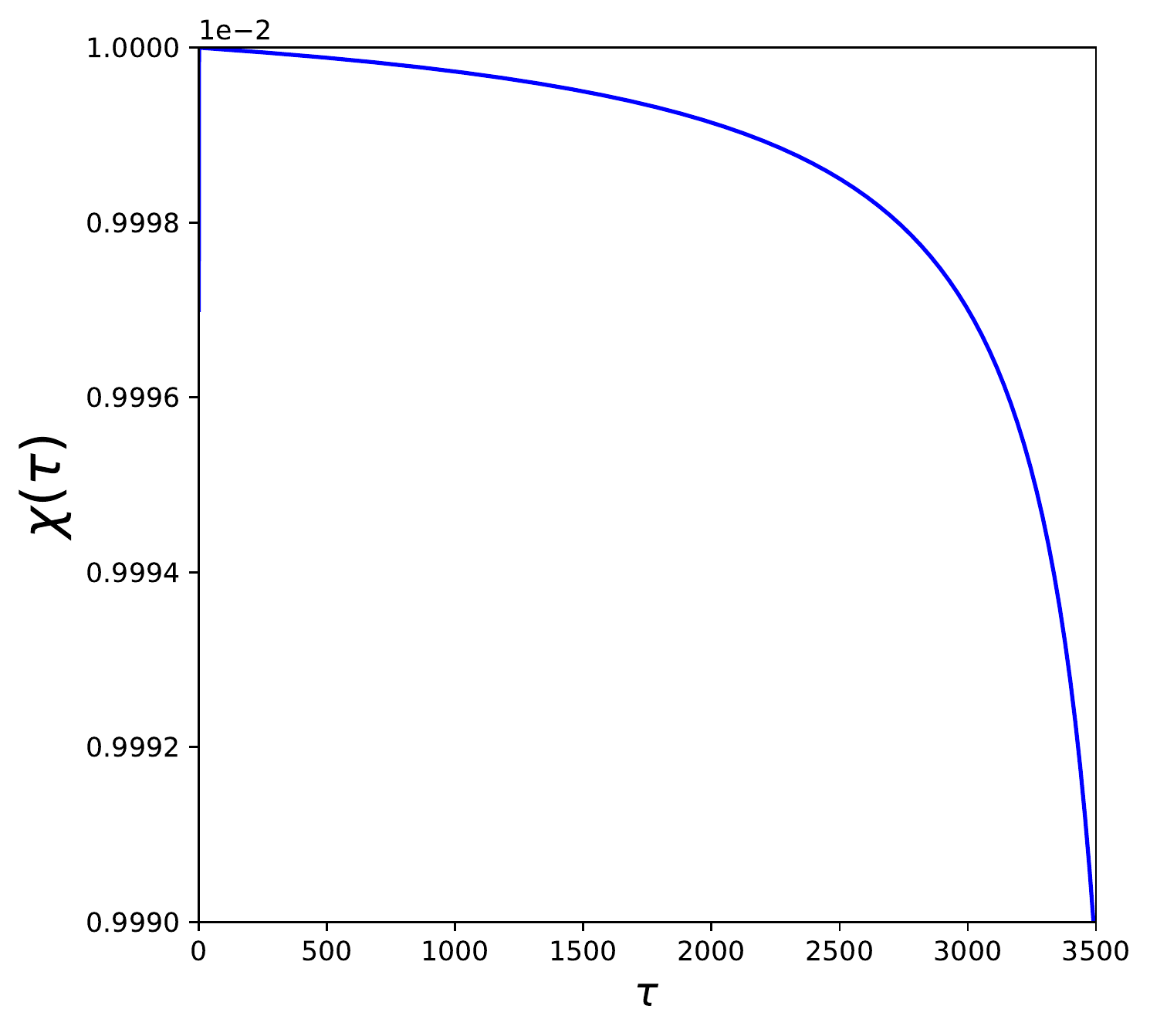}  {\hskip 2em}
  \includegraphics[scale =0.4]{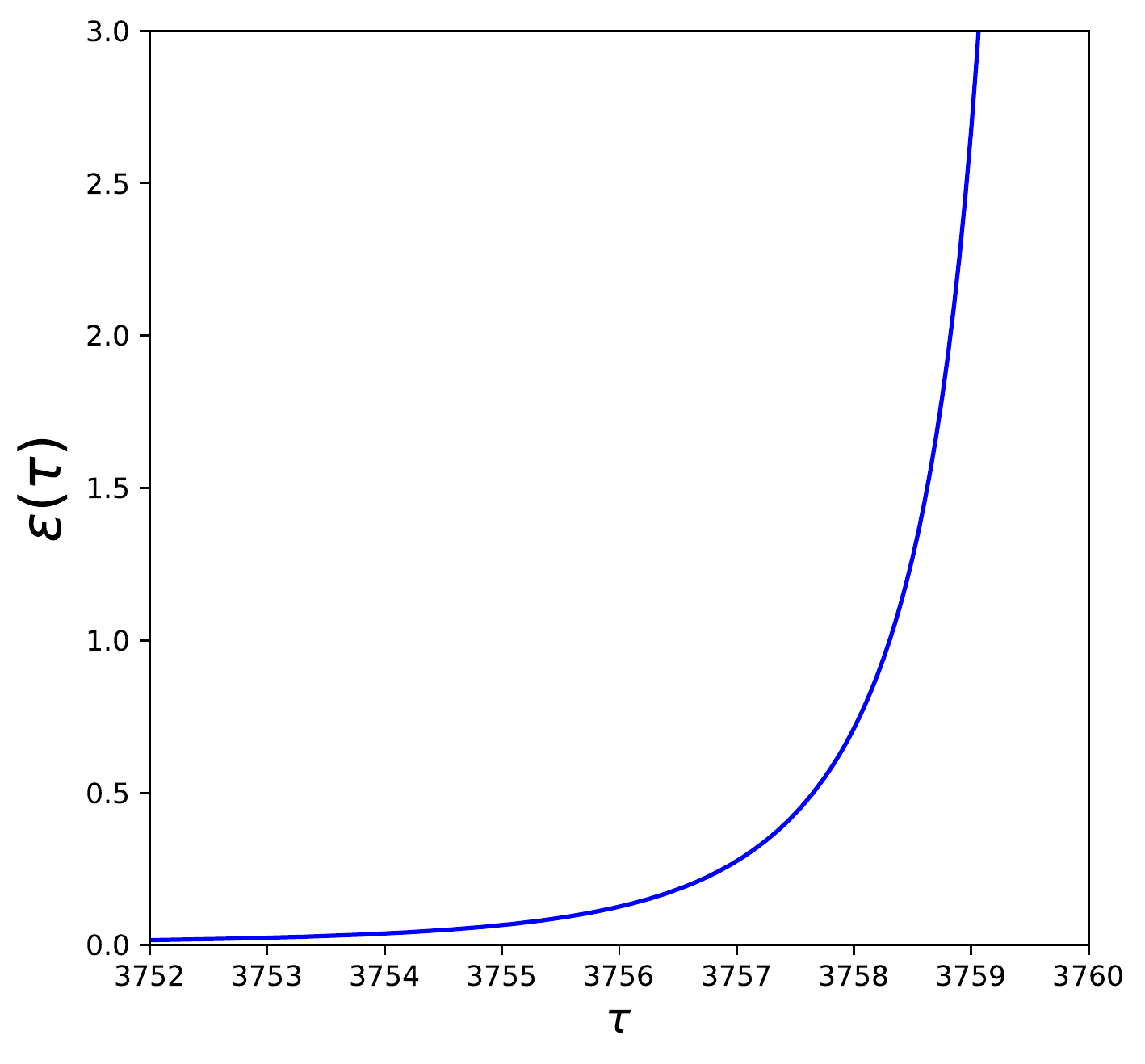}   
\caption[Plots Hubble parameter and the first  slow roll parameter  ($\epsilon$) for case 1]{Plots Hubble parameter and the first  slow roll parameter  ($\epsilon$) for $\chi_{in} = \frac{1}{100}$ and $ f(X) = \frac{X}{1-X} $. } \label{epiom}
\end{figure}

\paragraph{}
 Figure \ref{earlyalpha} and \ref{earlybeta} show well agreement between numerical and analytical result for values of $ \tau = $ 0 to 3500 and $ \tau = $ 0 to 1000 respectively. Figure \ref{chiom} shows the compatibility between the expression given in equation (\ref{hm1}) and numerical result for $\chi_{in}- \chi(t)$ and $ \epsilon (t)$.
 \begin{figure}[!ht]
  \centering
    \includegraphics[scale =0.4]{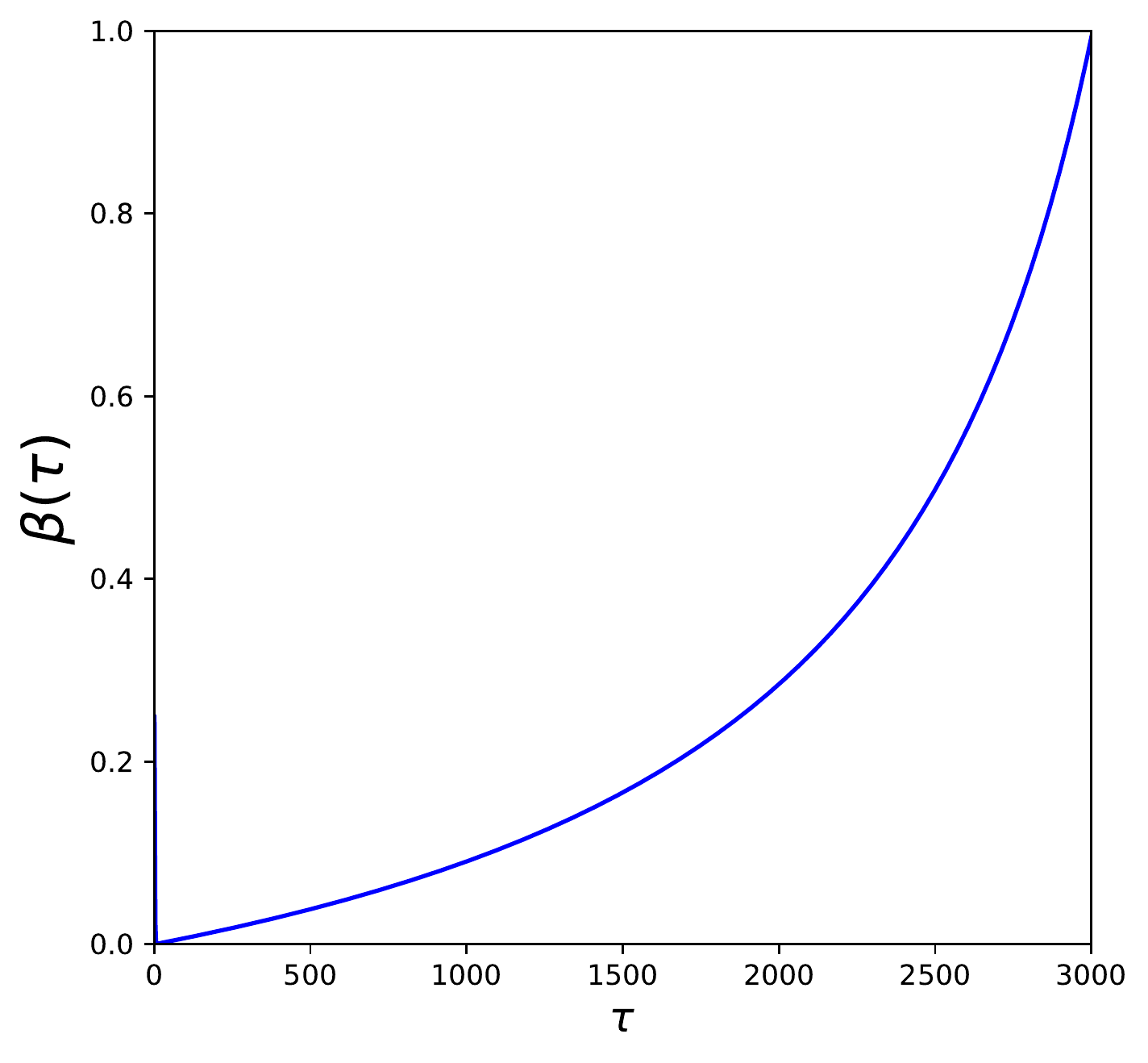}  {\hskip 2em}
  \includegraphics[scale =0.4]{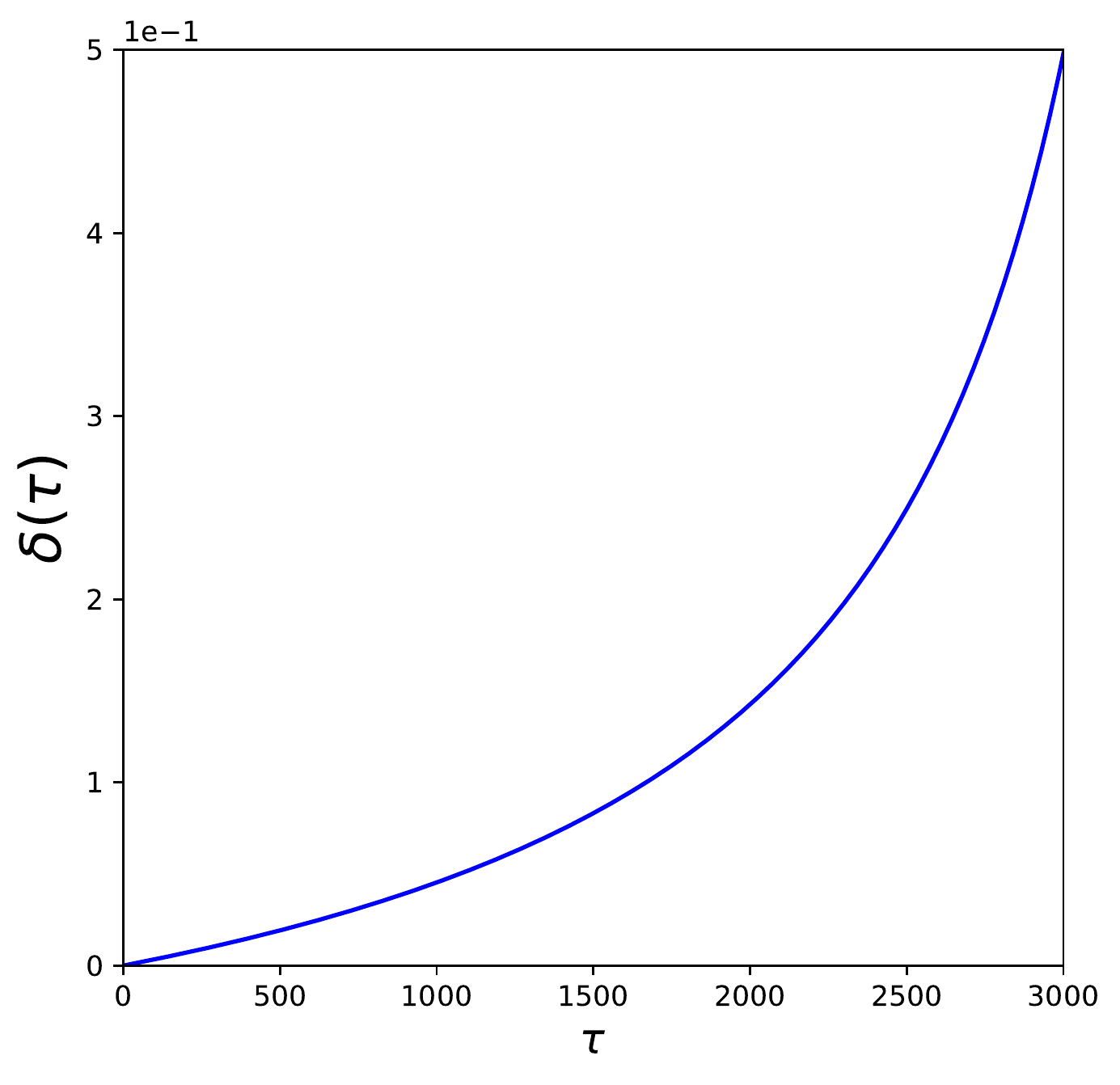}   
\caption[Plots for scalar fields $\beta(\tau)$ and $\delta (\tau)$ for case 1]{Plots for scalar fields $\beta(\tau)$ and $\delta (\tau)$ for $\chi_{in} = \frac{1}{100}$ and $ f(X) = \frac{X}{1-X} $. }   \label{betalateom}
\end{figure}
In  Figure \ref{betalateom} the curvature appears from coupling between auxiliary scalars which are small during the deSitter epoch . First coupling is of $\delta (\tau)$ to $ \gamma(\tau) $.Due to this coupling  $ \chi_{in}^{4}$ correction  for $\delta(\tau)$ in (\ref{earlygammaom}) is given by
 \begin{equation}
 \delta(\tau) \rightarrow \frac{1}{3}\chi_{in}^{2}\tau + \frac{8}{9} \chi_{in}^{4}\tau^{2}\hspace{0.3cm}.
 \end{equation}
 The curvature of $\beta (\tau)$ can be inferred from evolution of  $\delta(\tau)$,
 it can be easily observed that  $ \chi_{in}^{4}$ correction to $\beta(\tau)$ is 
 \begin{equation}
 \beta(\tau) \rightarrow \frac{2}{3}\chi_{in}^{2}\tau + \frac{16}{9}\chi_{in}^{4}\tau^{2}\hspace{0.3 cm}.
 \end{equation}
\paragraph{} The evolution of $ \epsilon (\tau) $ is affected by the curvature of $\beta(\tau) $ and $ \delta (\tau)$ which is evident in Figure \ref{epiom}. In turn, Hubble parameter decays much faster due to the curvature effects.
\paragraph{}
In this model, inflation ends around $\tau \simeq 3750$ and transit to a RD epoch as $\epsilon $ rises above 1. After that there is no mechanism which can prevent further rise in $\epsilon $ and crosses over to Matter dominated or late-time acceleration epoch in this model. 
\section{ Model \rom{2}} \label{sec2}
 Here $ X[g] = G \frac{1}{\Box_{c}}R\frac{1}{\Box}\left(\frac{1}{3}R^2 - R_{\mu\nu} R^{\mu\nu}\right) $.
The  equivalent Scalar tensor Lagrangian \cite{Nojiri:2007uq,Tsamis:2014hra} of action(\ref{main action}) 
\begin{equation}
\begin{split}
\mathcal{L} = & \Lambda^{2}h(GC)\sqrt{-g} + B\left[\Box A -\left(\frac{1}{3}R^2 - R_{\mu\nu} R^{\mu\nu}\right)\right]\sqrt{-g} \\&+ D[\Box_{c} C -RA]\sqrt{-g} {\hskip 1em},
\end{split}
\end{equation}
where A and C are auxilary scalar fields and B and D are lagrange multiplier in the action. \newline Varying the action w.r.t both auxilary and lagrange multiplier fields, we get
\begin{align}
&\frac{1}{\sqrt{-g}}\frac{\delta(\mathcal{L})}{\delta B} = \Box A -\left(\frac{1}{3}R^2 - R_{\mu\nu} R^{\mu\nu}\right) = 0{\hskip 1em},\\& \frac{1}{\sqrt{-g}}\frac{\delta(\mathcal{L})}{\delta D} = \Box_{c} C - RA = 0 {\hskip 1em}, 
\end{align}
\begin{align}
&\frac{1}{\sqrt{-g}}\frac{\delta(\mathcal{L})}{\delta A} = \Box B - RD = 0 {\hskip 1em}\Rightarrow B{\hskip 1em} = \frac{1}{\Box} RD {\hskip 1em}, \\& \frac{1}{\sqrt{-g}}\frac{\delta(\mathcal{L})}{\delta C} = \Box_{c} D + G \Lambda^{2}h^{'}(GC) = 0 {\hskip 1em} \Rightarrow {\hskip 1em} D  = - \frac{1}{\Box_{c}}G \Lambda^{2}h^{'}(GC) {\hskip 1em}, 
\end{align} 
The equation of motion for fields (A,B,C,D) for FRW metric become

\begin{eqnarray}
{\ddot A} &\!\! = \!\!&
- 3H{\dot A} 
- 12 (1 - \epsilon) H^4
\;\; , \label{eomAM2} \\
{\ddot B} &\!\! = \!\!&
- 3H{\dot B} 
- 6 (2 - \epsilon) H^2 D
\;\; , \\ \label{eomBM2}
{\ddot C} &\!\! = \!\!&
- 3H{\dot C} -(2-\epsilon)H^{2}C - 6 (2 - \epsilon) H^2 A
\;\; , \\ \label{eomCM2}
{\ddot D} &\!\! = \!\!&
- 3H{\dot D} + G \Lambda^2 h'(GC)-(2-\epsilon)H^{2}D
\;\; .  \label{eomDM2}
\end{eqnarray}

The $(00)$-component of Einstein field equations becomes
\begin{equation}
\begin{split}
\frac{3H^2}{16\pi G} + \frac{1}{2}\Lambda^{2}h(GC)  & -\frac{1}{2}(\dot{A}\dot{B} + \dot{C}\dot{D}) -6H^{3}\dot{B}   \\- &3(H\partial_{t} + H^{2})(\frac{1}{6}CD + AD) = \frac{\Lambda}{16 \pi G}{\hskip 1em}, \label{00m2}
\end{split}
\end{equation}
and $(11) $-component of Einstein equation provides
\begin{equation}
\begin{split}
&-(3-2\epsilon) \frac{H^{2}}{16 \pi G} - \frac{1}{2}\Lambda^{2}h(GC) + G\Lambda^{2}Ah'(GC)\Big( A + \frac{1}{6}C \Big) -\frac{1}{6}\dot{C}\dot{D} \\& -\frac{1}{2}\dot{A}\dot{B} + 2\dot{A}\dot{D}  {\hskip 1em}-2(1+2\epsilon)H^{3}\dot{B} -12(3- 2\epsilon)H^{4}D \\& -(H\partial_{t} + H^{2})(\frac{1}{6}CD + AD)  = - \frac{\Lambda}{16 \pi G}{\hskip 1em},
\end{split}   \label{11m2}
\end{equation}
Adding equation(\ref{00m2}) and (\ref{11m2}), we get
\begin{equation}
\begin{split}
 & \frac{2\epsilon H^{2}}{16 \pi G} + G\Lambda^{2}Ah'(GC)\Big(A + \frac{C}{6}\Big)- \frac{2}{3}\dot{C}\dot{D} -\dot{A}\dot{B} + 2\dot{A}\dot{D} -4(2 + \epsilon)H^{3}\dot{B} \\& {\hskip 1em} -12(3-2\epsilon)H^{4}D -4(H\partial_{t} + H^{2})(\frac{1}{6}CD + AD) = 0.
\end{split}  \label{00+11m2}
\end{equation}
In order to study the cosmological dynamics of this model, we use same set of dimensional parameters as Model \rom{1}
\begin{align}
 & H^{2} \equiv \frac{\chi^{2}}{G}{\hskip 2em} \Rightarrow{\hskip 2em} H^{2}_{in} \equiv \frac{\chi^{2}_{in}}{G}{\hskip 2em} , {\hskip 2em} \Lambda \equiv 3H_{in}^{2} \equiv \frac{3\chi^{2}_{in}}{G}{\hskip 1em} , \\& A \equiv -\frac{3\alpha}{G}{\hskip 2em},{\hskip 2em} B  \equiv -3\beta {\hskip 2em},{\hskip 2em} C \equiv \frac{9\gamma}{G}{\hskip 2em}, {\hskip 2em} D \equiv \delta {\hskip 1em},\\& h(GC) = h(9\gamma)\equiv f(\gamma){\hskip 1em}.
\end{align}
The set of dimensionles parameters we wish to solve for is $\left\lbrace\alpha ,\beta,\gamma,\delta,\chi,\epsilon\right\rbrace $ and is subjected to following initial conditions  at $ \tau = \tau_{in}$ :
\begin{align}
 & \alpha = \alpha' = \beta = \gamma = \gamma^{'} = \delta^{'} = \delta = 0 ,\beta^{'} = 1.0 {\hskip 2em},\\ & \chi = \chi_{in}, \epsilon = 0 {\hskip 1em}.
\end{align} 

Now we write (\ref{eomAM2})-(\ref{eomDM2}) in terms of dimensionless parameters 
\begin{align}
 & \alpha^{''} + 3\frac{\chi}{\chi_{in}}\alpha^{'}  = 4(1-\epsilon)\frac{\chi^{4}}{\chi^{2}_{in}} {\hskip 1em},\\&  \beta^{''} + 3\frac{\chi}{\chi_{in}}\beta^{'}  = 2(2-\epsilon)\frac{\chi^{2}}{\chi^{2}_{in}}\delta {\hskip 1em}, \\& \gamma^{''} + 3\frac{\chi}{\chi_{in}}\gamma{'}+(2-\epsilon)\frac{\chi^{2}}{\chi_{in}^{2}}\gamma = 2(2-\epsilon)\frac{\chi^{2}}{\chi^{2}_{in}}\alpha {\hskip 1em},\\& \delta^{''} + 3\frac{\chi}{\chi_{in}} \delta^{'} +(2-\epsilon)\frac{\chi^{2}}{\chi_{in}^{2}}\delta = \chi_{in}^{2}f'(\gamma) {\hskip 1em},\label{eomdeltam1}
\end{align}
Furthermore, the variable $ \chi $ is obtained from the constraint equation (\ref{00m2}) which takes the following form
\begin{equation}
\begin{split}
 \left[2\chi_{in}\beta^{'}\right]\chi^{3} & + \left[\frac{1}{3} -\frac{1}{2}\gamma \delta + \alpha \delta \right]\chi^{2} + \chi_{in}\partial_{t}\left[- \frac{1}{2}\gamma \delta + \alpha \delta \right] \chi \\& -\chi^{2}_{in}\left[\frac{1}{3} -\frac{1}{2}\chi_{in}^{2}f(\gamma) + \frac{1}{2}(\alpha^{'}\beta^{'} + \gamma^{'}\delta^{'})\right] = 0{\hskip 1em}.
\end{split}
\end{equation}
Then, we solve  $\epsilon $ in terms of dimensionless parameters from the equation (\ref{00+11m2})
\begin{equation}
\begin{split}
 &\left[2\chi^{2} +   12 \chi_{in}\chi^{3}\beta^{'} + 
24\chi^{4} \delta \right]\epsilon =
 3 \Bigg\{ \chi_{in}^{4} f'(\gamma)(\alpha  - \frac{1}{2}\gamma) + \chi_{in}^{2}(3\alpha' \beta' + 2\gamma'\delta'  \\&+ 2\alpha'\delta') + 12\delta\chi^{4} -8\chi_{in}\chi^{3}\beta' + 4(\chi_{in}\chi\partial_{t} + \chi^{2})\left(\frac{1}{2}\gamma \delta -\alpha \delta \right) \Bigg\} \label{eps2m2}
\end{split}
\end{equation}
Finally, we can further simplify equation (\ref{eps2m2}) and we arrive at:
\begin{equation}
\begin{split}
\epsilon\chi^{2} & = \frac{3}{2 + 12 \beta'\chi_{in}\chi + 24 \delta \chi^{2}}\times \\& \Bigg\{\Big[ f'(\gamma)(\alpha  -\frac{1}{2}\gamma) + 2f(\gamma)\Big]\chi_{in}^{4} -\frac{4}{3}\chi_{in}^{2} +\Big(\beta' +2\delta'\Big)\delta'\chi_{in}^{2} + \frac{4}{3}\chi^{2} + 12 \delta \chi^{4} \Bigg\}.
\end{split}
\end{equation}
{\vskip 2em} 
{\hskip 2em}
\subsection{Results}
For a long time period $(\tau)$, performing similar calculations as in (\ref{infm1}), scalar fields evolve as follows
\begin{eqnarray}
\alpha(\tau) &\!\! = \!\!& 
 \frac{4}{3}\chi_{in}^{2}\Big(\tau + \frac{e^{-3\tau}}{3}-\frac{1}{3}\Big)  \longrightarrow \frac{4}{3}\chi_{in}^{2} \Big(\tau -\frac{1}{3}\Big)
\; 
\; , \label{earlyalpham1} \\
\beta(\tau) &\!\! = \!\!& 
\frac43 \chi_{\rm in}^2 \Bigl( \tau \!-\! \frac{11}{6}
\!+\! 3 e^{-\tau} \!-\! \frac32 e^{-2\tau} 
\!-\! \frac13 e^{-3\tau} \Bigr) 
\; \longrightarrow \;
\frac43 \chi_{\rm in}^2 \Bigl( \tau \!-\! \frac{11}{6} \Bigr) 
\; , \label{earlybetam1} \\
\gamma(\tau) &\!\! = \!\!& 
\frac83 \chi_{\rm in}^2 \Bigl( \tau \!-\! \frac{11}{6}
\!+\! 3 e^{-\tau} \!-\! \frac32 e^{-2\tau} 
\!-\! \frac13 e^{-3\tau} \Bigr) 
\; \longrightarrow \;
\frac83 \chi_{\rm in}^2 \Bigl( \tau \!-\! \frac{11}{6}
\Bigr) \; , \label{earlygammam1} \\
\delta(\tau) &\!\! = \!\!& 
\frac{1}{2}\chi_{in}^{2}\bigg(1-e^{-\tau}\bigg)^{2} \longrightarrow \frac{1}{2}\chi_{in}^{2}
\; . \label{earlydeltam1}
\end{eqnarray}
\begin{figure}[!ht]
  \centering
   \includegraphics[scale =0.4]{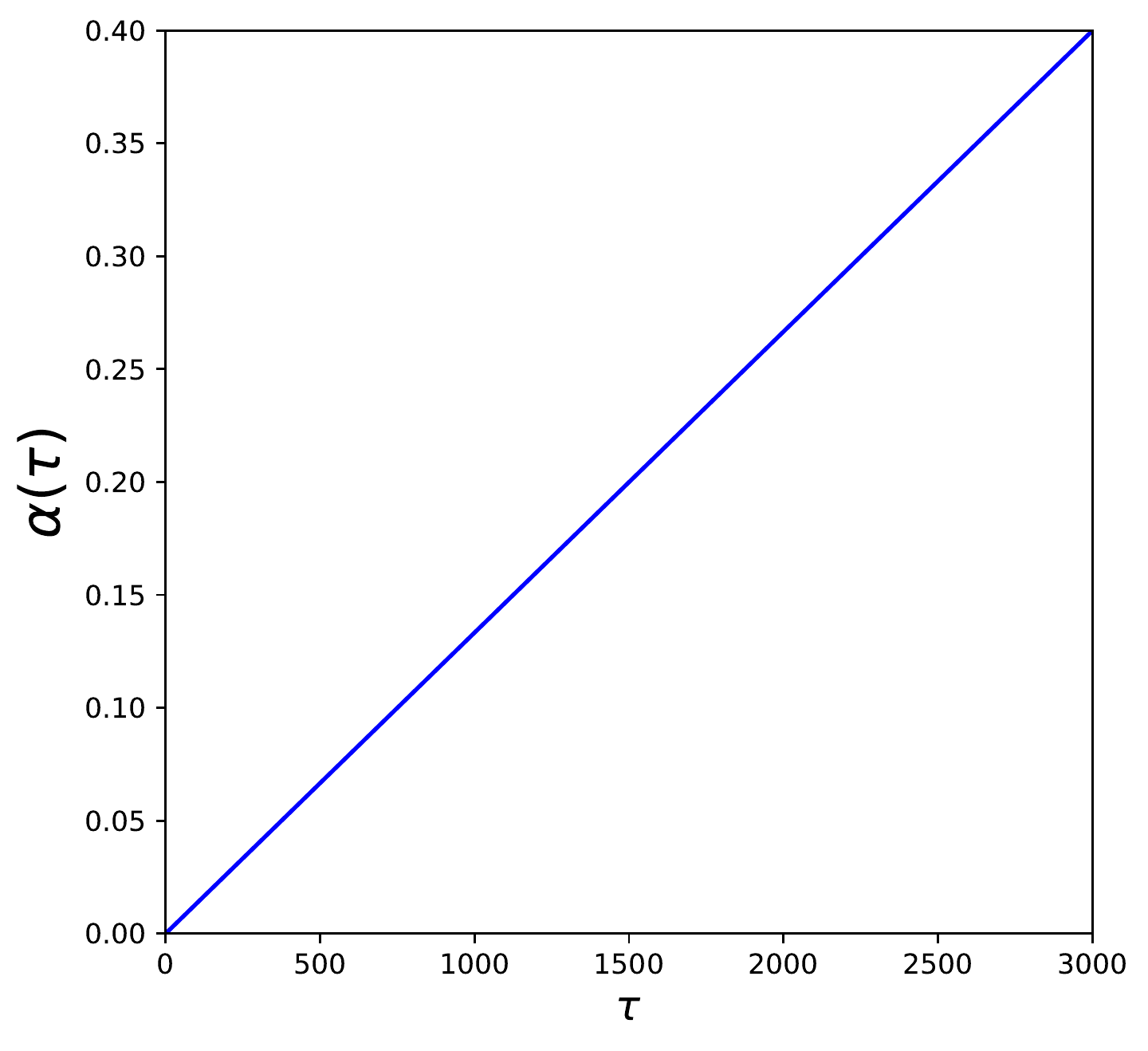}  {\hskip 2em}
  \includegraphics[scale =0.4]{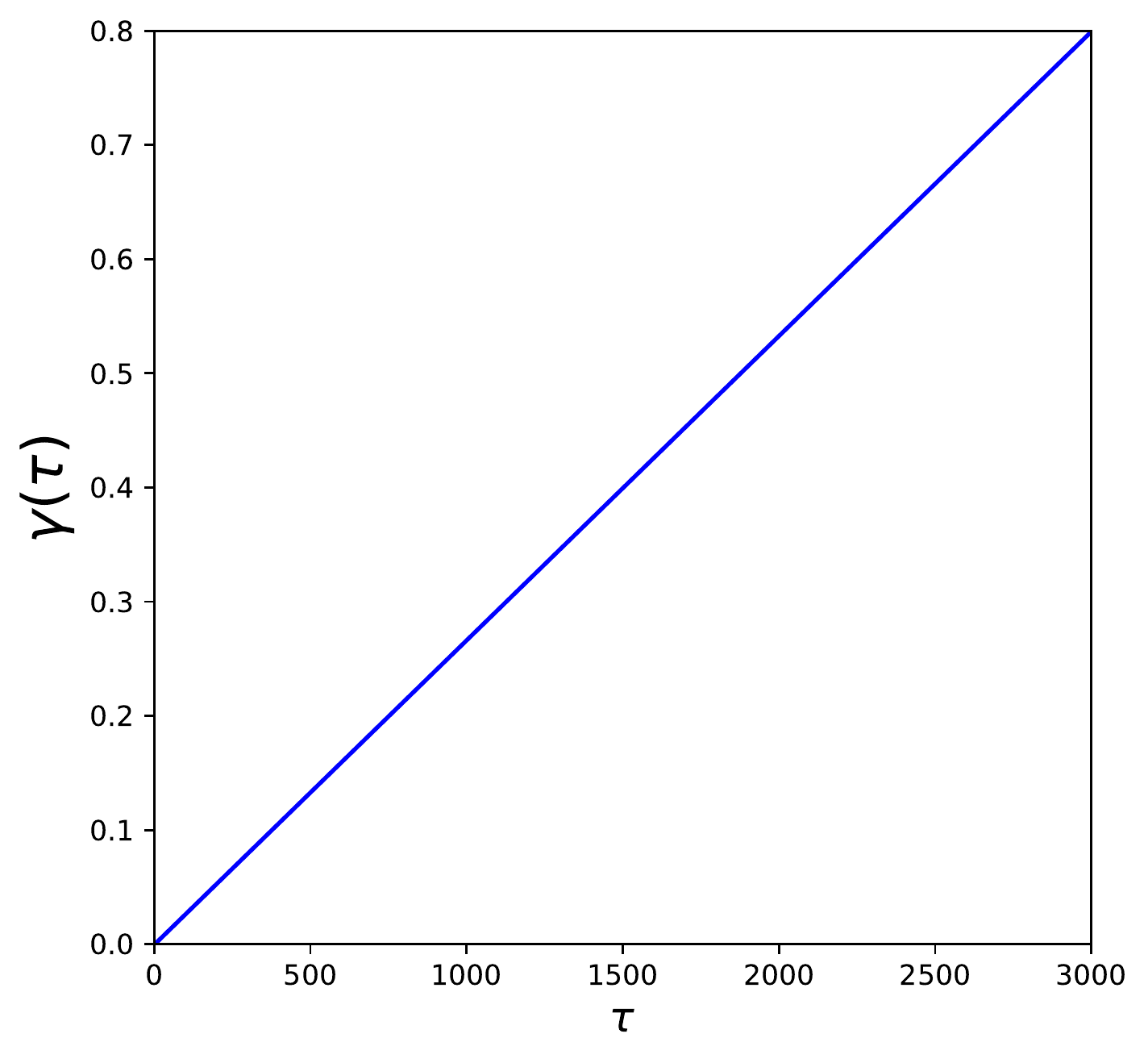}  
\caption[Plots for scalar field $\alpha(\tau)$ and $\gamma(\tau)$ for case 2]{Plots for scalar field $\alpha(\tau)$ and $\gamma(\tau)$ for $\chi_{in} = \frac{1}{100}$ and $ f(X) = \frac{X}{1-X} $ during the early epoch of de-Sitter expansion.} \label{plotalphagamma3000m1} 
\end{figure} 

From the equations $(\ref{earlyalpham1}) - (\ref{earlydeltam1})$, derivative of auxilary fields during this phase can be derived as 
\begin{equation}
\alpha'(\tau) \longrightarrow 
\frac{4}{3} \chi_{in}^{2} \;\; , \;\; 
\beta'(\tau) \longrightarrow 
\frac{4}{3} \chi_{in}^{2} \;\; , \;\;
\gamma'(\tau) \longrightarrow 
\frac83 \chi_{\rm in}^2 \;\; , \;\;
\delta'(\tau) \longrightarrow 
0 \; .
\end{equation}
\begin{figure}[!ht]
  \centering
    \includegraphics[scale =0.4]{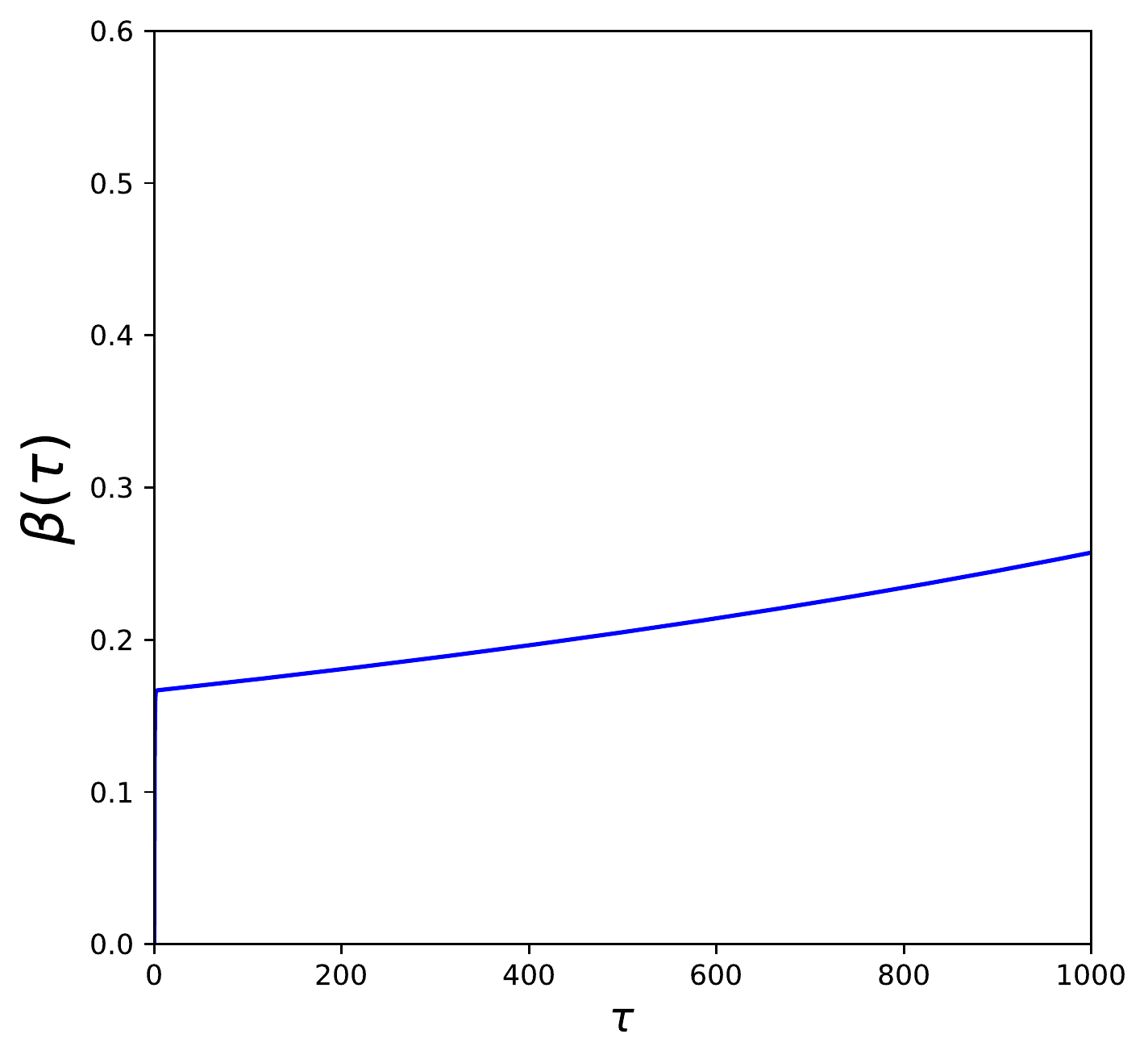}  {\hskip 2em}
  \includegraphics[scale =0.4]{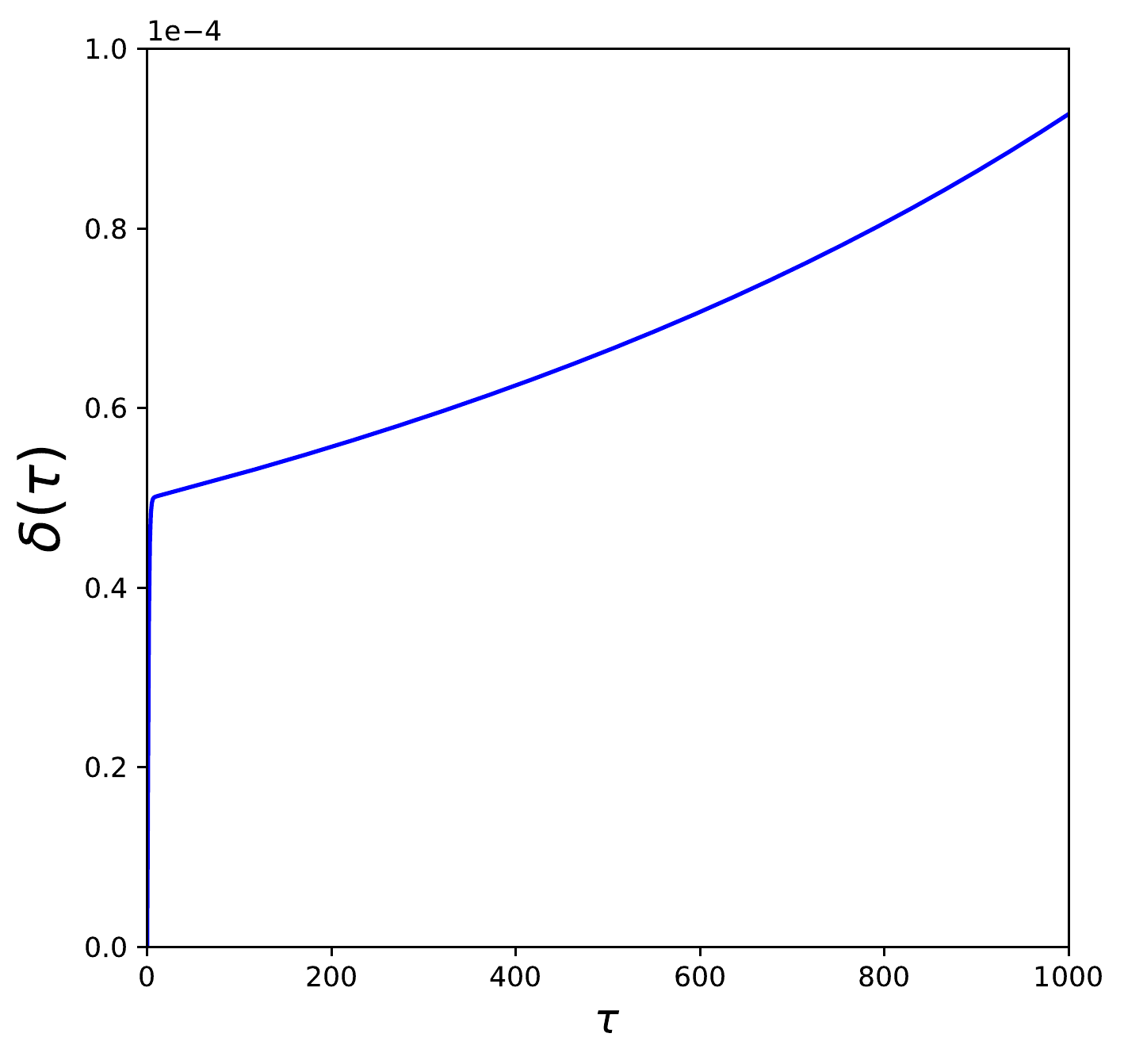} 
\caption[Plots for scalar field $\beta(\tau)$ and $\delta(\tau)$ for case 2]{Plots for scalar field $\beta(\tau)$ and $\delta(\tau)$ for $\chi_{in} = \frac{1}{100}$ and $ f(X) = \frac{X}{1-X} $ during the early epoch of de-Sitter expansion.}\label{plotbetadeltainm1}
\end{figure}

\begin{figure}[!ht]
  \centering
    \includegraphics[scale = 0.4]{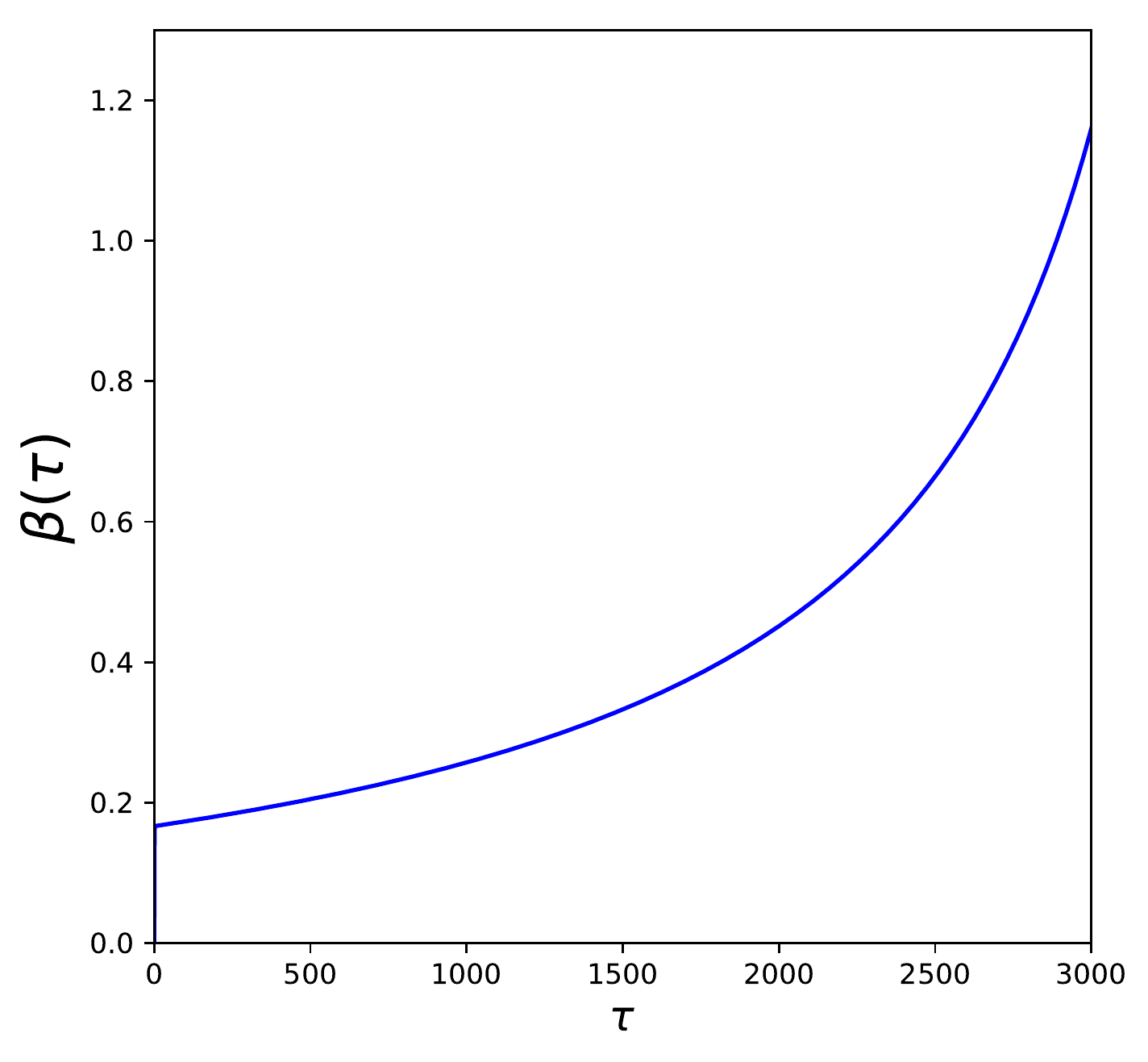}  {\hskip 2em}
  \includegraphics[scale = 0.4]{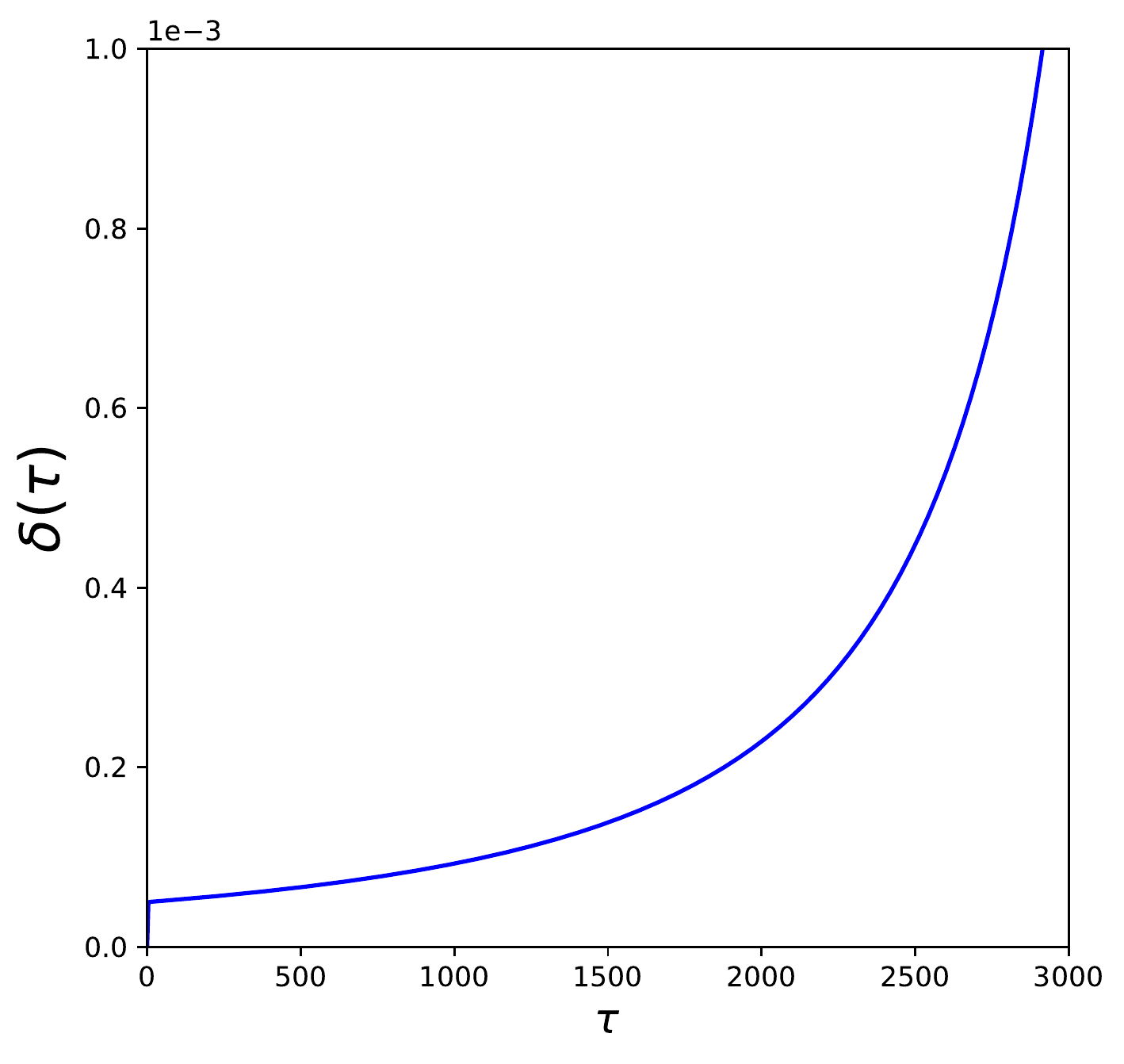} 
\caption[Plots for scalar field $\beta(\tau)$ and $\delta(\tau)$ during the early epoch of de-Sitter expansion]{Plots for scalar field $\beta(\tau)$ and $\delta(\tau)$ for $\chi_{in} = \frac{1}{100}$ and $ f(X) = \frac{X}{1-X} $ during the early epoch of de-Sitter expansion.}\label{plotbetadeltainem1}
\end{figure}

Using expression of field variables and their derivatives, corresponding Hubble parameter and the   effective equation of state $(w_{eff})$ evolve as follows:

\begin{equation}
 \chi(\tau) \longrightarrow \chi_{in}(1 - 2\chi_{in}^{4}\tau) \hspace{1 cm},\hspace{1 cm} w_{eff} \equiv -1 + \frac23 \epsilon  \hspace{0.3 cm}. \label{apprchiweffm1}
\end{equation}
\paragraph{} Numerically we plot evolution of field variables in Figure \ref{plotalphagamma3000m1} and \ref{plotbetadeltainm1} from $\tau = 0$ to 3000 and $ \tau = 0$ to 1000 respectively.At long times, from Figure {\ref{plotalphagamma3000m1}}, we observe that the expressions (\ref{earlyalpham1}) and (\ref{earlygammam1}) match with numerical plot. Similarly, from Figure \ref{plotbetadeltainm1}, it is clear that expressions for field variables (\ref{earlybetam1}) and (\ref{earlydeltam1}) also agree with our numerical calculation roughly for short time period. We plot $\beta $ and $ \delta $ in Figure \ref{plotbetadeltainem1}  and obtained curvature in their plot can be explained by the coupling of auxilary scalar fields. 
   
\paragraph{} There are two coupling present in the system. First is of  $ \beta(\tau)$ to $\delta(\tau)$ which can be seen as
\begin{equation}
\beta'' = -3\frac{\chi}{\chi_{in}}\beta'  + 2(2-\epsilon)\frac{\chi^{2}}{\chi_{in}^{2}}\delta  \longrightarrow -3\beta' + 4\delta     \label{chi4correctionm1}
\end{equation} 
 Solving  equation (\ref{chi4correctionm1}), which provides order $ \chi_{[n}^{4} $corrections to the field $ \beta $.
\begin{equation}
 \beta (\tau) \longrightarrow  \chi_{in}^{4}\tau^2 \hspace{0.5 cm}.\label{totalcorrectedbetam1}
\end{equation} 
\newline
While equation (\ref{totalcorrectedbetam1}) shows a rough agreemment to left hand plot of Figure \ref{plotbetadeltainem1} , one can find a perfect fit by choosing the exponent of $ \tau $ as $ 2.24 $ of $ \chi_{in}^{4} $ order.    
\paragraph{} 
 {\hskip 2 em}The variation of $ \delta (\tau)$ can be understood from the second coupling of $ \delta(\tau) $ to $\gamma(\tau)$ from  equation (\ref{eomdeltam1}) while assuming $ f(X) = \frac{X}{1-X}$ \hspace{0.3 cm}. 
 \begin{equation}
 \delta'' = -3\frac{\chi}{\chi^{2}_{in}}\delta' - (2 -\epsilon)\frac{\chi^{2}}{\chi_{in}^{2}} \delta + \chi_{in}^{2}\hspace{0.1 cm}\textit{f} \hspace{0.1 cm}(X) \longrightarrow -3\delta' + \chi_{in}^{2}(1 + 2\gamma)\hspace{0.5 cm}.
 \end{equation}
Now the $\chi_{in}^{6} $ correction appears in $\delta (\tau)$ as 
\begin{equation}
\delta(\tau) \longrightarrow \frac{1}{2}\chi_{in}^{2} + \frac{4}{27}\chi_{in}^{6}\tau^{3} \hspace{0.5 cm}.\label{totalcorrectedeltam1}
\end{equation}
{\hskip 2em} R.H.S. plot of Figure \ref{plotbetadeltainem1} shows rough agreement to equation  (\ref{totalcorrectedeltam1}) .
\begin{figure}[!ht]
  \centering
 \includegraphics[scale=0.4]{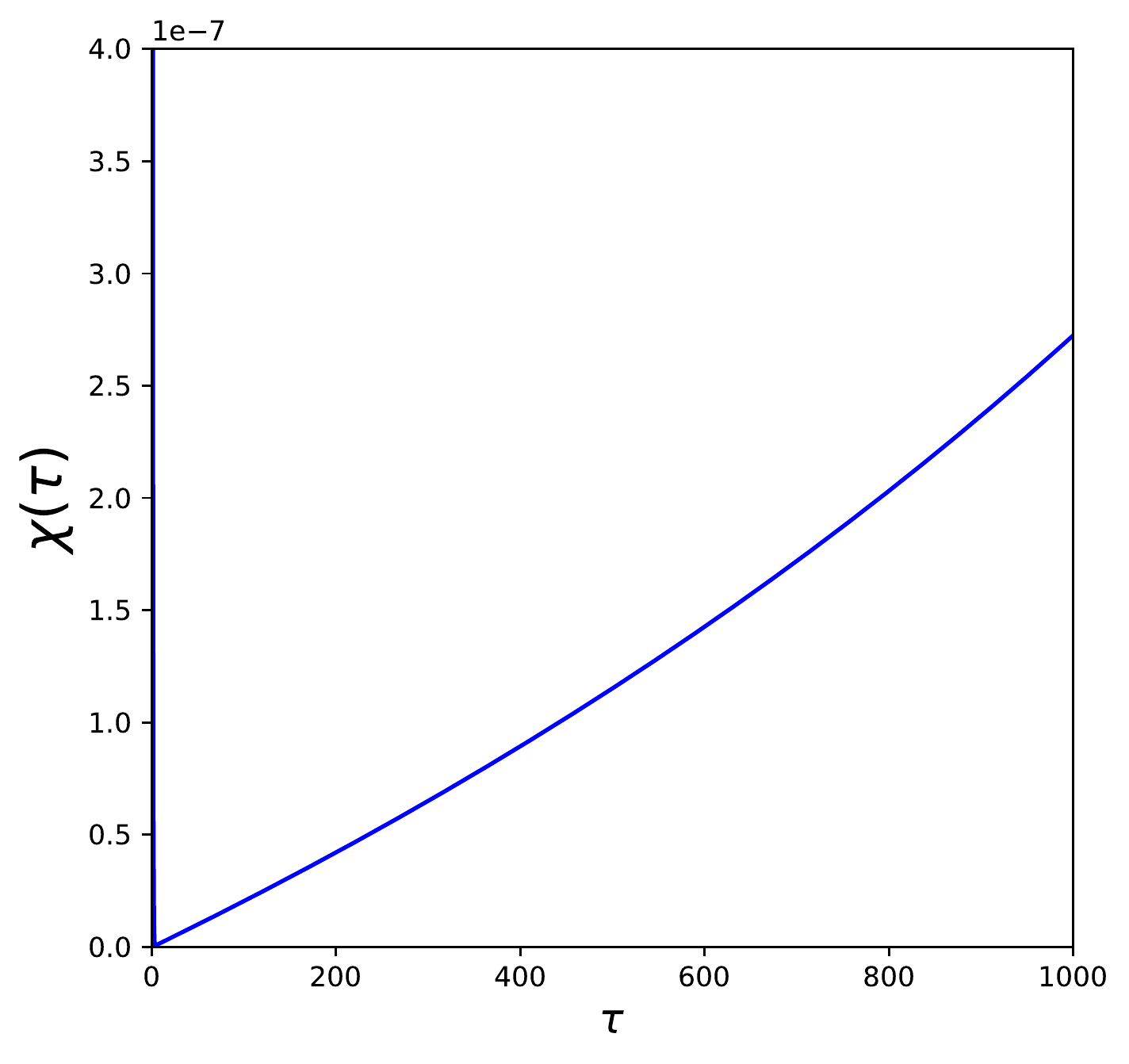}  {\hskip 2em} 
\includegraphics[scale = 0.4]{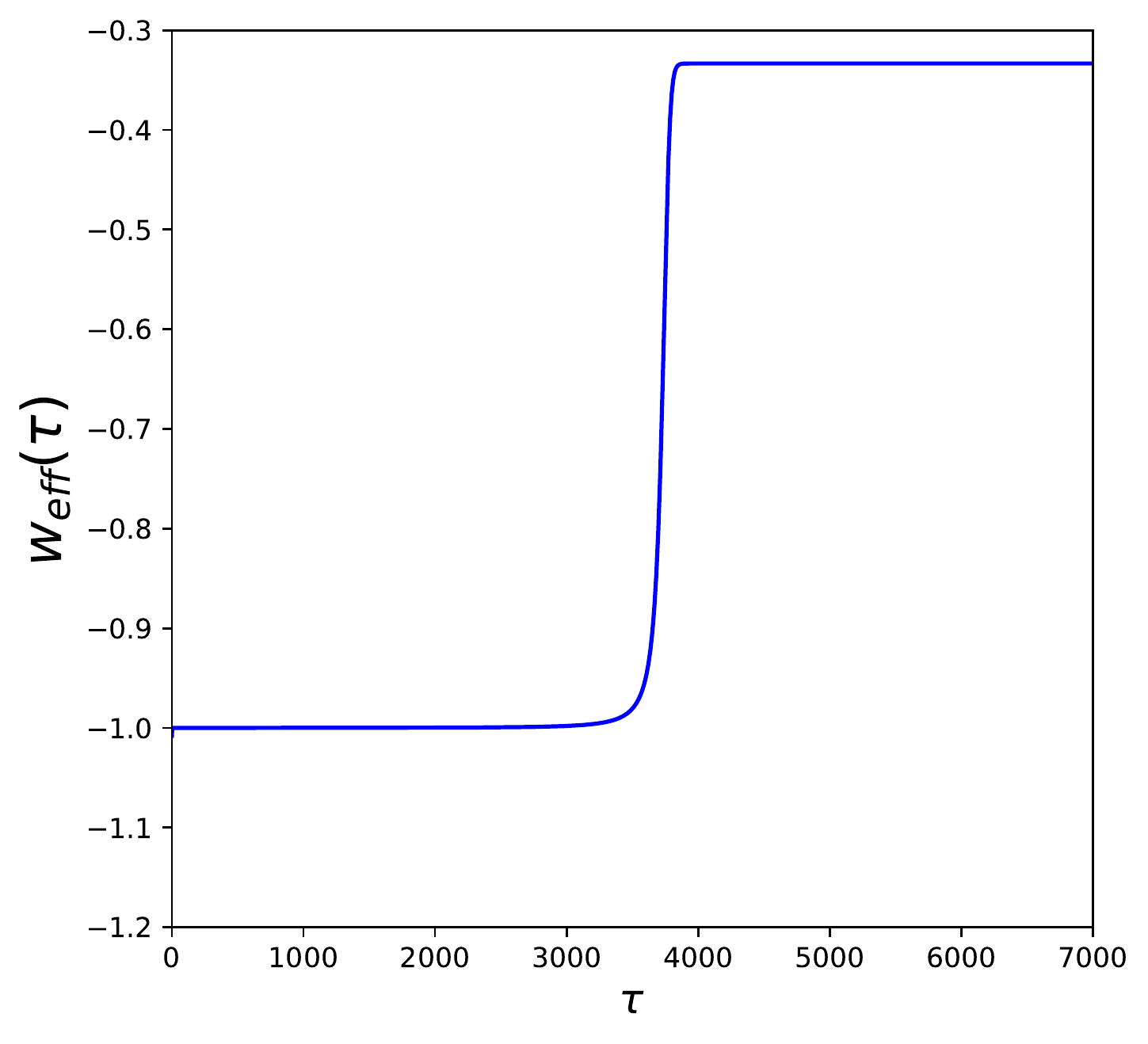}  
\caption[Plots for  $\chi_{in}^{2}-\chi $ and $ w_{eff}$ for case 2]{Plots for geometric quantities $\chi_{in}^{2}-\chi $ and $ w_{eff}$ for $\chi_{in} = \frac{1}{100}$ and $ f(X) = \frac{X}{1-X} $ during the early epoch of de-Sitter expansion.}\label{plotchiweffm1}
\end{figure} 
\paragraph{} The effect of curvature of $ \beta(\tau)$ and $\delta(\tau)$ is to evolve the values of $\epsilon $ = 0 to 1 at late-times.
From Figure \ref{plotchiweffm1}, it can be observed that $ w_{eff} = -1 $ stays for a long time period and eventually it increases to $ w_{eff} = -0.33$ and stays there forever.
\begin{figure}[!ht]
  \centering
 \includegraphics[scale=0.4]{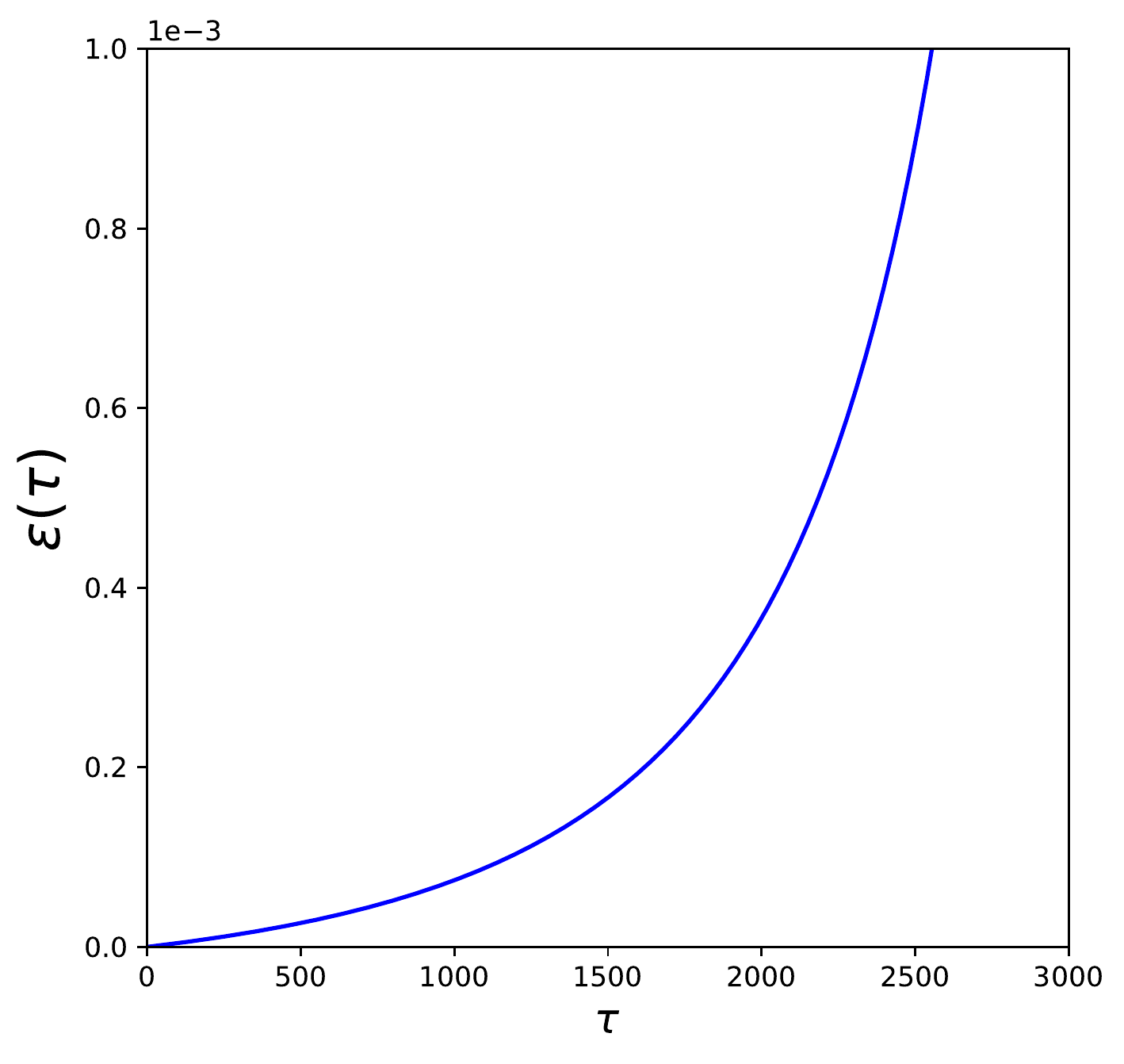}  {\hskip 2em}
\includegraphics[scale=0.4]{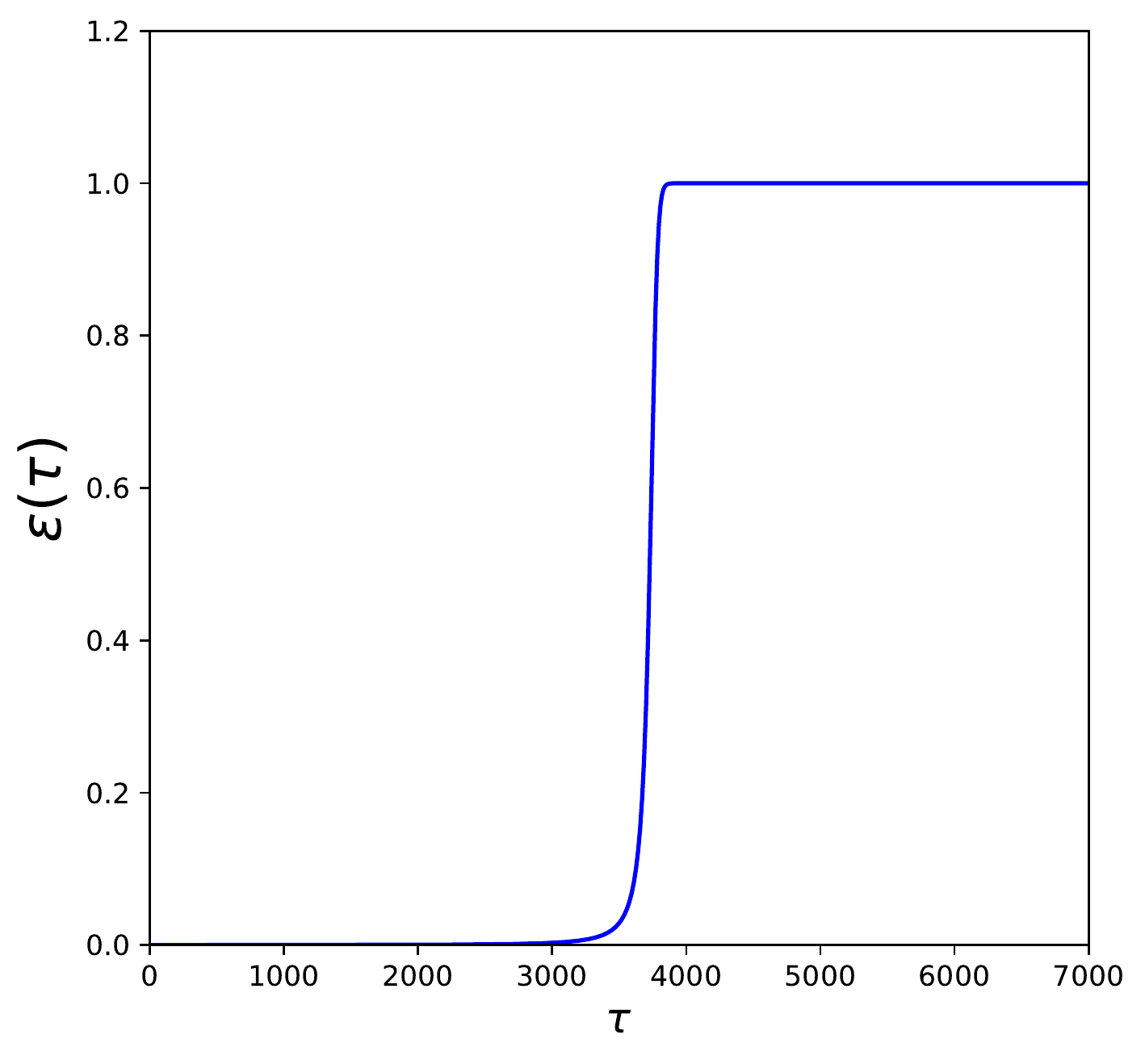}  
\caption[Plots for  $\epsilon (\tau)$ for case 2]{Plots for geometric quantities $\epsilon (\tau)$ for $\chi_{in} = \frac{1}{100}$ and $ f(X) = \frac{X}{1-X} $ Left hand plot shows the variation of $ \epsilon(\tau)$ during the early epoch of nearly de Sitter expansion. Right hand  plot shows the variation after that epoch and up to $\tau = 7000 $. }\label{plotepsilonintwodowmains}
\end{figure}
\begin{figure}[!ht]
  
  \includegraphics[scale=0.4]{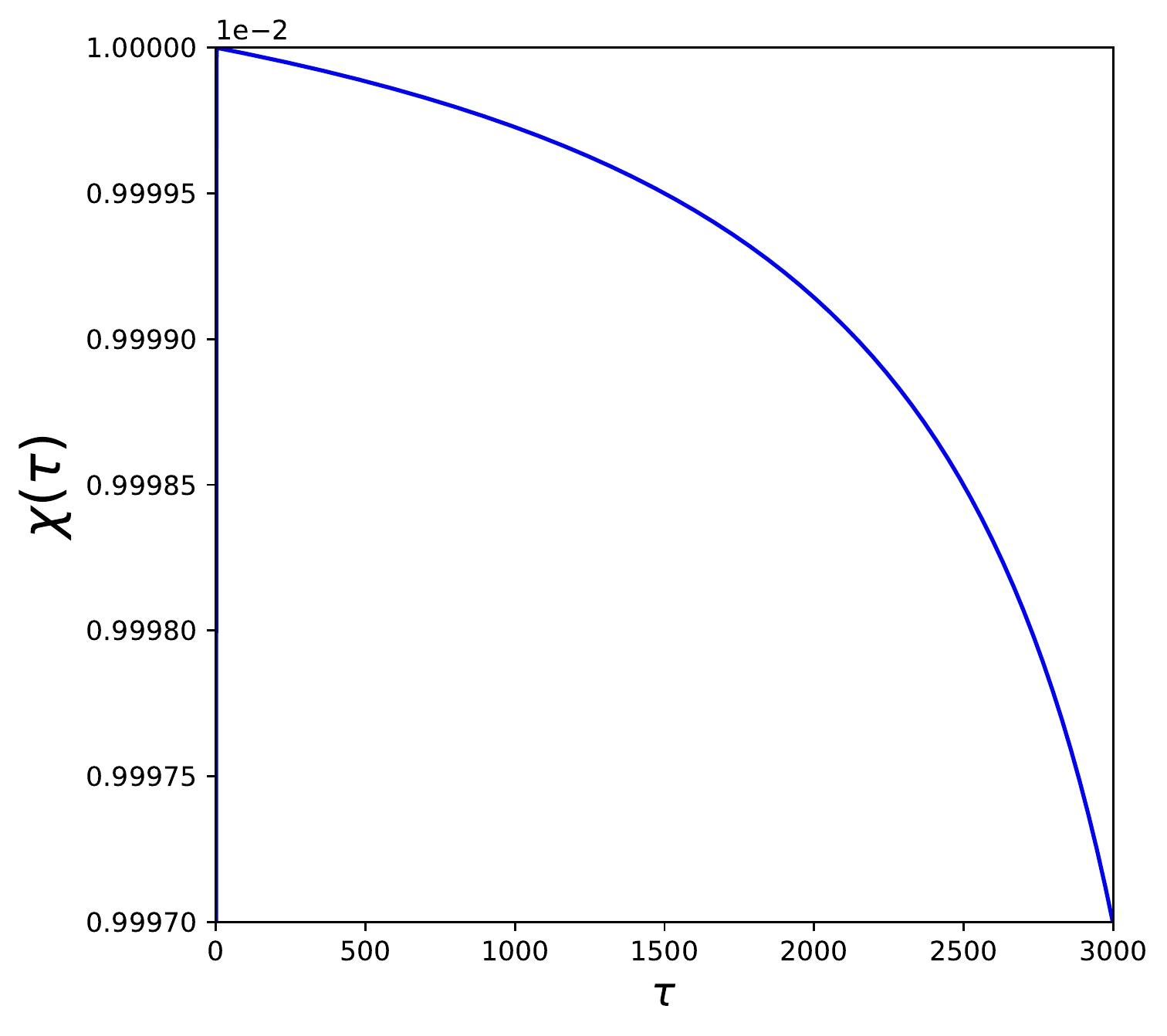}  {\hskip 2em}
  \includegraphics[scale=0.4]{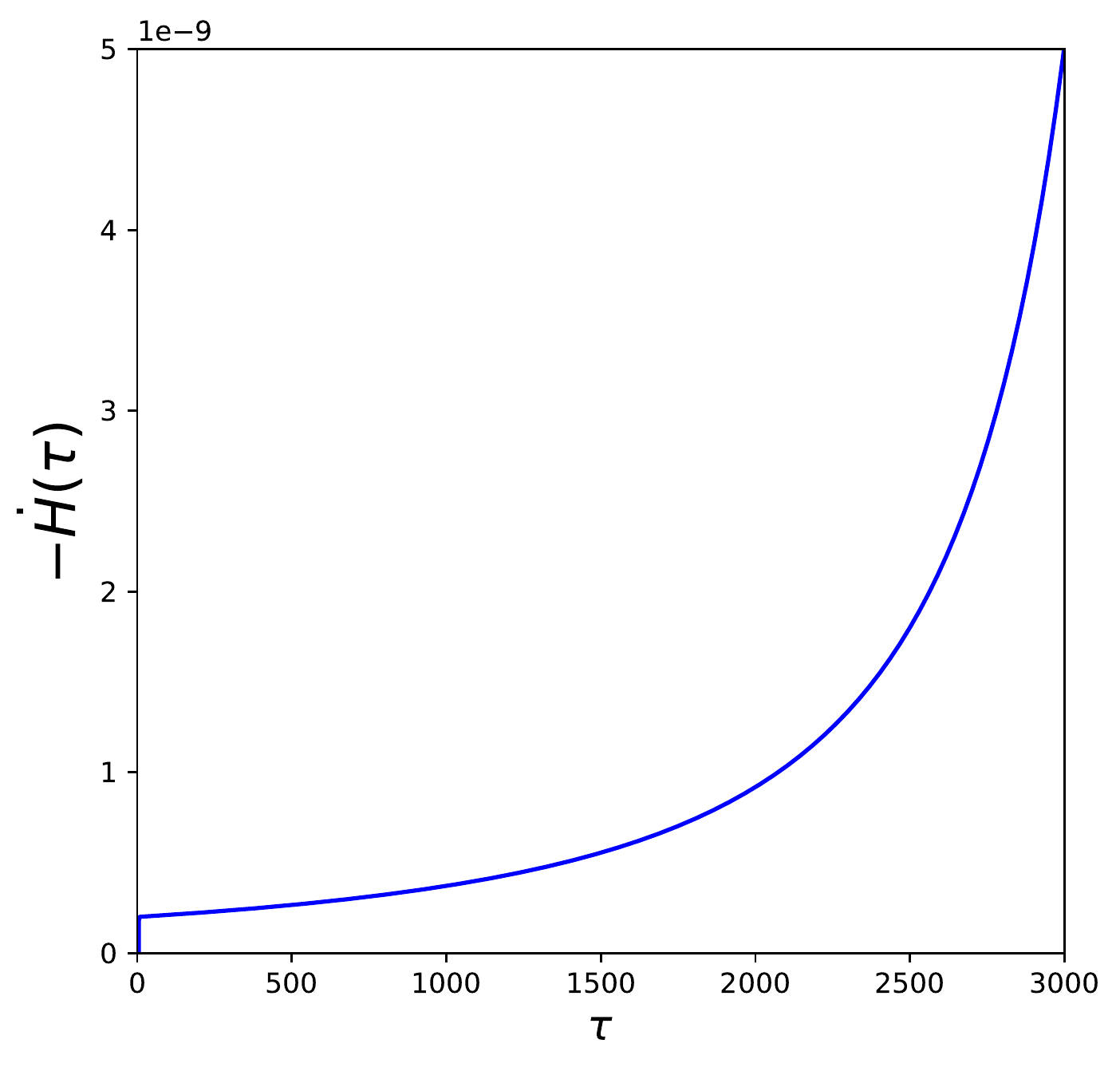} 
\caption[Plots for $H (\tau)$  and $ \dot{H(\tau)}$ for case 2]{Plots for geometric quantities $H (\tau)$  and its first derivative  for $\chi_{in} = \frac{1}{100}$ and $ f(X) = \frac{X}{1-X} $.  }\label{plotshubbleandfirstderivativem1}
 Figure \ref{plotshubbleandfirstderivativem1} shows the Hubble parameter and its first derivative . 

\end{figure}

\section{Model \rom{3}} \label{sec3}
In this case $X[g]$ becomes
\begin{equation}
X[g] = G \frac{1}{\Box_{c}}\left(\frac{1}{3}R^2 - R_{\mu\nu} R^{\mu\nu}\right)\frac{1}{\Box}R
\end{equation} \newline Equivalent scalar tensor lagrangian can be written as follows
\begin{equation}
\begin{split}
\mathcal{L} = & \Lambda^{2}h(GC)\sqrt{-g} + B\left[\Box A - R \right]\sqrt{-g} \\&+ D[\Box_{c} C -\left(\frac{1}{3}R^2 - R_{\mu\nu} R^{\mu\nu}\right)A]\sqrt{-g} {\hskip 1em},
\end{split},    \label{model3}
\end{equation}
where A and C are auxilary scalar fields and B and D are lagrange multiplier in the action. \newline Varying the action w.r.t both auxilary and lagrange multiplier fields, we get
\begin{align}
&\frac{1}{\sqrt{-g}}\frac{\delta(\mathcal{L})}{\delta B} = \Box A -R = 0{\hskip 1em},\\& \frac{1}{\sqrt{-g}}\frac{\delta(\mathcal{L})}{\delta D} = \Box_{c} C - \left(\frac{1}{3}R^2 - R_{\mu\nu} R^{\mu\nu}\right)A = 0 {\hskip 1em}, 
\end{align}
\begin{align}
&\frac{1}{\sqrt{-g}}\frac{\delta(\mathcal{L})}{\delta A} = \Box B - \left(\frac{1}{3}R^2 - R_{\mu\nu} R^{\mu\nu}\right)D = 0 {\hskip 1em}\Rightarrow B{\hskip 1em} = \frac{1}{\Box} \left(\frac{1}{3}R^2 - R_{\mu\nu} R^{\mu\nu}\right)D {\hskip 1em}, \\& \frac{1}{\sqrt{-g}}\frac{\delta(\mathcal{L})}{\delta C} = \Box_{c} D + G \Lambda^{2}h^{'}(GC) = 0 {\hskip 1em} \Rightarrow {\hskip 1em} D  = - \frac{1}{\Box_{c}}G \Lambda^{2}h^{'}(GC) {\hskip 1em}, 
\end{align} 
The equation of motion for fields(A,B,C,D) for FRW metric in this model becomes
\begin{eqnarray}
{\ddot A} &\!\! = \!\!&
- 3H{\dot A} 
- 6 (2 - \epsilon) H^2
\;\; , \label{eomAM3} \\
{\ddot B} &\!\! = \!\!&
- 3H{\dot B} 
- 12 (1 - \epsilon) H^4 D
\;\; , \\ \label{eomBM3}
{\ddot C} &\!\! = \!\!&
- 3H{\dot C} -(2-\epsilon)H^{2}C - 12 (1 - \epsilon) H^4 A
\;\; , \\ \label{eomCM3}
{\ddot D} &\!\! = \!\!&
- 3H{\dot D} + G \Lambda^2 h'(GC)-(2-\epsilon)H^{2}D
\;\; .  \label{eomDM3}
\end{eqnarray}
 The $(00)$-component of modified Einstein field equation is
\begin{equation}
\begin{split}
\frac{3H^2}{16\pi G} + \frac{1}{2}\Lambda^{2}h(GC)  & -\frac{1}{2}(\dot{A}\dot{B} + \dot{C}\dot{D}) -6H^{3}\dot{A}D -6H^{3}\dot{D}A   \\- &3(H\partial_{t} + H^{2})(\frac{1}{6}CD + B) = \frac{\Lambda}{16 \pi G}{\hskip 1em}, \label{00m3}
\end{split}
\end{equation}
and $(11) $-component of field equation is
\begin{equation}
\begin{split}
&-(3-2\epsilon) \frac{H^{2}}{16 \pi G} - \frac{1}{2}\Lambda^{2}h(GC) + G\Lambda^{2}Ah'(GC)\Big( 2A + \frac{1}{6}C \Big) -\frac{1}{6}\dot{C}\dot{D} -\frac{1}{2}\dot{A}\dot{B} + 4\dot{A}\dot{D}H^{2} \\& {\hskip 1em}-2(1+2\epsilon)H^{3}(\dot{A}D + \dot{D}A) -12(3- 2\epsilon)H^{4}D -(H\partial_{t} + H^{2})\frac{1}{6}CD -2\epsilon BH^{2} \\&-2(3-2\epsilon)H^{4}AD  + 3H^{2}B  -H\partial_{t}B  {\hskip 1em}= - \frac{\Lambda}{16 \pi G}{\hskip 1em}. \label{11m3}
\end{split}
\end{equation}
Adding equations (\ref{00m3}) and (\ref{11m3}), we obtain 
\begin{equation}
\begin{split}
 & \frac{2\epsilon H^{2}}{16 \pi G} + G\Lambda^{2}Ah'(GC)\Big(2A + \frac{C}{6}\Big)- \frac{2}{3}\dot{C}\dot{D} -\dot{A}\dot{B} + 4\dot{A}\dot{D}H^{2}  \\& {\hskip 1em} -12(3-2\epsilon)H^{4}D -4(H\partial_{t} + H^{2})\frac{1}{6}CD -4H\partial_{t}B -2\epsilon BH^{2} \\& -4(2+\epsilon)H^{3}(\dot{A}D + \dot{D}A) -2(3-2\epsilon)H^{4}AD  = 0
\end{split} \label{00+11m3}
\end{equation}
To convert equations (\ref{eomAM3}) - (\ref{00+11m3}) into dimensionless equations,  we use a different set of dimensionless parameters i.e. $\left\lbrace\alpha ,\beta,\gamma,\delta\right\rbrace $  as:
\begin{align}
 & A \equiv -3\alpha{\hskip 2em},{\hskip 2em} B  \equiv \frac{-3\beta}{G} {\hskip 2em},{\hskip 2em} C \equiv \frac{9\gamma}{G}{\hskip 2em}, {\hskip 2em} D \equiv \delta {\hskip 1em},\\& h(GC) = h(9\gamma)\equiv f(\gamma){\hskip 1em}.
\end{align}
We solve  equation for $\left\lbrace\alpha ,\beta,\gamma,\delta,\chi,\epsilon\right\rbrace $ numerically subjected to following initial conditions  at $ \tau = \tau_{in}$  :
\begin{align}
 & \alpha = \alpha' = \beta = \gamma = \gamma^{'} = \delta^{'} = \delta = 0 ,\beta^{'} = 0 {\hskip 2em},\\ & \chi = \chi_{in} {\hskip 2em},{\hskip 2em} \epsilon = 0 {\hskip 1em}.
\end{align} 

 Using these dimensionless variables, equation of motion for $\left\lbrace\alpha ,\beta,\gamma,\delta \right\rbrace $   cast as :
\begin{align}
 & \alpha^{''} + 3\frac{\chi}{\chi_{in}}\alpha^{'}  = 2(2-\epsilon)\frac{\chi^{2}}{\chi^{2}_{in}} {\hskip 1em},\\&  \beta^{''} + 3\frac{\chi}{\chi_{in}}\beta^{'}  = 4(1-\epsilon)\frac{\chi^{4}}{\chi^{2}_{in}}\delta {\hskip 1em}, \\& \gamma^{''} + 3\frac{\chi}{\chi_{in}}\gamma{'}+(2-\epsilon)\frac{\chi^{2}}{\chi_{in}^{2}}\gamma = 4(1-\epsilon)\frac{\chi^{4}}{\chi^{2}_{in}}\alpha {\hskip 1em},\\& \delta^{''} + 3\frac{\chi}{\chi_in} \delta^{'} +(2-\epsilon)\frac{\chi^{2}}{\chi_{in}^{2}}\delta = \chi_{in}^{2}f'(\gamma) {\hskip 1em},
\end{align}
 The constraint equation for $\chi$ in terms of dimensionless variables is derived from equation  (\ref{00m3}) as 
\begin{equation}
\begin{split}
 2\chi_{in}\left[\alpha'\delta + \delta'\alpha \right]\chi^{3} & + \left[\frac{1}{3} -\frac{1}{2}\gamma \delta + \beta \right]\chi^{2} - \chi_{in}\partial_{t}\left[ \frac{1}{2}\gamma \delta - \beta \right] \chi \\& -\chi^{2}_{in}\left[\frac{1}{3} -\frac{1}{2}\chi_{in}^{2}f(\gamma) + \frac{1}{2}(\alpha^{'}\beta^{'} + \gamma^{'}\delta^{'})\right] = 0{\hskip 1em},
\end{split} \label{chi1m3}
\end{equation}
 We solve for $\epsilon $ from the equation (\ref{00+11m3}) as
\begin{equation}
\begin{split}
 & \epsilon \Big[ 2\chi^{2} + 6\beta \chi^{2} -12\chi^{4}\Big( \alpha \delta -2\delta \Big)+ 12\Big(\alpha'\delta- \alpha \delta' \Big)\chi_{in}\chi^{3}\Big] = 3\Bigg\{ \chi^{4}_{in}f'(\gamma)\Big(3\alpha -\frac{3}{2}\delta\Big) \\& + \chi_{in}^{2}\{3\alpha'\beta' +2\gamma'\delta' + 4\alpha'\delta'\chi^{2}\} -8\chi_{in}\chi^{3}\left[\alpha'\delta + \alpha\delta'\right]  -6\chi^{4}(\alpha \delta -2\delta) -4\chi \chi_{in}\beta'  \\& + 2\Big( \chi\chi_{in}\partial_{t} + \chi^{2}\Big)\gamma\delta\Bigg\}. 
\end{split} \label{chim3}
\end{equation}
Finally, we can further simplify equation (\ref{chim3}) to arrive at
\begin{equation}
\begin{split}
\epsilon \chi^{2}& = \frac{3}{\Big[ 2 + 6\beta  -12\chi^{2}\Big( \alpha \delta -2\delta \Big)+ 12\Big(\alpha'\delta -\alpha \delta' \Big)\chi_{in}\chi\Big]} \times  \\ &\Bigg\{ \chi^{4}_{in}[f'(\gamma) (2\alpha -\frac{3}{2}\delta) + 2f(\gamma)] + \chi_{in}^{2}(\alpha' \beta' + 4\alpha' \delta' \chi^{2})\\& + \frac{4}{3}(\chi^{2} -\chi^{2}_{in}) + (4\chi^{2} + \chi_{in}\chi \partial_{t})\beta - 6\chi^{4}(\alpha \delta -2\delta ) \Bigg\}.
 \end{split}  \label{epsilonm3}
\end{equation} 
\subsection{Results}
Here we perform a similar analysis campared to model \rom{1} and \rom{2}. In this model scalar fields are approximated, for large values of $ \tau $, as 
\begin{eqnarray}
\alpha(\tau) &\!\! = \!\!& 
 \frac{4}{3}\chi_{in}^{2}\Big(\tau + \frac{e^{-3\tau}}{3}-\frac{1}{3}\Big)  \longrightarrow \frac{4}{3}\chi_{in}^{2} \Big(\tau -\frac{1}{3}\Big)
\; 
\; , \label{earlyalpham3} \\
\beta(\tau) &\!\! = \!\!& 
\frac43 \chi_{\rm in}^2 \Bigl( \tau \!-\! \frac{11}{6}
\!+\! 3 e^{-\tau} \!-\! \frac32 e^{-2\tau} 
\!-\! \frac13 e^{-3\tau} \Bigr) 
\; \longrightarrow \;
\frac43 \chi_{\rm in}^2 \Bigl( \tau \!-\! \frac{11}{6} \Bigr) 
\; , \label{earlybetam3} \\
\gamma(\tau) &\!\! = \!\!& 
\frac83 \chi_{\rm in}^2 \Bigl( \tau \!-\! \frac{11}{6}
\!+\! 3 e^{-\tau} \!-\! \frac32 e^{-2\tau} 
\!-\! \frac13 e^{-3\tau} \Bigr) 
\; \longrightarrow \;
\frac83 \chi_{\rm in}^2 \Bigl( \tau \!-\! \frac{11}{6}
\Bigr) \; , \label{earlygammam3} \\
\delta(\tau) &\!\! = \!\!& 
\frac{1}{2}\chi_{in}^{2}\bigg(1-e^{-\tau}\bigg)^{2} \longrightarrow \frac{1}{2}\chi_{in}^{2}
\; . \label{earlydeltam3}
\end{eqnarray}

\begin{figure}[!ht]
  \centering
   \includegraphics[scale = 0.4]{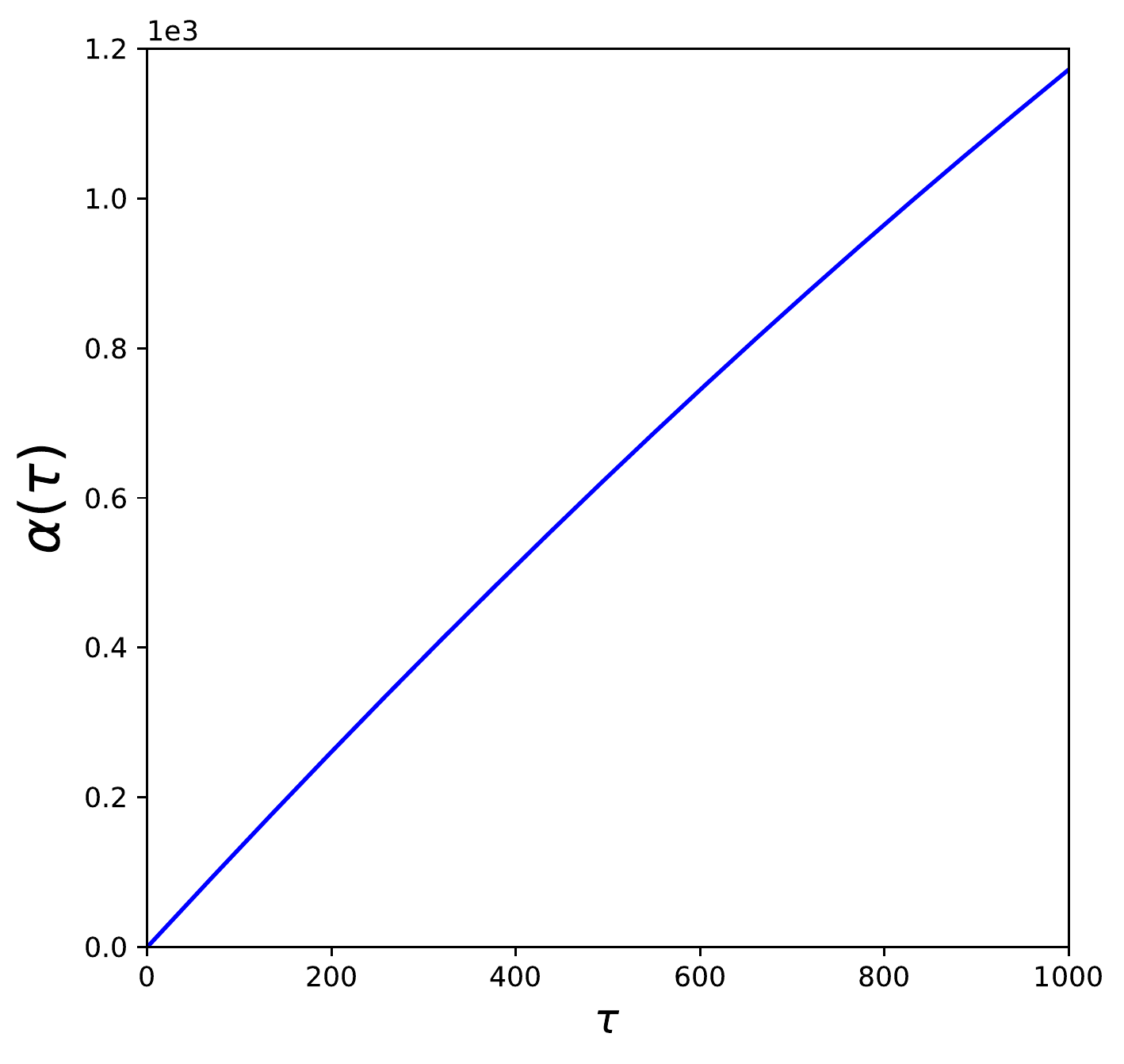}  {\hskip 2em} 
 \includegraphics[scale = 0.4]{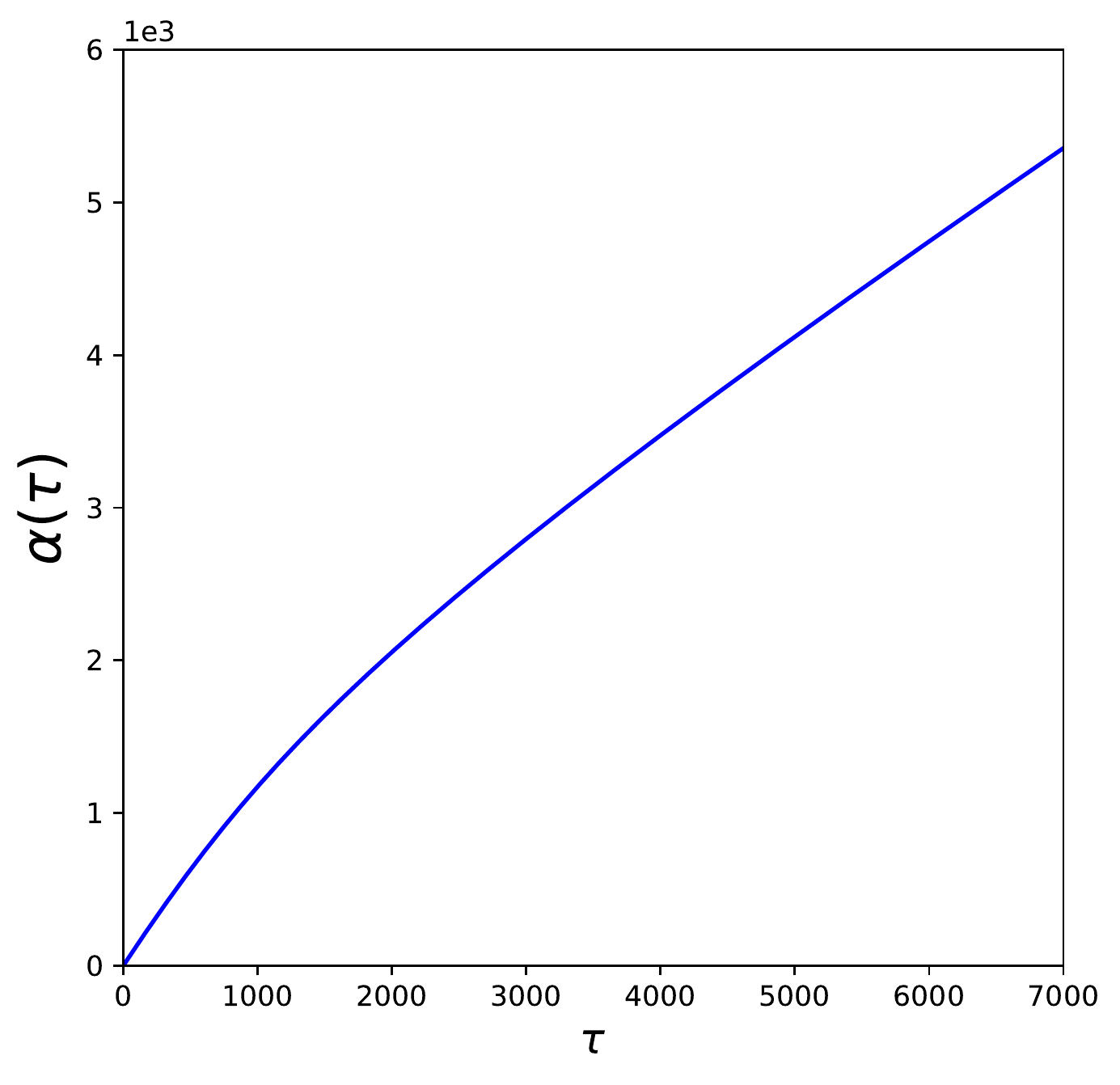} 
\caption[Plots for scalar field $\alpha(\tau)$  for case 3]{Plots for scalar field $\alpha(\tau)$  for $\chi_{in} = \frac{1}{100}$ and $ f(X) = \frac{X}{1-X} $. } \label{plotalphadelta1000m3} 
\end{figure}
\paragraph{} Similar to our previous analysis, Figure \ref{plotalphadelta1000m3} shows the matching of our numerical result with the expression for $ \alpha(\tau)$ in equation (\ref{earlyalpham3}).
\begin{figure}[!ht]
  \centering
    \includegraphics[scale = 0.4]{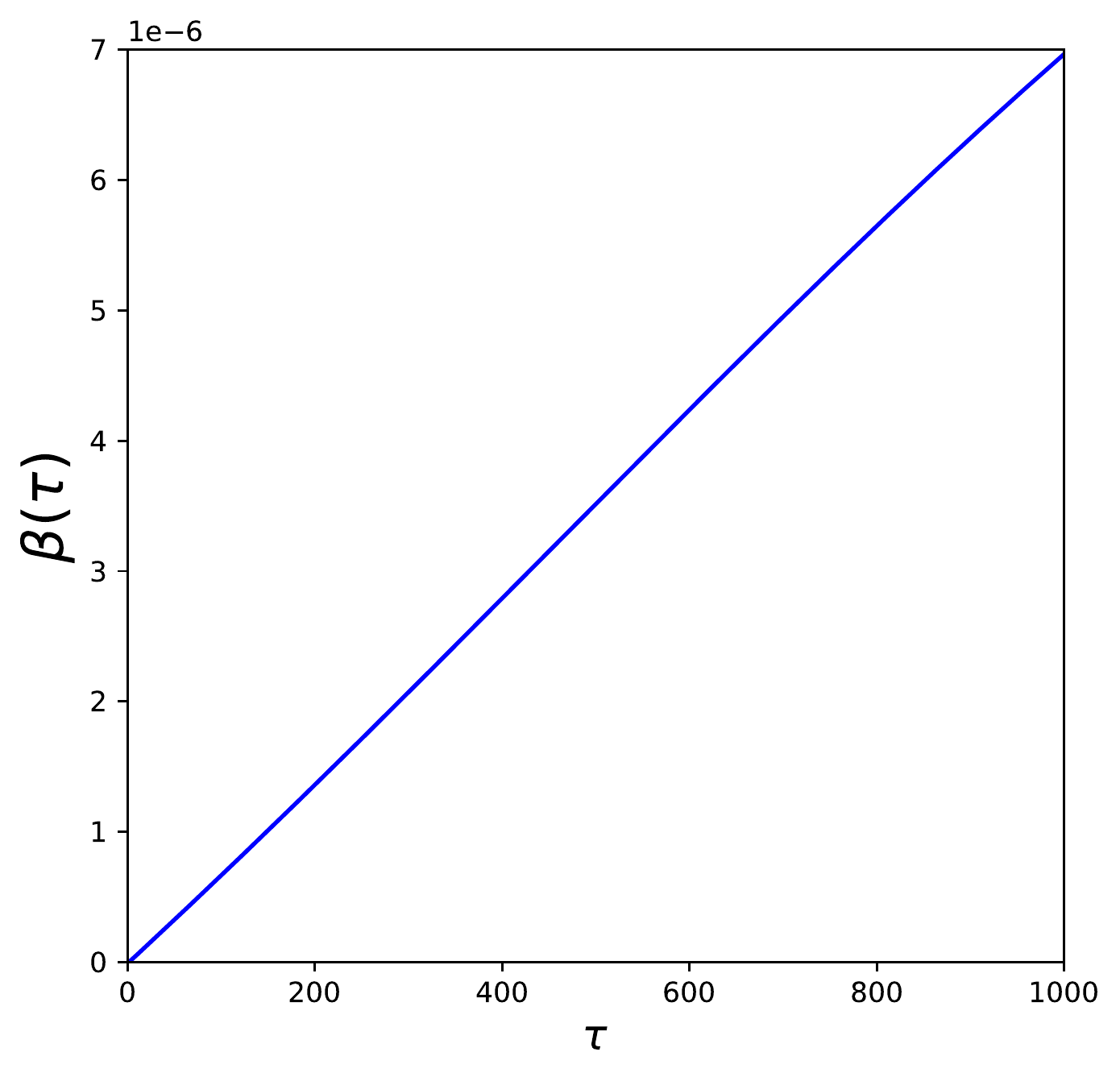} {\hskip 2em}   
  \includegraphics[scale = 0.4]{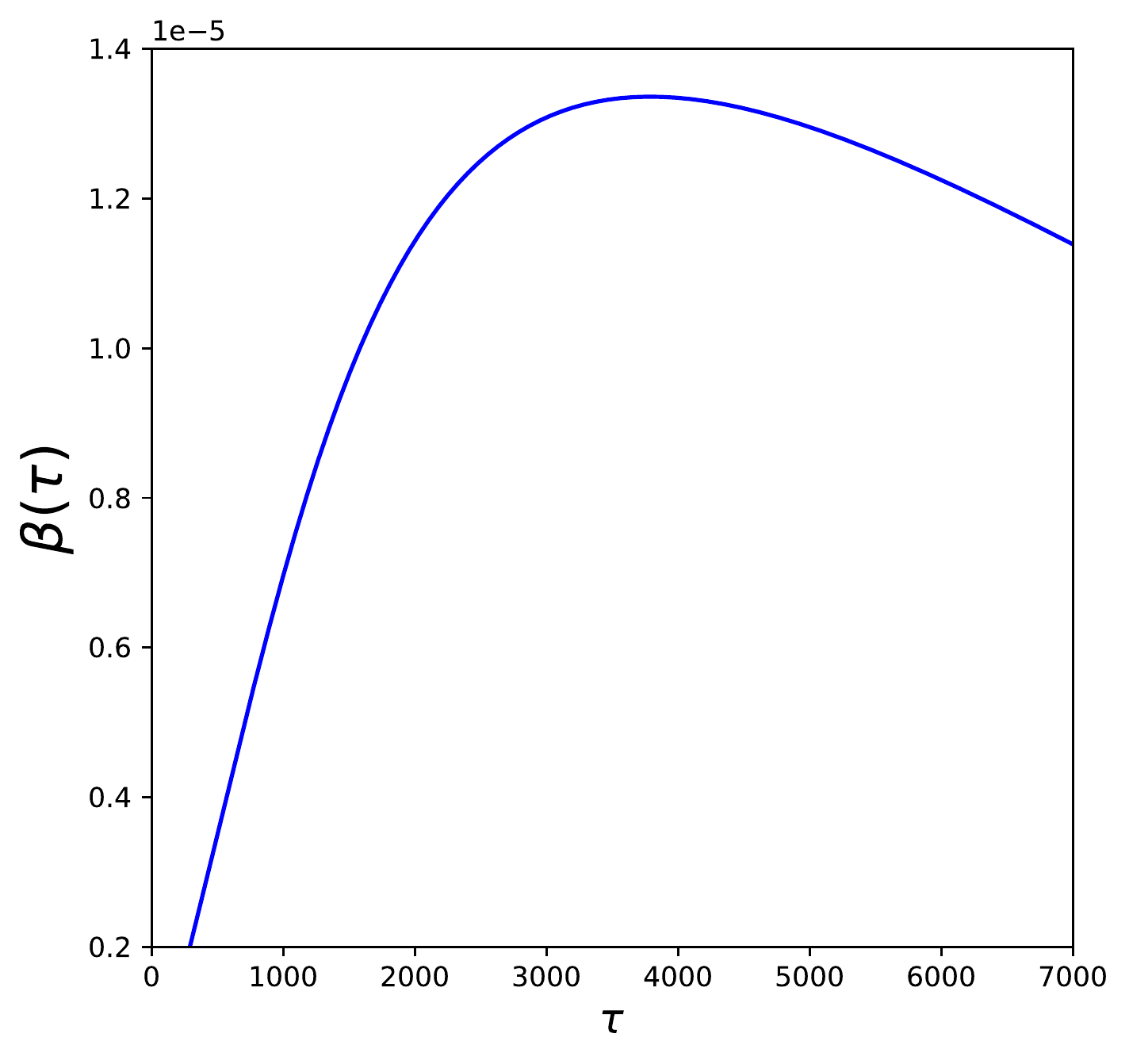}  
\caption[Plots for scalar $\beta(\tau)$ for case 3]{Plots for scalar field $\beta(\tau)$ for $\chi_{in} = \frac{1}{100}$ and $ f(X) = \frac{X}{1-X} $. }\label{plotbetam3} 
\end{figure}
\paragraph{} Similarly plots for $\beta (\tau) $ in Figure \ref{plotbetam3} also match with our numerical and analytical expression from $ \beta (\tau)$ in equation (\ref{earlybetam3}). Curvature in $\beta(\tau)$ shows up for values of $\tau > 1000$.
\begin{figure}[!ht]
  \centering
   \includegraphics[scale = 0.4]{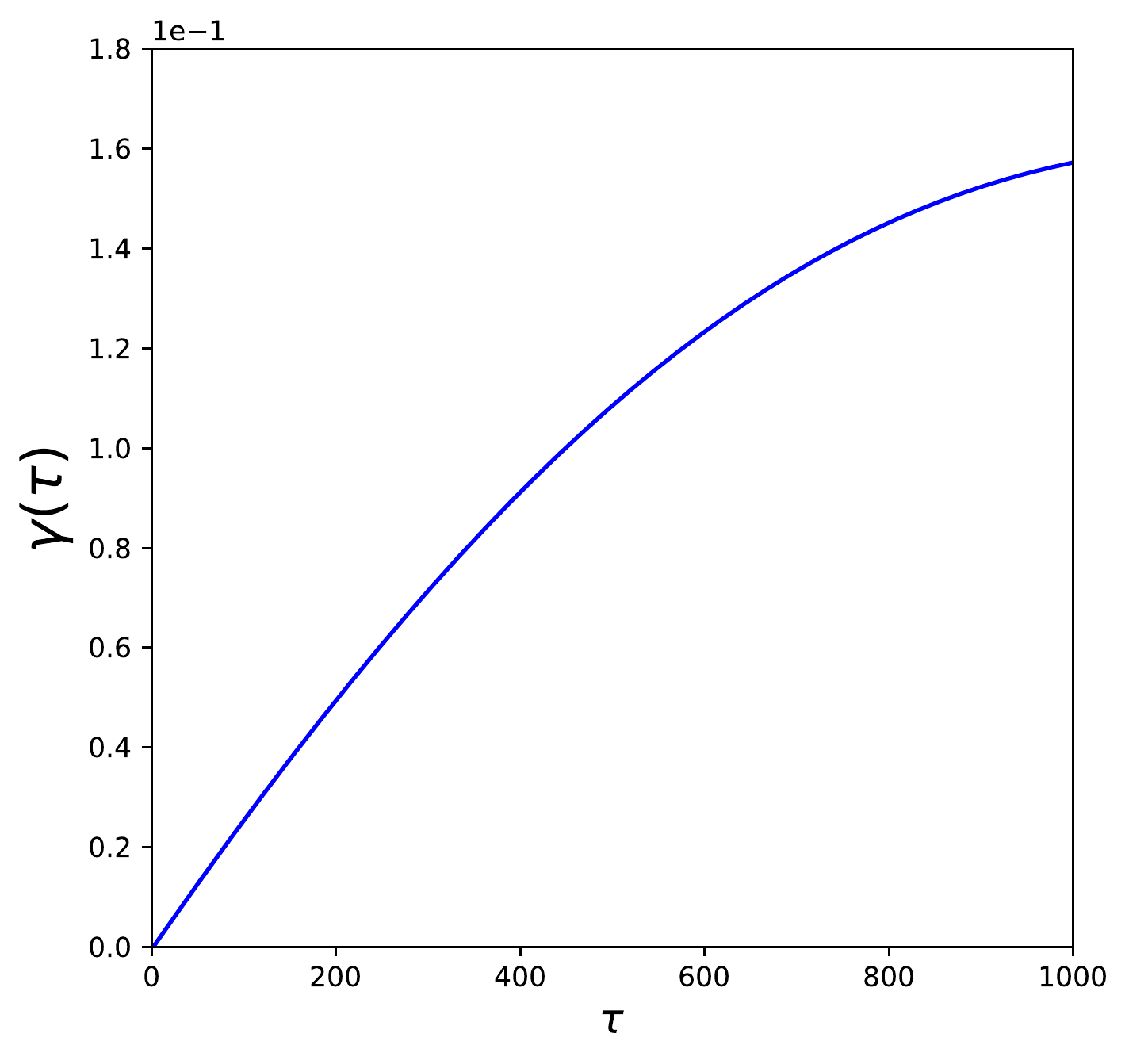} {\hskip 2em}    
  \includegraphics[scale = 0.4]{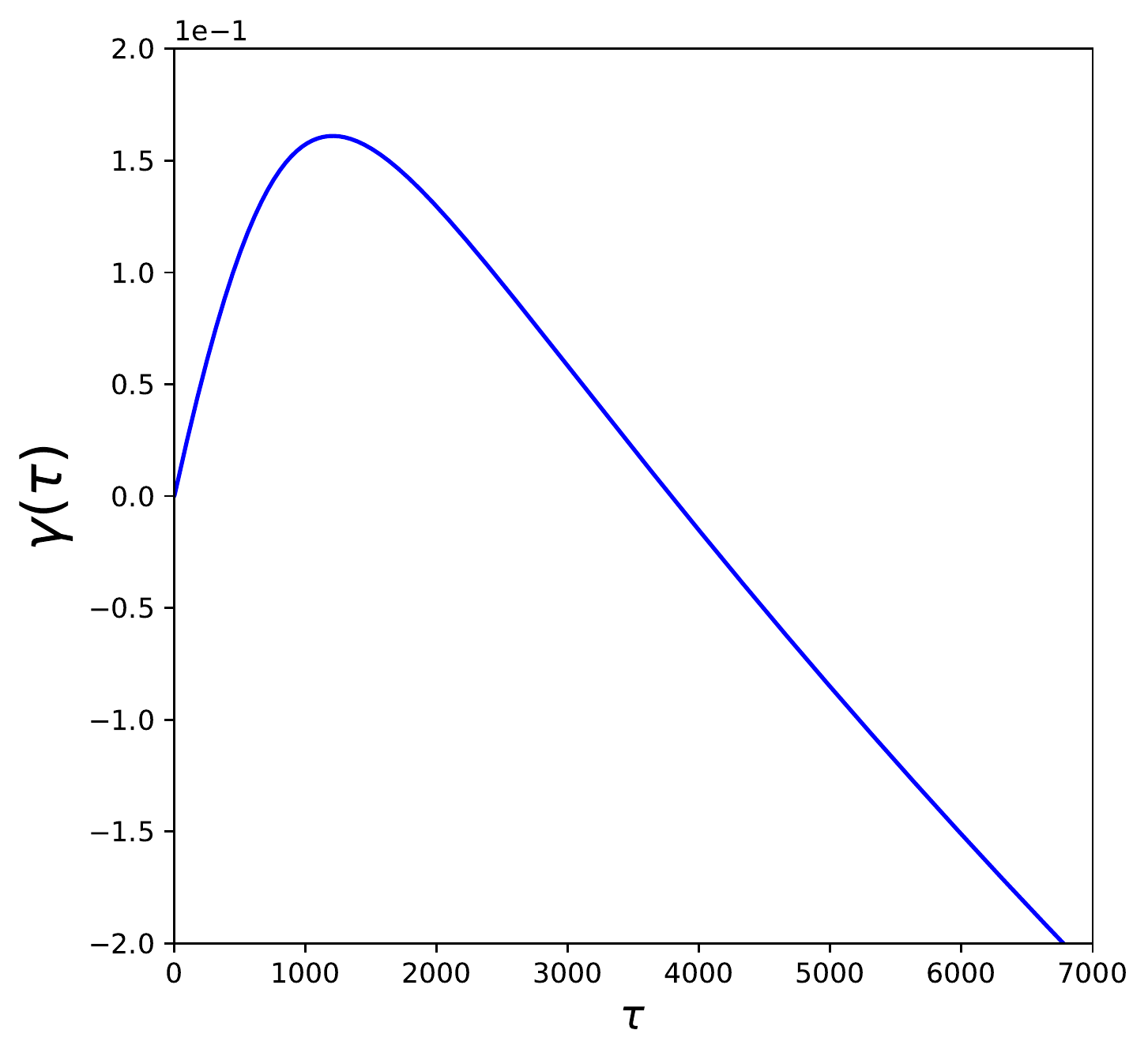}  
\caption[Plots for scalar  $\gamma(\tau)$ for case 3]{Plots for scalar field $\gamma(\tau)$ for $\chi_{in} = \frac{1}{100}$ and $ f(X) = \frac{X}{1-X} $. }\label{plotgammam3} 
\end{figure}
\paragraph{} 
$\gamma (\tau)$ initially increases with $\tau $ and around $ \tau \simeq 1300 $ it shows a decaying trend as seen in Figure \ref{plotgammam3}. Using the fact that $ f(X[g]) = \frac{X}{1-X}$ becomes singular at X = 1 where inflation ends as shown in the case of Model \rom{1} \cite{Tsamis:2016boj}. In current model this value X = 1 corresponds to a value of $\tau = 3752.33 $.
\begin{figure}[!ht]
  \centering
   \includegraphics[scale = 0.4]{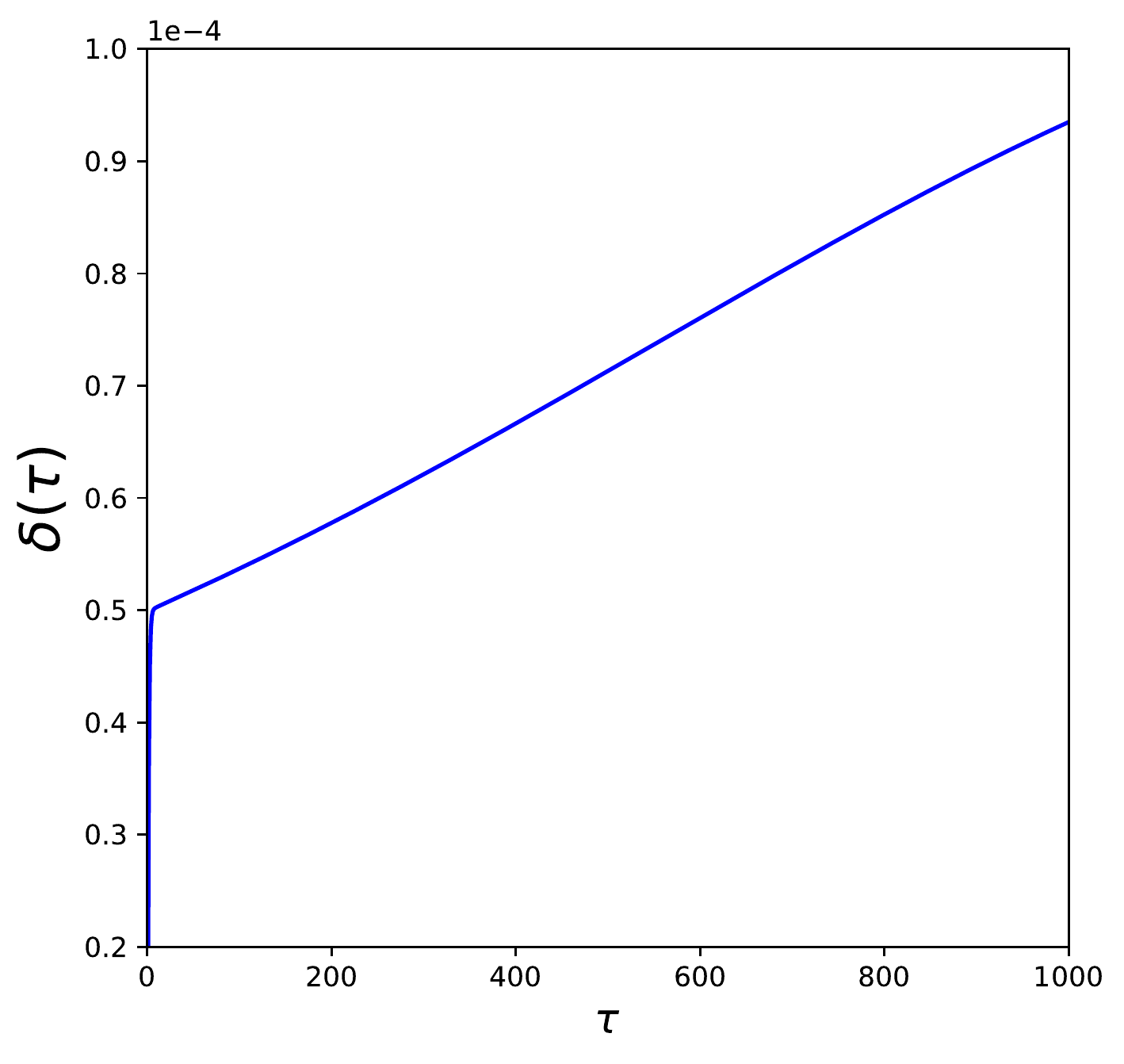} {\hskip 2em}    
 \includegraphics[scale = 0.4]{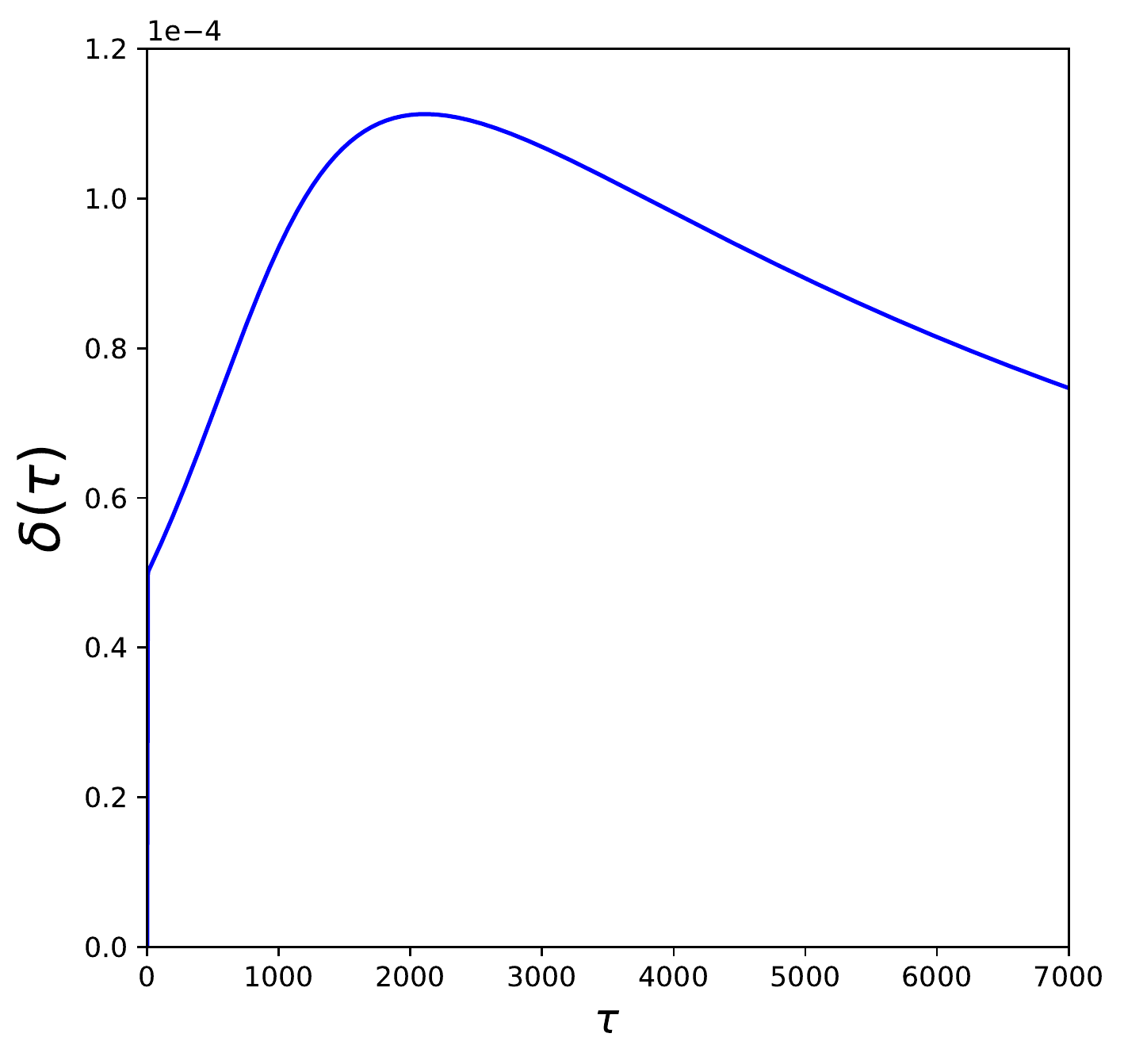}  
\caption[Plots for scalar $\delta(\tau)$ for case 3]{Plots for scalar field $\delta(\tau)$  for $\chi_{in} = \frac{1}{100}$ and $ f(X) = \frac{X}{1-X} $. }\label{plotdeltam3} 
\end{figure}
\paragraph{}
Figure \ref{plotdeltam3} shows the behaviour of scalar field $ \delta(\tau)$. It attains its maximum at $ \tau = 2110.27 $ and decreases beyond this value. In order to see the late time behaviour, we determine the derivatives of auxilary fields during de-Sitter epoch as 
 \begin{equation}
\alpha'(\tau) \longrightarrow 
\frac{4}{3} \;\; , \;\; 
\beta'(\tau) \longrightarrow 
\frac{4}{3} \chi_{in}^{4} \;\; , \;\;
\gamma'(\tau) \longrightarrow 
\frac83 \chi_{\rm in}^2 \;\; , \;\;
\delta'(\tau) \longrightarrow 
0 \; .
\end{equation}
{\hskip 2em}  Using these derivatives and equations (\ref{chi1m3})  and (\ref{epsilonm3}) to the expressions for Hubble parameter and the the first slow roll parameter as
\begin{equation}
\chi(\tau) \longrightarrow \chi_{in}(1- \chi_{in}^{5}\tau)  \hspace{0.5cm},\hspace{0.5 cm} \epsilon(\tau) \longrightarrow  5\chi_{in}^{2}\tau  \label{chiepsilonm3}
\end{equation}
\begin{figure}[!ht]
  \centering
   \includegraphics[scale = 0.4]{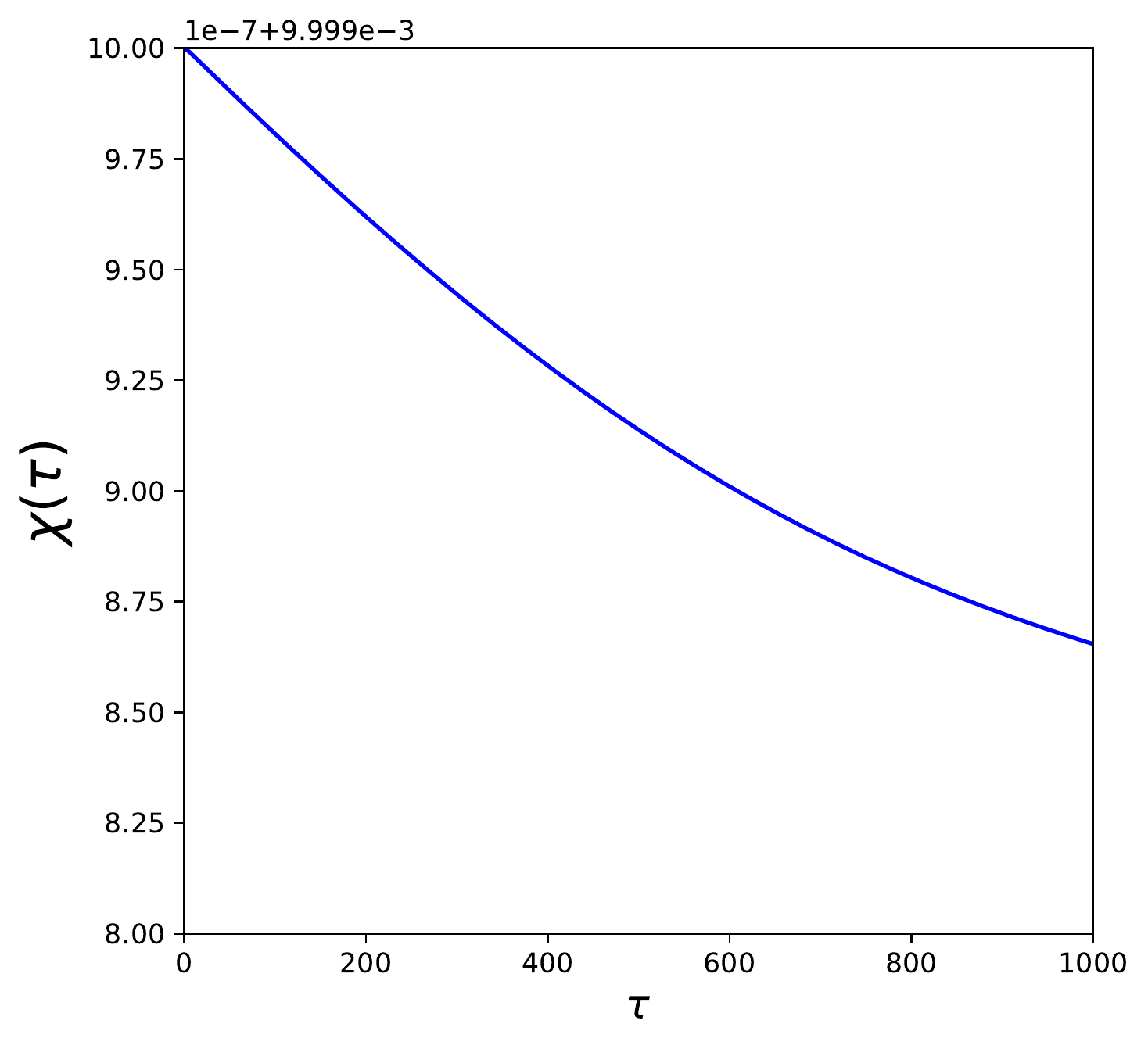} {\hskip 2em}      
 \includegraphics[scale = 0.4]{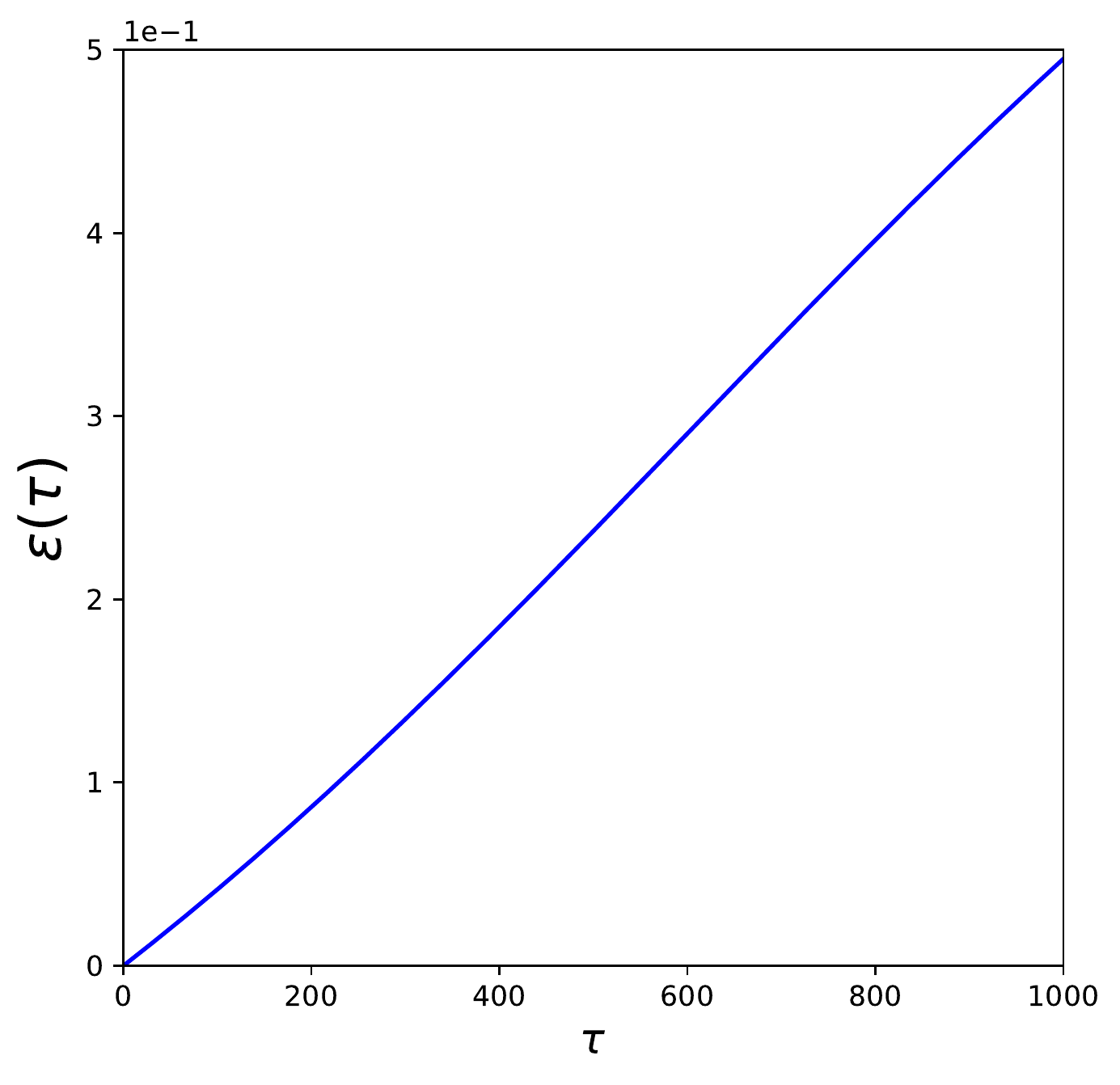}
\caption[Plots for $\chi(\tau)$  and $ \epsilon(\tau)$ for case 3]{Plots for Geometric quantities  $\chi(\tau)$  and $ \epsilon(\tau)$ for $\chi_{in} = \frac{1}{100}$ and $ f(X) = \frac{X}{1-X} $. }\label{plotchiepsilonm3} 

\end{figure}
\begin{figure}[!ht]
  \centering
    \includegraphics[scale = 0.4]{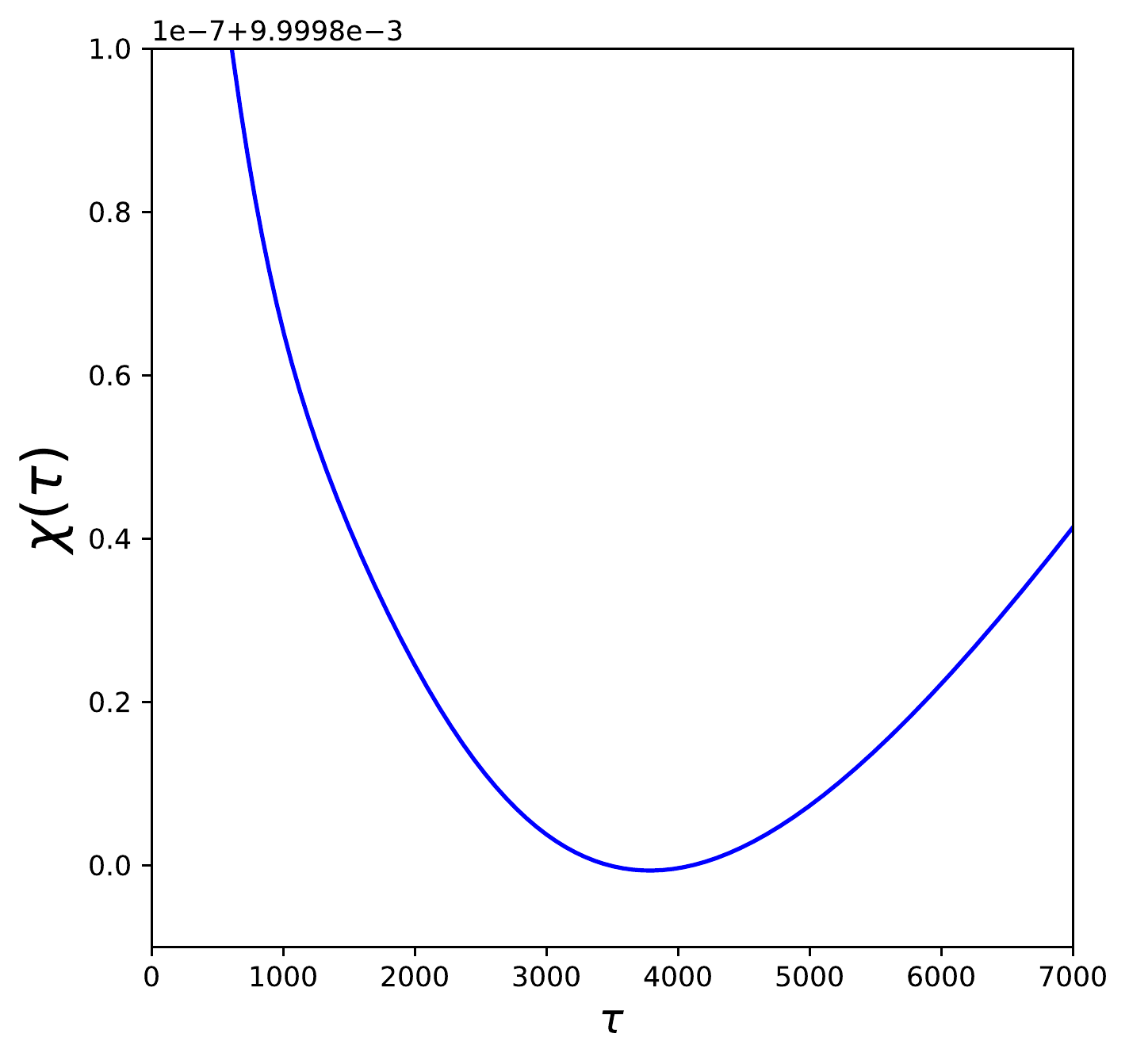}  {\hskip 2em}     
 \includegraphics[scale = 0.4]{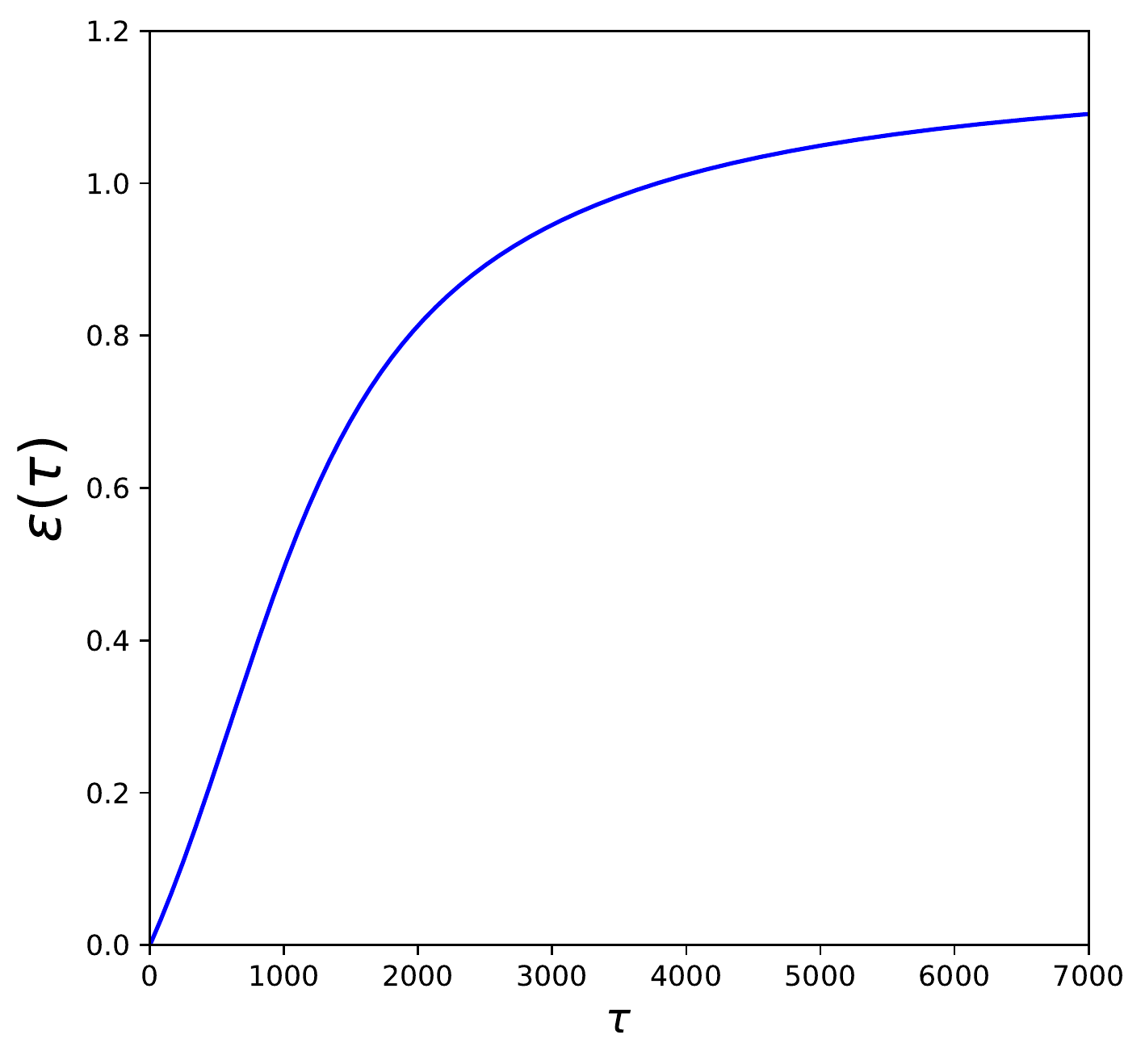}  
\caption[Plots for $\chi(\tau)$  and $ \epsilon(\tau)$ for case 3]{Plots for Geometric quantities  $\chi(\tau)$  and $ \epsilon(\tau)$ for $\chi_{in} = \frac{1}{100}$ and $ f(X) = \frac{X}{1-X} $. }\label{plotchiepsilonlatem3} 
\end{figure}
\paragraph{} The behaviour of $ \chi(\tau)$ and $\epsilon (\tau)$ are plotted in Figure \ref{plotchiepsilonm3} and \ref{plotchiepsilonlatem3} for $ \tau $ from $0 $ to $ 1000$ and for $\tau $ from 0 to 7000 respectively. It turns out that at late-times $\epsilon \longrightarrow 1$ and remains to that value thereafter. This corresponds to equation of state $w_{eff} = -\frac{1}{3}$. The exit from inflation to Radiation Dominated  epoch will never be achieved in this case.
\section{Model \rom{4}} \label{sec4}
Model \rom{4} has $X[g] $ of form 
\begin{equation} 
X[g] = GR \frac{1}{\Box}\frac{1}{\Box_{c}}\left(\frac{1}{3}R^2 - R_{\mu\nu} R^{\mu\nu}\right)
\end{equation} \newline
 In this case we will restrict ourself to particular form of $h(X[g]) = X[g]$ . Any other form of  $ h(X[g]) $ will lead to higher derivaive in the equation of motion. \newline For $h(X[g]) = X[g] $, lagrangian is written as
\begin{equation}
\begin{split}
\mathcal{L} = & \Lambda^{2}GRC \sqrt{-g} + B\left[\Box_{c}A -\left(\frac{1}{3}R^2 - R_{\mu\nu} R^{\mu\nu}\right)\right]\sqrt{-g} \\&+ D[\Box C -A]\sqrt{-g} {\hskip 1em}, \label{lag}
\end{split}
\end{equation}
To write langrangian (\ref{lag}) in equivalent scalar-tensor lagrangian, we introduce two auxilary scalar fields A and C  and two lagrange multiplier B and D which obeys following equations of motion 

\begin{align}
&\frac{1}{\sqrt{-g}}\frac{\delta(\mathcal{L})}{\delta B} = \Box_{c}A -\left(\frac{1}{3}R^2 - R_{\mu\nu} R^{\mu\nu}\right) = 0{\hskip 1em},\\& \frac{1}{\sqrt{-g}}\frac{\delta(\mathcal{L})}{\delta D} = \Box C - A = 0 {\hskip 1em}, 
\end{align}
\begin{align}
&\frac{1}{\sqrt{-g}}\frac{\delta(\mathcal{L})}{\delta A} = \Box_{c}B - D = 0 {\hskip 1em}\Rightarrow B{\hskip 1em} = \frac{1}{\Box_c} D {\hskip 1em}, \\& \frac{1}{\sqrt{-g}}\frac{\delta(\mathcal{L})}{\delta C} = \Box D +  \Lambda^{2}GRC = 0 {\hskip 1em} \Rightarrow {\hskip 1em} D  = - \frac{1}{\Box}G \Lambda^{2}R {\hskip 1em}, 
\end{align} 
 Equation of motion for fields (A,B,C,D) using FRW metric can be as

\begin{eqnarray}
{\ddot A} &\!\! = \!\!&
- 3H{\dot A} -(2-\epsilon)H^{2}A
- 12 (1 - \epsilon) H^4
\;\; , \label{eomAM4} \\
{\ddot B} &\!\! = \!\!&
- 3H{\dot B} 
-  (2 - \epsilon) H^2 B -D
\;\; , \\ \label{eomBM4}
{\ddot C} &\!\! = \!\!&
- 3H{\dot C} - A
\;\; , \\ \label{eomCM4}
{\ddot D} &\!\! = \!\!&
- 3H{\dot D} + G \Lambda^2 R
\;\; .  \label{eomDM4}
\end{eqnarray}
Now taking the variation of lagrangian with respect to metric and using the F.R.W. geometry, $(00)$-component of field equations becomes
\begin{equation}
\begin{split}
\frac{3H^2}{16\pi G} + 3H^{2}(\Lambda^{2}G)C & + 3H\partial_{t}(\Lambda^{2}GC)    -\frac{1}{2}(\dot{A}\dot{B} + \dot{C}\dot{D}) -6H^{3}\dot{B}   \\- &3(H\partial_{t} + H^{2})\frac{1}{6}AB + \frac{1}{2} AD = \frac{\Lambda}{16 \pi G}{\hskip 1em}, \label{00m4}
\end{split}
\end{equation}
and $(11) $-component becomes 
\begin{equation}
\begin{split}
&-(3-2\epsilon) \frac{H^{2}}{16 \pi G} -(3 -2\epsilon)H^{2}(\Lambda^{2}GC)+ H\partial_{t}(\Lambda^{2}GC)+ G\Lambda^{2}A -\frac{1}{6}\dot{A}\dot{B} -\frac{1}{2}\dot{C}\dot{D}  \\& {\hskip 1em}-2(1+2\epsilon)H^{3}\dot{B} -2(3- 2\epsilon)H^{4}B -(H\partial_{t} + H^{2})\frac{1}{6}AB -2DH^{2}- \frac{1}{3} AD  \\&{\hskip 1em}= \frac{\Lambda}{16 \pi G}{\hskip 1em},  \label{11m4}
\end{split}
\end{equation}
 Now adding equations (\ref{00m4}) and (\ref{11m4}), we find
\begin{equation}
\begin{split}
 & \frac{2\epsilon H^{2}}{16 \pi G}+ 2\epsilon H^{2}(\Lambda^{2}GC)+ 4H\partial_{t}(\Lambda^{2}GC)+\Lambda^{2}GA - \frac{2}{3}\dot{A}\dot{B} -\dot{C}\dot{D}  \\& {\hskip 1em} -2(3-2\epsilon)H^{4}B -4(2 + \epsilon)H^{3}\dot{B}-4(H\partial_{t} + H^{2})\frac{1}{6}AB - 2DH^{2}- \frac{1}{6} AD = 0  \label{00+11m4}
\end{split}
\end{equation}
To convert equations (\ref{eomAM4}) -(\ref{00+11m4}) in to dimensionless equations,  we use different set of dimensionless parameters i.e. $\left\lbrace\alpha ,\beta,\gamma,\delta\right\rbrace $  defined as:
\begin{align}
 & A \equiv \frac{-3\alpha}{G} {\hskip 2em},{\hskip 2em} B  \equiv -3\beta {\hskip 2em},{\hskip 2em} C \equiv 3\gamma {\hskip 2em}, {\hskip 2em} D \equiv \frac{9\delta}{G} {\hskip 1em},.
\end{align} 
The set of dimensionles parameters we wish to solve for are $\left\lbrace\alpha ,\beta,\gamma,\delta,\chi,\epsilon\right\rbrace $, which are subjected to following initial conditions  at $ \tau = \tau_{in}$  :
\begin{align}
 & \alpha = \alpha' = \beta = \gamma = \gamma^{'} = \delta^{'} = \delta = 0 ,\beta^{'} = 10 {\hskip 2em},\\ & \chi = \chi_{in} {\hskip 2em},{\hskip 2em} \epsilon = 0 {\hskip 1em}.
\end{align} 
{\hskip 2em} Using these dimensionless variables, equation of motion for $\left\lbrace\alpha ,\beta,\gamma,\delta \right\rbrace $  can be written in following forms :
\begin{equation}
 \alpha^{''} + 3\frac{\chi}{\chi_{in}}\alpha^{'} + (2 -\epsilon)\frac{\chi^{2}}{\chi^{2}_{in}}\alpha = 4(1-\epsilon)\frac{\chi^{4}}{\chi^{2}_{in}} {\hskip 1em} \label{alpham4}
\end{equation}
\begin{equation}
\beta^{''} + 3\frac{\chi}{\chi_{in}}\beta^{'} + (2 -\epsilon)\frac{\chi^{2}}{\chi^{2}_{in}}\beta = \frac{3\delta}{\chi^{2}_{in}} {\hskip 1em}, \label{betam4}
\end{equation}
\begin{equation}
\gamma^{''} + 3\frac{\chi}{\chi_{in}}\gamma{'} = \frac{\alpha}{\chi^{2}_{in}} {\hskip 1em} ,\label{gammam4}
\end{equation}
\begin{equation}
\delta^{''} + 3\frac{\chi}{\chi_in} \delta^{'} = 6 \chi_{in}^{2}\chi^{2}(2-\epsilon) {\hskip 1em}, \label{deltam4}
\end{equation}

Furthermore, the variable $ \chi $ is obtained from equation (\ref{00m4});
\begin{equation}
\begin{split}
 \left[2\chi_{in}\beta^{'}\right]\chi^{3} & + \left[\frac{1}{3}+ 9\gamma \chi_{in}^{4} -\frac{1}{2}\alpha \beta  \right]\chi^{2} + \chi_{in}\partial_{t}\left[- \frac{1}{2}\alpha \beta + 9\chi_{in}^{4}\gamma \right] \chi \\& -\chi^{2}_{in}\left[\frac{1}{3} 
 + \frac{1}{2}\alpha^{'}\beta^{'} + \frac{3}{2}\gamma^{'}\delta^{'} -\frac{3}{2}\alpha\delta\right] = 0{\hskip 1em}, \label{chim4}
\end{split}
\end{equation}
Moreover, we solve for  $\epsilon $ from the constraint equation (\ref{00+11m4}) and we get :

\begin{equation}
\begin{split}
\epsilon\chi^{2} & = \frac{3}{2 + 54\chi_{in}^{2}\gamma + 12\beta\chi^{2} -12\chi\chi_{in}\beta'}\times \\& \Bigg\{ 9\chi_{in}^{2}\alpha -9\alpha \delta - 18\chi^{2}\beta +18\chi^{2}\delta +\frac{4}{3}\chi^{3} -\frac{4}{3}\chi_{in}^{2} + 36\chi_{in}^{4}\chi^{4}\gamma + 3\chi_{in}\gamma'\delta' \Bigg\}.
\end{split} \label{epsilonm4}
\end{equation}
\subsection{Results}

\begin{figure}[!ht]
  \centering
 \includegraphics[scale=0.4]{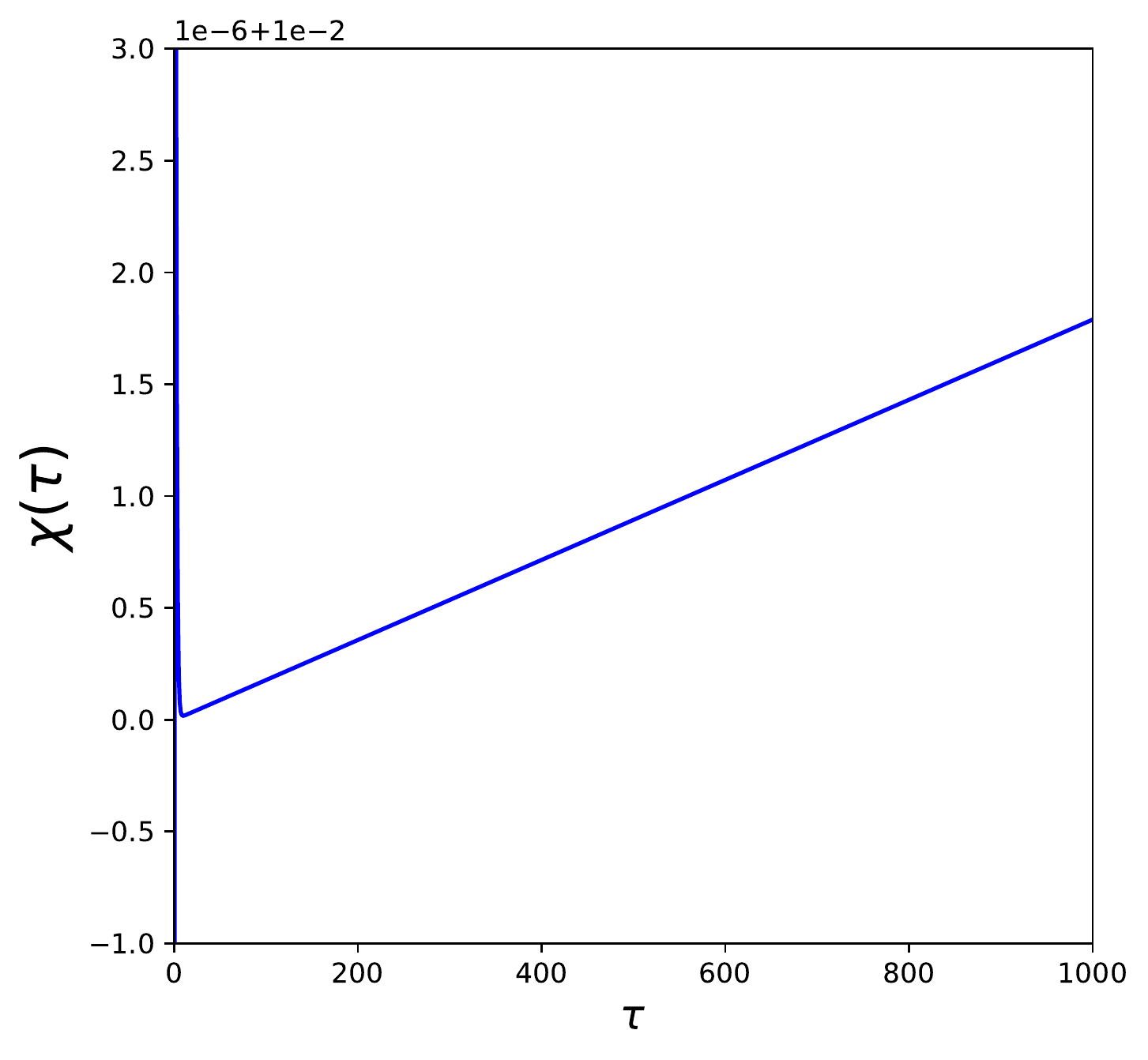} 
 \includegraphics[scale=0.4]{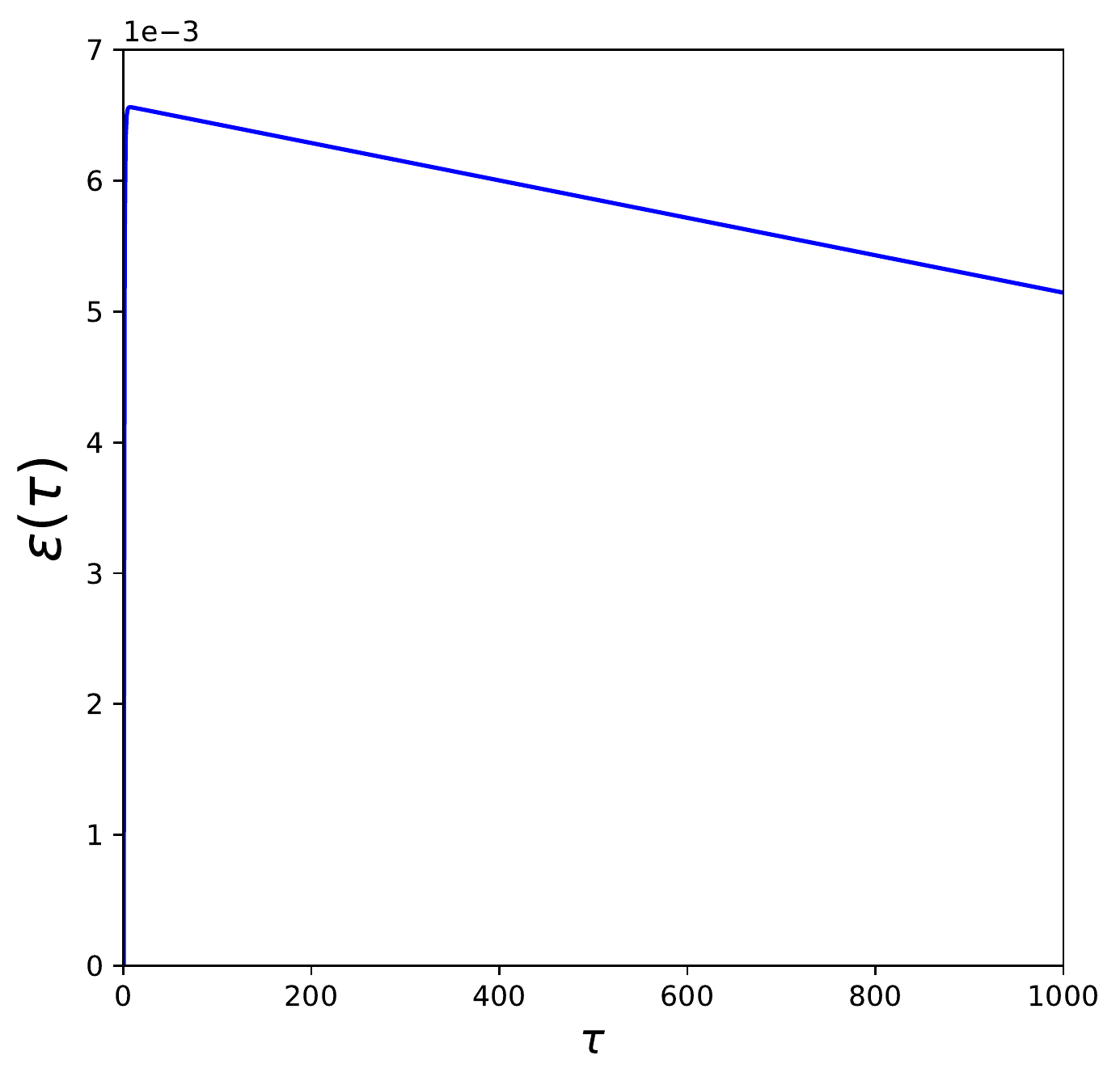}
  
\caption[Plots for  $\chi(\tau)$ ,$ \epsilon(\tau)$  and $w_{eff}$  for case 4]{Plots for Geometric quantities $\chi(\tau)$ and $ \epsilon(\tau)$  and $w_{eff}$ for $\chi_{in} = \frac{1}{100}$ and $ f(X) = X $. } \label{plotchi1000m4} 
\end{figure}

Here we perform a similar analysis campared to model \rom{1} and \rom{2}. In this model scalar fields are approximated, for large values of $ \tau $, as 
\begin{eqnarray}
\alpha(\tau) &\!\! = \!\!& 
2 \chi_{\rm in}^2 \Bigl( 1 \!-\! e^{-\tau}\Bigr)^2 
\; \longrightarrow \; 
2 \chi_{\rm in}^2 
\; , \label{earlyalpham4} \\
\beta(\tau) &\!\! = \!\!& 
\frac{1}{6}\chi_{\rm in}^2 \Bigl( \tau \!-\! \frac{3}{2}
\!+\! 3 e^{-\tau} \!-\! \frac32 e^{-2\tau} 
 \Bigr) 
\; \longrightarrow \;
\frac{1}{6} \chi_{\rm in}^2 \Bigl( \tau \!-\! \frac{3}{2} \Bigr) 
\; , \label{earlybetam4} \\
\gamma(\tau) &\!\! = \!\!& 
\frac23  \Bigl( \tau \!-\! \frac{11}{6}
\!+\! 3 e^{-\tau} \!-\! \frac32 e^{-2\tau} 
\!-\! \frac13 e^{-3\tau} \Bigr) 
\; \longrightarrow \;
\frac23  \Bigl( \tau \!-\! \frac{11}{6}
\Bigr) \; , \label{earlygammam4} \\
\delta(\tau) &\!\! = \!\!& 
4\chi_{\rm in}^4 \Bigl( \tau \!-\! \frac{1}{3}
\!+\! \frac{1}{3} e^{-3\tau} \Bigr) 
\; \longrightarrow \; 
4 \chi_{\rm in}^2 
\Bigl( \tau \!-\! \frac{1}{3} \Bigr) 
\; . \label{earlydeltam4}
\end{eqnarray}
\begin{figure}[ht]
 \includegraphics[width = .45\textwidth, height=0.3\textheight]{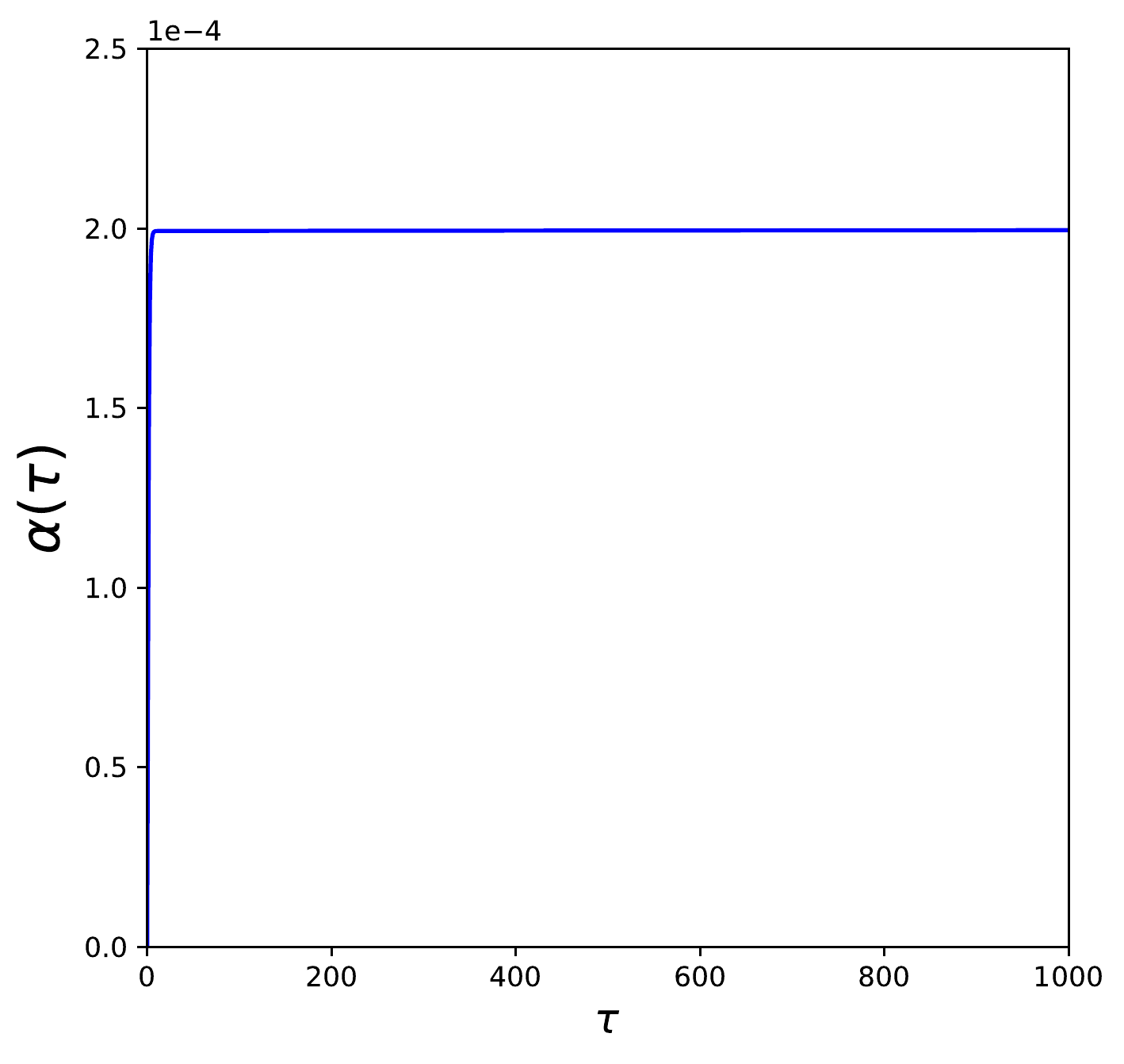} {\hskip 2em}
 \includegraphics[width = .45\textwidth, height=0.3\textheight]{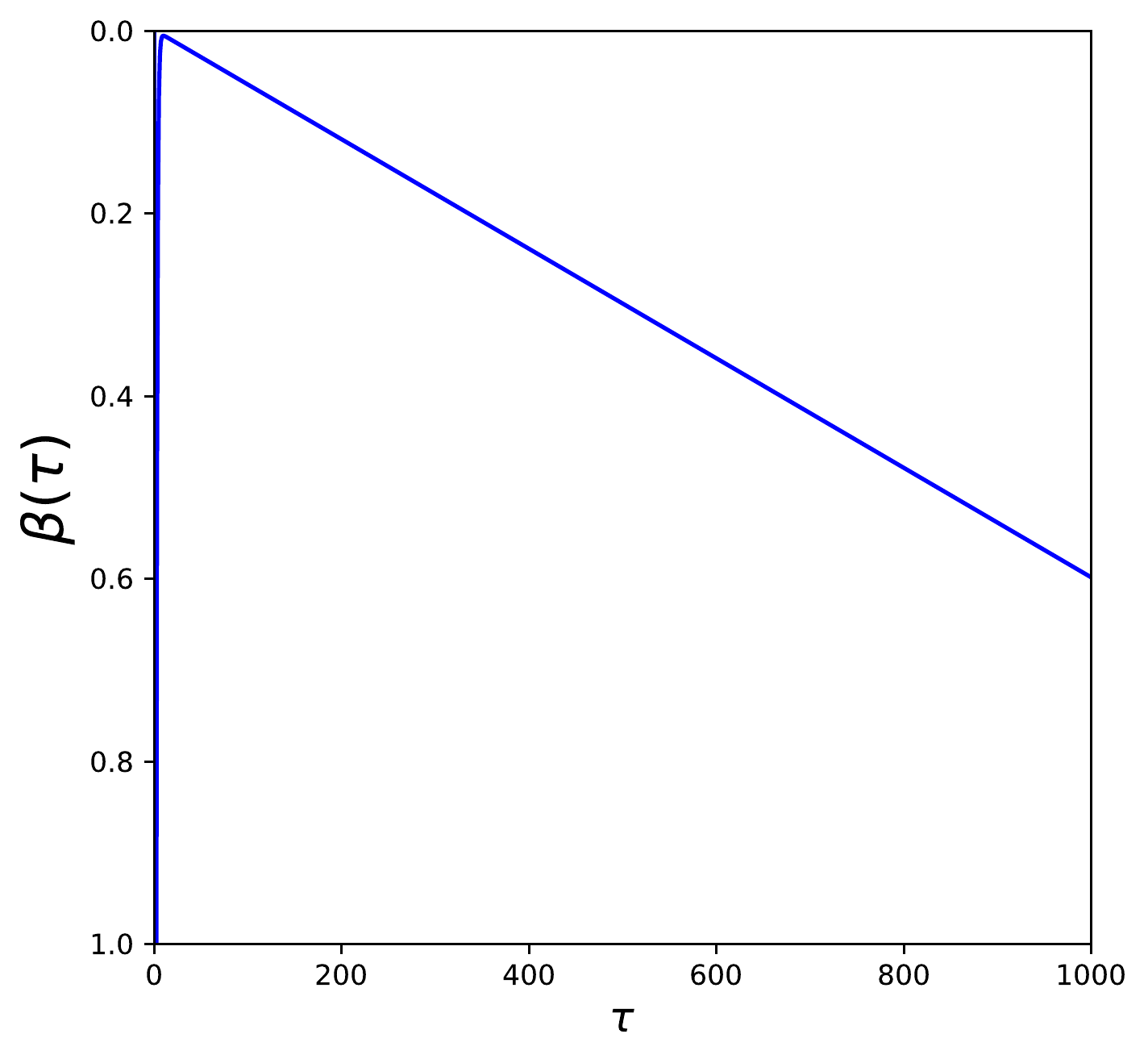} 
\caption[Plots for scalars $\alpha(\tau)$ and $ \beta(\tau)$ for case 4]{Plots for scalar field $\alpha(\tau)$ and $ \beta(\tau)$  for $\chi_{in} = \frac{1}{100}$ and $ f(X) = X $. } \label{plotalpha1000m4} 
\end{figure}
\begin{figure}[!ht]
  \centering
 \includegraphics[scale=0.4]{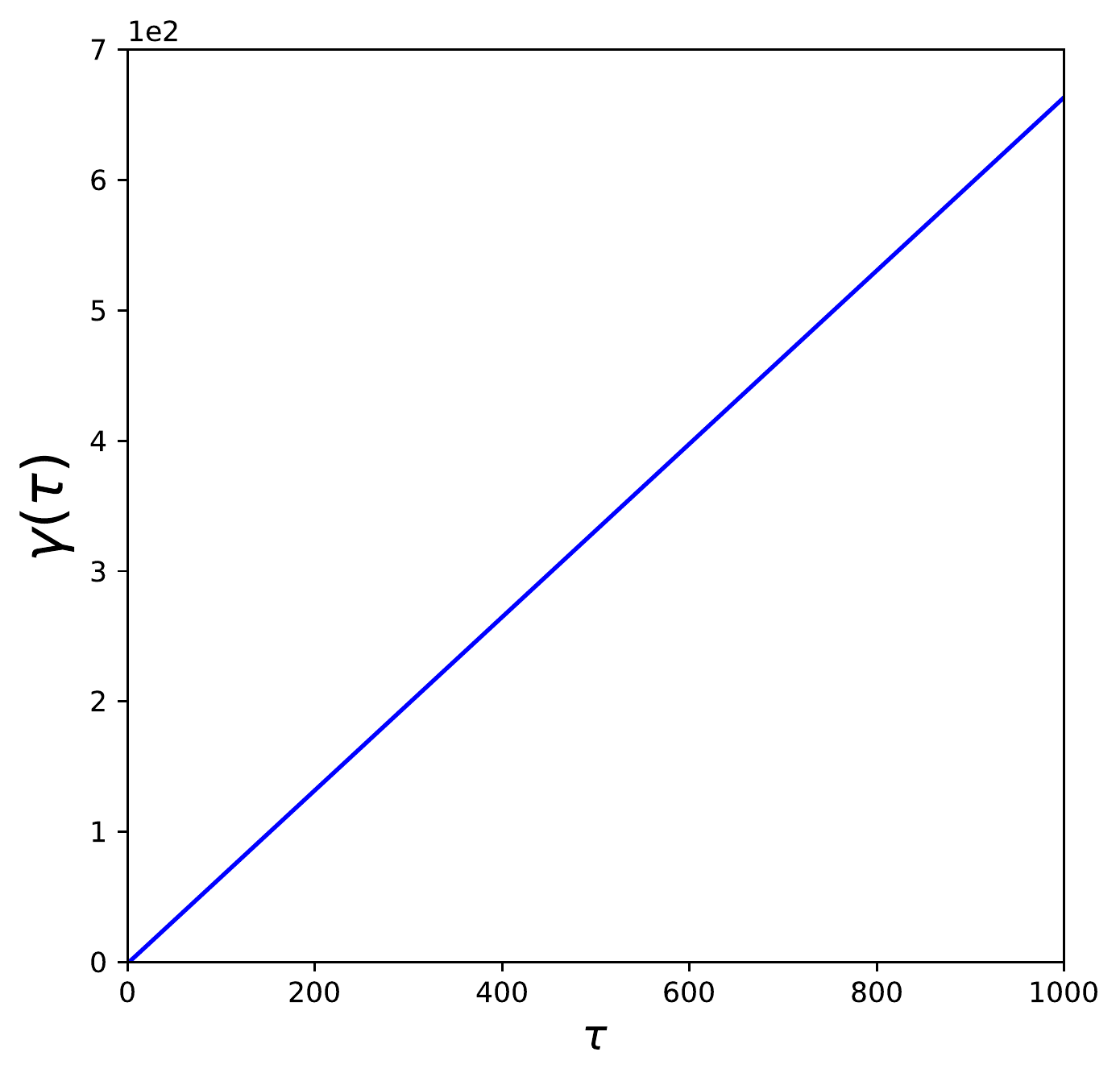} \hspace{0.5 cm}
 \includegraphics[scale=0.4]{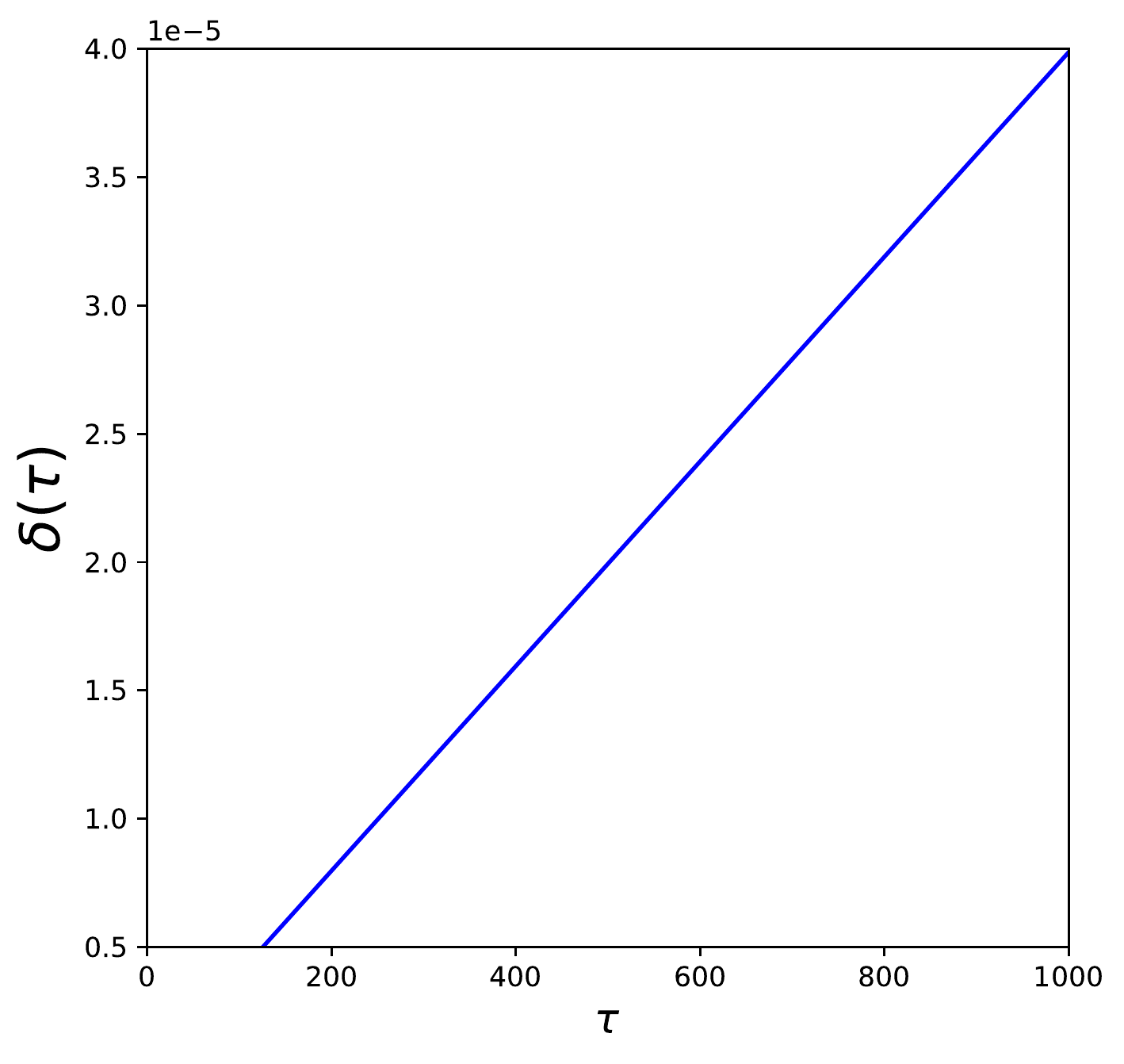} 
\caption[Plots for scalar $\gamma(\tau)$ and $ \delta(\tau)$  for case 4]{Plots for scalar field $\gamma(\tau)$ and $ \delta(\tau)$  for $\chi_{in} = \frac{1}{100}$ and $ f(X) = X $. } \label{plotdelta1000m4} 
\end{figure}
\paragraph{}
Figure \ref{plotalpha1000m4} and Figure \ref{plotdelta1000m4} shows a rough agreement with the (\ref{earlyalpham4}) - (\ref{earlydeltam4}). Unlike other models here the effect of the coupling between $\gamma (\tau)$ and $ \alpha (\tau)$ ,$\beta (\tau)$ and  between $\delta (\tau)$ is not significant enough to provide any curvature in the behaviour of scalar fields as seen in Figure \ref{plotalpha1000m4} and \ref{plotdelta1000m4}. In this case derivative of auxilary scalar fields $\alpha,\beta,\gamma,\delta $ becomes
\begin{equation}
\alpha'(\tau) \longrightarrow 
0 \;\; , \;\; 
\beta'(\tau) \longrightarrow 
 \frac{1}{6}\chi_{\rm in}^2 \;\; , \;\;
\gamma'(\tau) \longrightarrow 
\frac23  \;\; , \;\;
\delta'(\tau) \longrightarrow 
4 \chi_{\rm in}^4 \; .
\end{equation}
{\hskip 2em} From these derivatives, we can draw conclusions regarding the behaviour of $\chi(\tau )$ and $ \epsilon(\tau)$. It is evident in Figure \ref{plotchi1000m4} that $\epsilon \longrightarrow $ 1 after a very long $ \tau $. In our case this happens for $\tau = 10^{6}$. Similar to Model \rom{3}, exit from inflation to RD era is also not possible in this model.
\begin{figure}[!ht]
  \centering
 
  \includegraphics[scale=0.4]{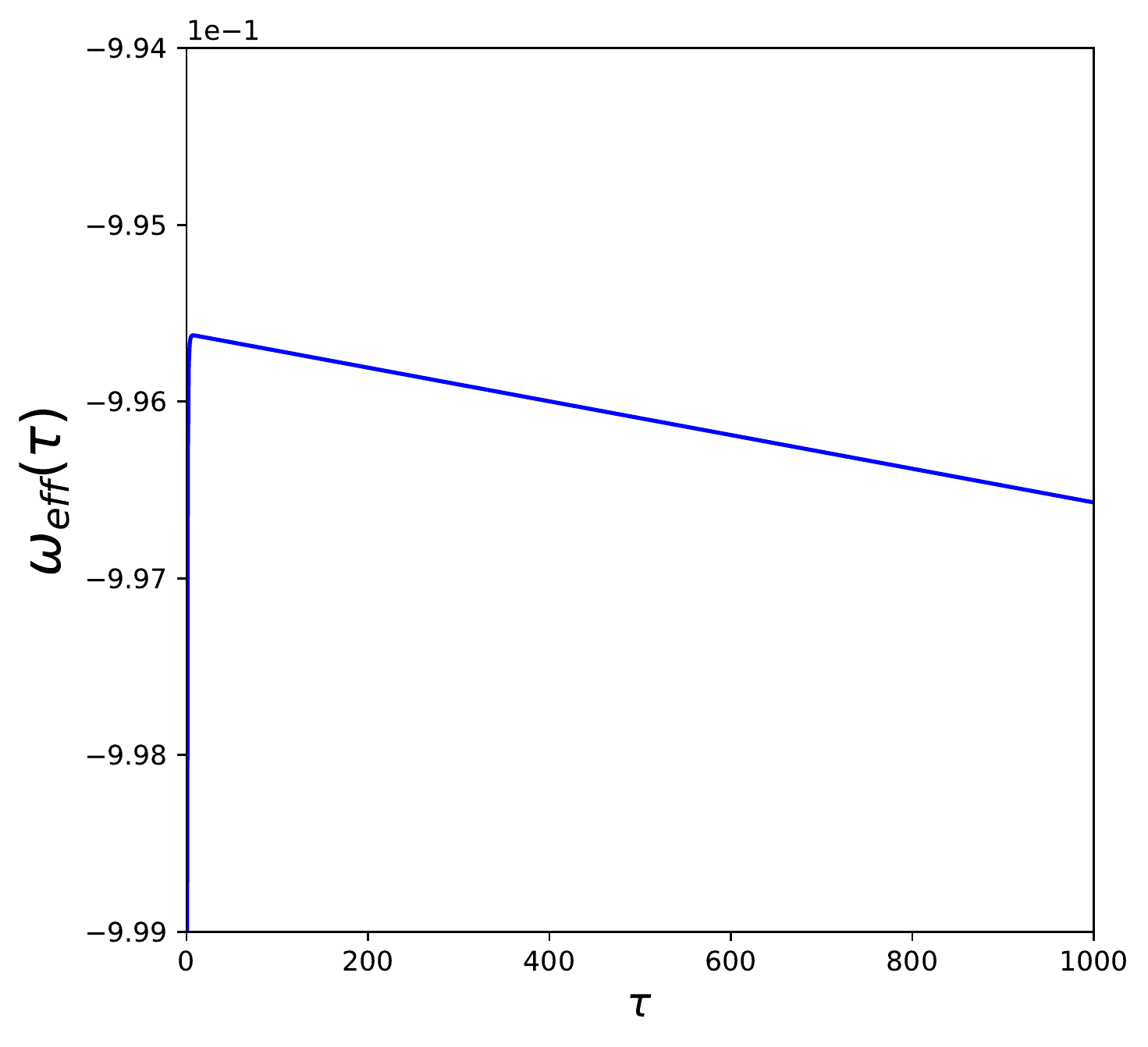}
 \includegraphics[scale=0.4]{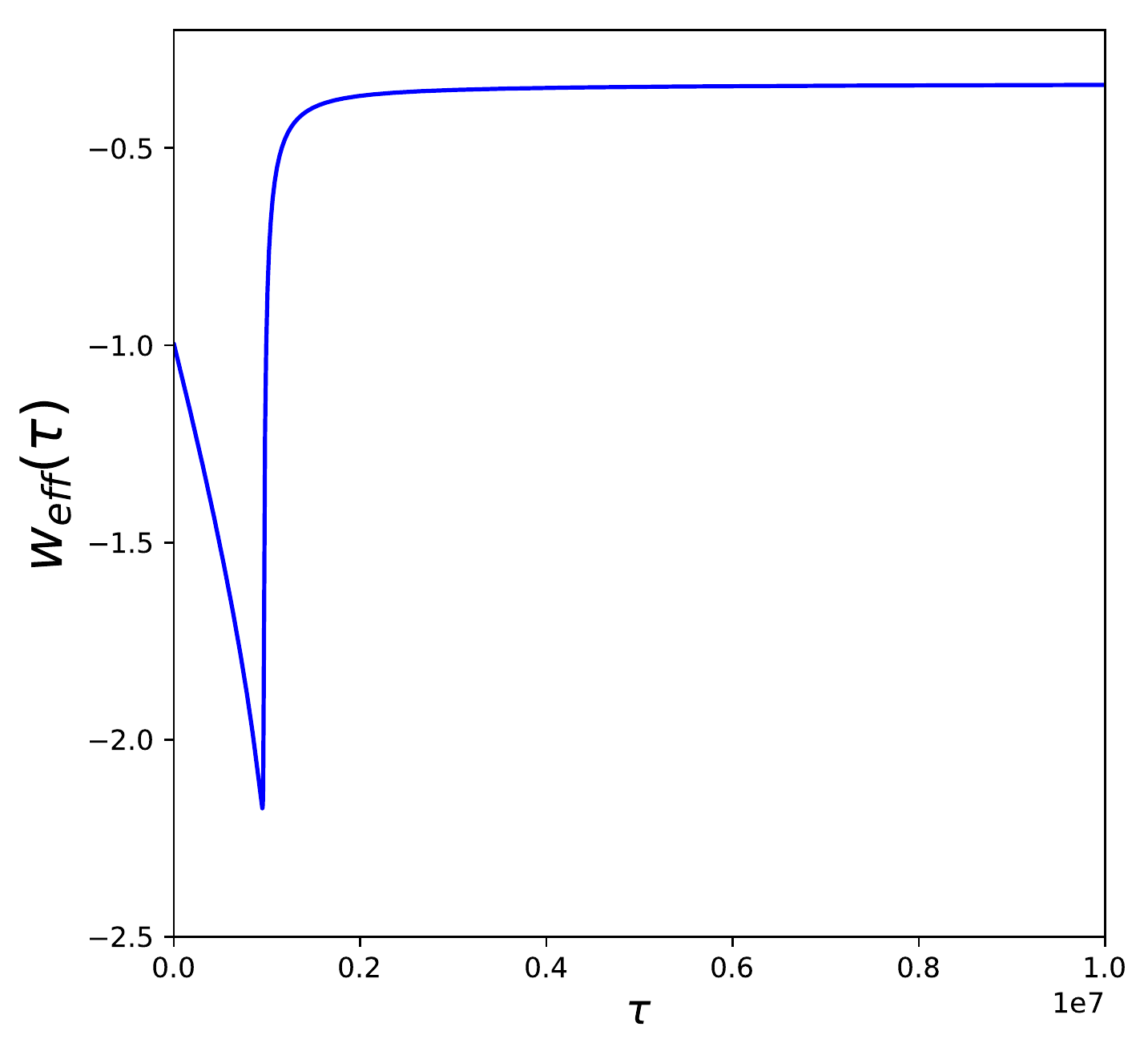}  
\caption[Plots for  $\chi(\tau)$ ,$ \epsilon(\tau)$  and $w_{eff}$  for case 4]{Plots for Geometric quantities $\chi(\tau)$ and $ \epsilon(\tau)$  and $w_{eff}$ for $\chi_{in} = \frac{1}{100}$ and $ f(X) = X $. } \label{plotchi7000m4} 
\end{figure}
\section{Model \rom{5}} \label{sec5}
Here, we consider following form of  $ X[g] $ 
\begin{equation}
X[g] = G\left(\frac{1}{3}R^2 - R_{\mu\nu} R^{\mu\nu}\right) \frac{1}{\Box}\frac{1}{\Box_{c}}R
\end{equation}
In this case we will also restrict ourself to particular form of $h(X[g]) = X[g]$ .  Similar to Model ($\rom{4}$)any other form of  $ h(X[g]) $ will give rise to higher derivative in the equation of motion. \newline For $h(X[g]) = X[g] $, lagrangian is written as
\begin{equation}
\begin{split}
\mathcal{L} = & \Lambda^{2}h(GC)\left(\frac{1}{3}R^2 - R_{\mu\nu} R^{\mu\nu}\right)\sqrt{-g} + B\left[\Box_{c}A - R \right]\sqrt{-g} \\&+ D[\Box C -A]\sqrt{-g} {\hskip 1em},  \label{model 5}
\end{split}
\end{equation}
Here we introduce two auxilary scalar fields A, C and two lagrange multiplier B, D , to convert nonlocal lagrangian into local one, which obeys following equation of motion  :
\begin{align}
&\frac{1}{\sqrt{-g}}\frac{\delta(\mathcal{L})}{\delta B} = \Box_{c}A -R = 0{\hskip 1em},\\& \frac{1}{\sqrt{-g}}\frac{\delta(\mathcal{L})}{\delta D} = \Box C - A = 0 {\hskip 1em}, 
\end{align}
\begin{align}
&\frac{1}{\sqrt{-g}}\frac{\delta(\mathcal{L})}{\delta A} = \Box_{c}B - D = 0 {\hskip 1em}\Rightarrow B{\hskip 1em} = \frac{1}{\Box_{c}} D {\hskip 1em}, \\& \frac{1}{\sqrt{-g}}\frac{\delta(\mathcal{L})}{\delta C} = \Box D +  \Lambda^{2}G\left(\frac{1}{3}R^2 - R_{\mu\nu} R^{\mu\nu}\right) = 0 {\hskip 1em} \Rightarrow {\hskip 1em} D  = - \frac{1}{\Box_{c}}\Lambda^{2}G\left(\frac{1}{3}R^2 - R_{\mu\nu} R^{\mu\nu}\right) {\hskip 1em}, 
\end{align} 
The equation of motion for fields for FRW metric can be written as 

\begin{eqnarray}
{\ddot A} &\!\! = \!\!&
- 3H{\dot A} -(2-\epsilon)H^{2}A
- 6 (2 - \epsilon) H^2
\;\; , \label{eomam5} \\
{\ddot B} &\!\! = \!\!&
- 3H{\dot B} 
-  (2 - \epsilon) H^2B -D 
\;\; , \\ \label{eombm5}
{\ddot C} &\!\! = \!\!&
- 3H{\dot C} - A
\;\; , \\ \label{eomcm5}
{\ddot D} &\!\! = \!\!&
- 3H{\dot D} +   12 \Lambda^{2}G(1 -\epsilon)H^4
\;\; .  \label{eomdm5}
\end{eqnarray}
Now taking the variation of lagrangian with respect to metric and using the F.R.W. geometry $(00)$ component of field equations is
\begin{equation}
\begin{split}
\frac{3H^2}{16\pi G} + 6H^{3}\dot{C}(\Lambda^{2}G)  & -\frac{1}{2}(\dot{A}\dot{B} + \dot{C}\dot{D})    \\- &3(H\partial_{t} + H^{2})(\frac{1}{6}AB + B) -\frac{1}{2}AD = \frac{\Lambda}{16 \pi G}{\hskip 1em}, \label{00m5}
\end{split}
\end{equation}
and $(11) $-component is :
\begin{equation}
\begin{split}
-(3-2\epsilon) \frac{H^{2}}{16 \pi G} &+ 2H^{3}(1+2\epsilon)\dot{C}(\Lambda^{2}G) -2H^{2}A(\Lambda^{2}G) -\frac{1}{6}\dot{A}\dot{B} -\frac{1}{2}\dot{C}\dot{D} \\&-(H^{2} + H\partial_{t})(\frac{1}{6}AB + B)- \frac{1}{6}AD -D = - \frac{\Lambda}{16\pi G},
\end{split} \label{11m5}
\end{equation}
 Now addition of (\ref{00m5}) and (\ref{11m5}) leads to  :
\begin{equation}
\begin{split}
 & \frac{2\epsilon H^{2}}{16 \pi G} + 4H^{3}(2+\epsilon)(\Lambda^{2}G)\dot{C} -2H^{2}A(\Lambda^{2}G) -\frac{2}{3}\dot{A}\dot{B} -\dot{C}\dot{D} \\& {\hskip 2em} -4(H^{2} + H\partial_{t})(\frac{1}{6}AB + B) -\frac{1}{6}AD  = 0
\end{split}  \label{00+11m5}
\end{equation}
To convert equations (\ref{eomam5}) -(\ref{00+11m5}) into dimensionaless equations,  we use  set of dimensionless parameters i.e. $\left\lbrace\alpha ,\beta,\gamma,\delta\right\rbrace $  defined as:
\begin{align}
 & A \equiv -3\alpha,{\hskip 2em} B  \equiv \frac{-9\beta}{G} {\hskip 2em},{\hskip 2em} C \equiv 3\gamma G {\hskip 2em}, {\hskip 2em} D \equiv \frac{9\gamma}{G^{2}} {\hskip 1em},.
\end{align}
The set of dimensionles parameters we wish to solve for $\left\lbrace\alpha ,\beta,\gamma,\delta,\chi,\epsilon\right\rbrace $ is subjected to following initial conditions  at $ \tau = \tau_{in}$  :
\begin{align}
 & \alpha = \alpha' = \beta = \beta^{'} = \gamma = \delta^{'} = \delta = 0 ,\gamma^{'} = 10 ,\\ & \chi = \chi_{in},\epsilon = 0.
\end{align} 

{\hskip 2em} Using these dimensionless variables equation of motion for $\left\lbrace\alpha ,\beta,\gamma,\delta \right\rbrace $  can be cast as 
\begin{align}
 & \alpha^{''} + 3\frac{\chi}{\chi_{in}}\alpha^{'} + (2-\epsilon)\frac{\chi^{2}}{\chi^{2}_{in}} \alpha  = 2(2-\epsilon)\frac{\chi^{2}}{\chi^{2}_{in}} {\hskip 1em},\\&  \beta^{''} + 3\frac{\chi}{\chi_{in}}\beta^{'} + (2 -\epsilon)\frac{\chi^{2}}{\chi_{in}^{2}}\beta  = \frac{\delta}{\chi_{in}^{2}} {\hskip 1em}, \\& \gamma^{''} + 3\frac{\chi}{\chi_{in}}\gamma{'} = \frac{\alpha}{\chi^{2}_{in}} {\hskip 1em},\\& \delta^{''} + 3\frac{\chi}{\chi_in} \delta^{'}  = 12(1-\epsilon)\chi_{in}^{2} \chi^{4}{\hskip 1em},
\end{align}
Furthermore, the variable $ \chi $ is solved from the dimensionless form of the $ (00)$ component of field equation 
\begin{equation}
\begin{split}
 \left[6\chi_{in}^{5}\gamma^{'}\right]\chi^{3} & + \left[\frac{1}{9} -\frac{1}{2}\alpha \beta +\beta \right]\chi^{2} + \chi_{in}\partial_{t}\left[- \frac{1}{2}\alpha \beta + \beta \right] \chi \\& -\chi^{2}_{in}\left[\frac{1}{9} - \frac{1}{2}(\alpha^{'}\beta^{'} + \gamma^{'}\delta^{'})\right] = 0{\hskip 1em},
\end{split}
\end{equation}
Moreover, we solve for $\epsilon $ from the equation (\ref{00+11m5}) and we get :
\begin{equation}
\begin{split}
\epsilon \chi^{2}& {\hskip 2em} = \frac{9}{\Big[ 2 + 108\chi_{in}^{5}\chi\gamma'\Big]} \times  \\ &\Bigg\{ \frac{4}{9}\chi^{2}-\frac{4}{9}\chi_{in}^{2} -2\chi_{in}^{2}\alpha'\beta'-6\chi_{in}^{4}\chi^{2}\alpha + 3\chi_{in}^{2}\gamma'\delta' + \delta +2\delta \alpha ) \Bigg\}.
 \end{split}
\end{equation}
\subsection{Results} 

Here we perform a similar analysis compared to previous models. For this model, scalar fields are approximated, for large values of $ \tau $, as 

\begin{eqnarray}
\alpha(\tau) &\!\! = \!\!& 
2 \!+\! 2 e^{-2\tau}\!-\! 4 e^{-\tau}
\; \longrightarrow \; 
2 
\; , \label{earlyalpham5} \\
\beta(\tau) &\!\! = \!\!& 
2\chi_{\rm in}^4 \Bigl( \tau \!-\! \frac{11}{6}
\!-\! \frac{1}{3} e^{-3\tau} \!-\! \frac{3}{2} e^{-2\tau} 
\!+\! 3 e^{-\tau} \Bigr) 
\; \longrightarrow \;
2 \chi_{\rm in}^4 \Bigl( \tau \!-\! \frac{11}{6} \Bigr) 
\; , \label{earlybetam5} \\
\gamma(\tau) &\!\! = \!\!& 
\frac23 \chi_{in}^{-2} \Bigl( \tau \!-\! \frac{11}{6}
\!+\! \frac13 e^{-3\tau} \!-\! \frac{3}{2}e^{-2\tau} 
\!+\!3 e^{-\tau}  \Bigr) 
\; \longrightarrow \;
\frac23 \chi_{in}^{-2} \Bigl( \tau \!-\! \frac{11}{6}
\Bigr) \; , \label{earlygammam5} \\
\delta(\tau) &\!\! = \!\!& 
4\chi_{\rm in}^6 \Bigl( \tau \!-\! \frac{1}{3}
\!+\! \frac{1}{3} e^{-3\tau} \Bigr) 
\; \longrightarrow \; 
4 \chi_{\rm in}^6 
\Bigl( \tau \!-\! \frac{1}{3} \Bigr) 
\; . \label{earlydeltam5}
\end{eqnarray}

\begin{figure}[!ht]
  \centering
  \includegraphics[scale=.4]{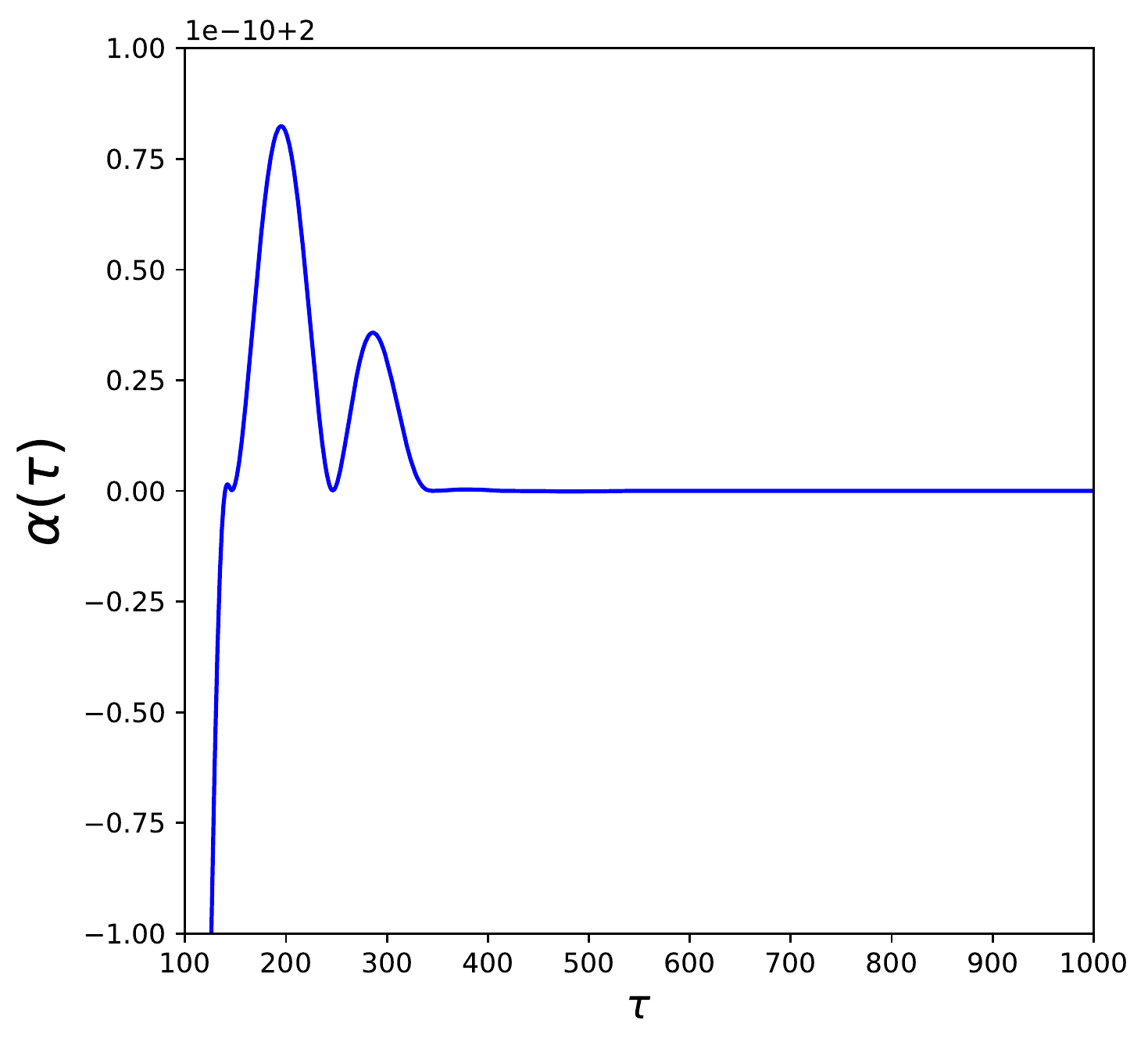} 
   \includegraphics[scale=.4]{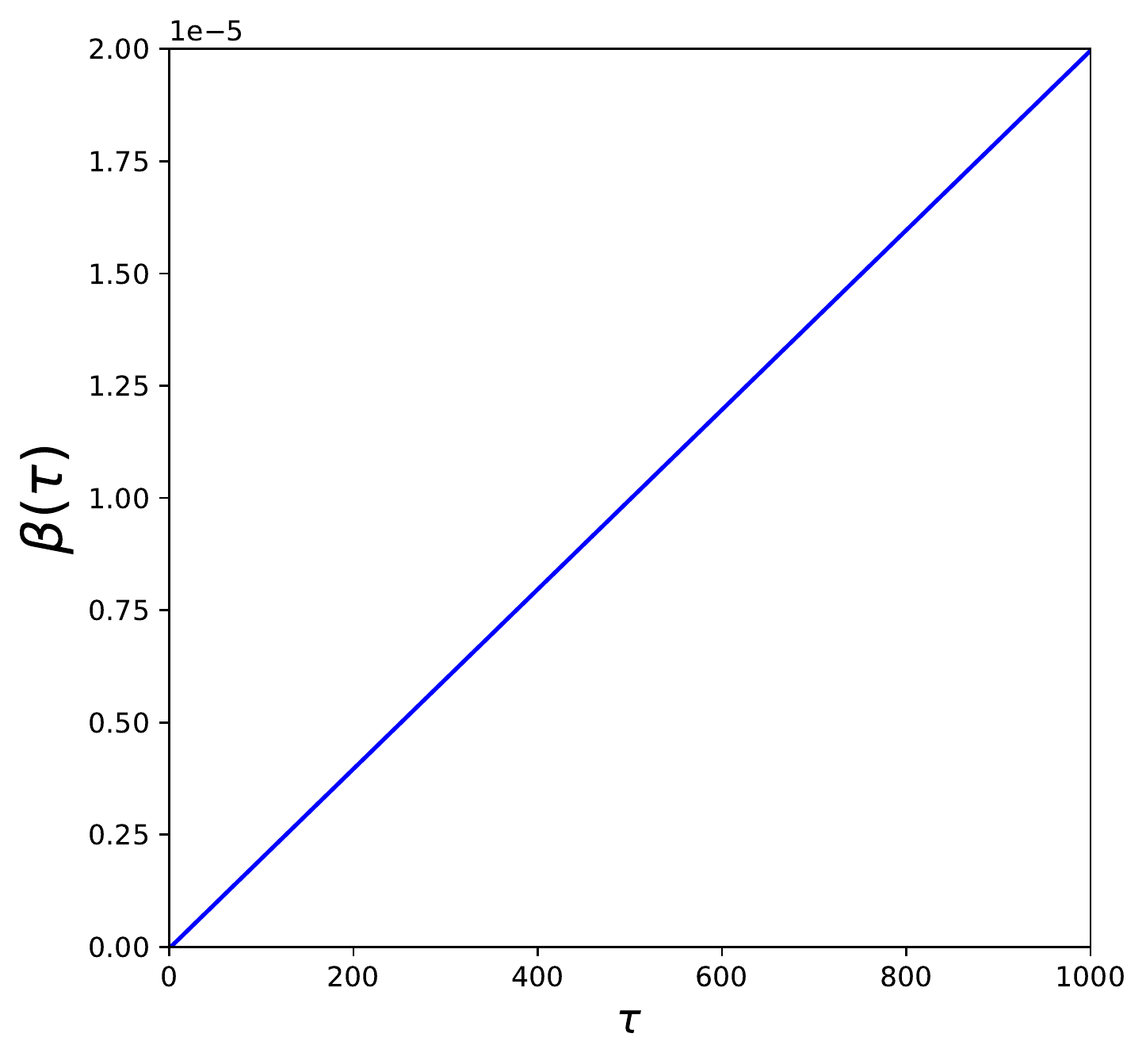}  
\caption[Plots for scalar $\alpha(\tau)$ and $ \beta(\tau)$ for case 5]{Plots for scalar field $\alpha(\tau)$ and $ \beta(\tau)$  for $\chi_{in} = \frac{1}{100}$ and $ f(X) = X $. } \label{plotalpha1000m5} 
\end{figure}

\begin{figure}[!ht]
  \centering
  \includegraphics[scale=.4]{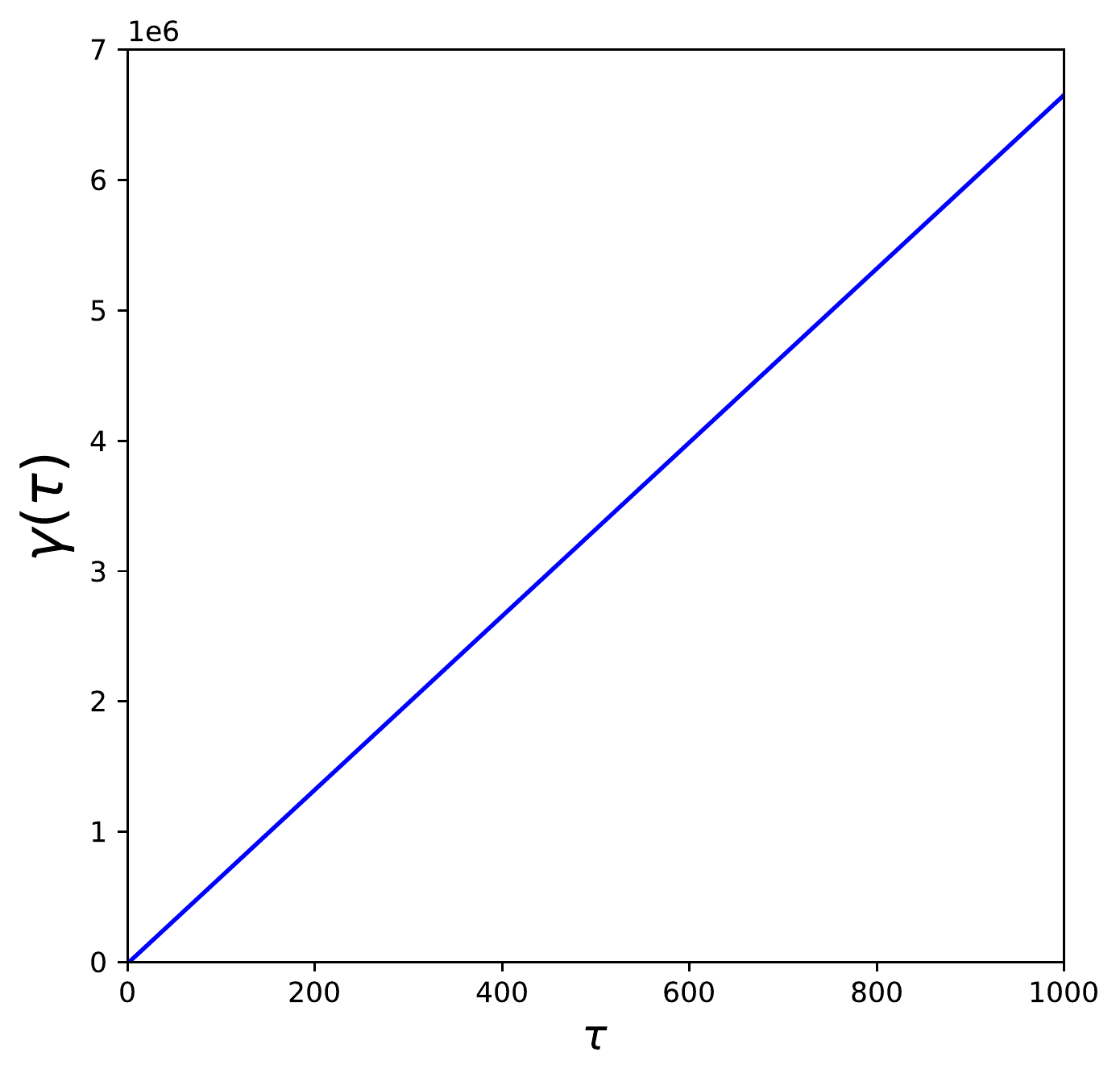} 
   \includegraphics[scale=.4]{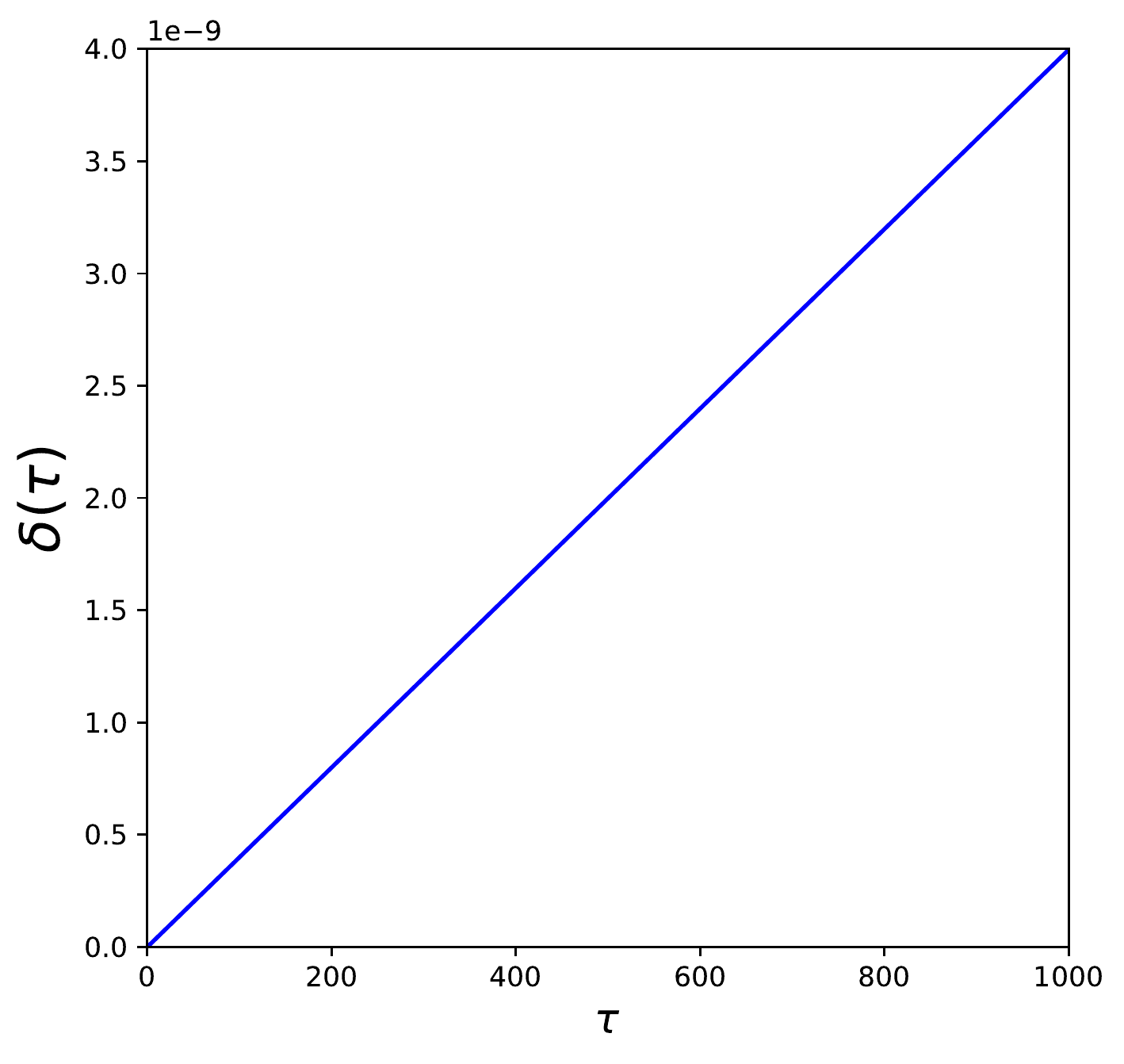} 
\caption[Plots for scalar $\gamma(\tau)$ and $ \delta(\tau)$ for case 5]{Plots for scalar field $\gamma(\tau)$ and $ \delta(\tau)$  for $\chi_{in} = \frac{1}{100}$ and $ f(X) = X $.  } \label{plotdelta1000m5} 
\end{figure}
Figure \ref{plotalpha1000m5} shows behaviour of auxilary scalars $\alpha(\tau)$ and $ \beta(\tau) $  in time range $0 \leq \tau \leq 1000$ . Examining the Figure \ref{plotalpha1000m5} in the time range  $ \tau \leq 400 $, we find that field $ \alpha(\tau)$ shows small peaks  around $ 2.0000005 $ while $ \beta(\tau)$ shows a linear dependence. Numerical results for sclar fields $\gamma (\tau)$ and $ \delta(\tau)$ match with analytical expressions (\ref{earlygammam5}) and (\ref{earlydeltam5}).
 
From relations (\ref{earlyalpham5}) -(\ref{earlydeltam5}) we compute  expressions for first derivative of auxilary salar fields during de Sitter epoch as
  \begin{equation}
\alpha'(\tau) \longrightarrow 
0 \;\; , \;\; 
\beta'(\tau) \longrightarrow 
 2\chi_{\rm in}^4 \;\; , \;\;
\gamma'(\tau) \longrightarrow 
\frac23 \chi_{in}^{-2}  \;\; , \;\;
\delta'(\tau) \longrightarrow 
4 \chi_{\rm in}^6 \; . \label{scalarderivativem5}
\end{equation}

\begin{figure}[!ht]
  \centering
  \includegraphics[scale=.4]{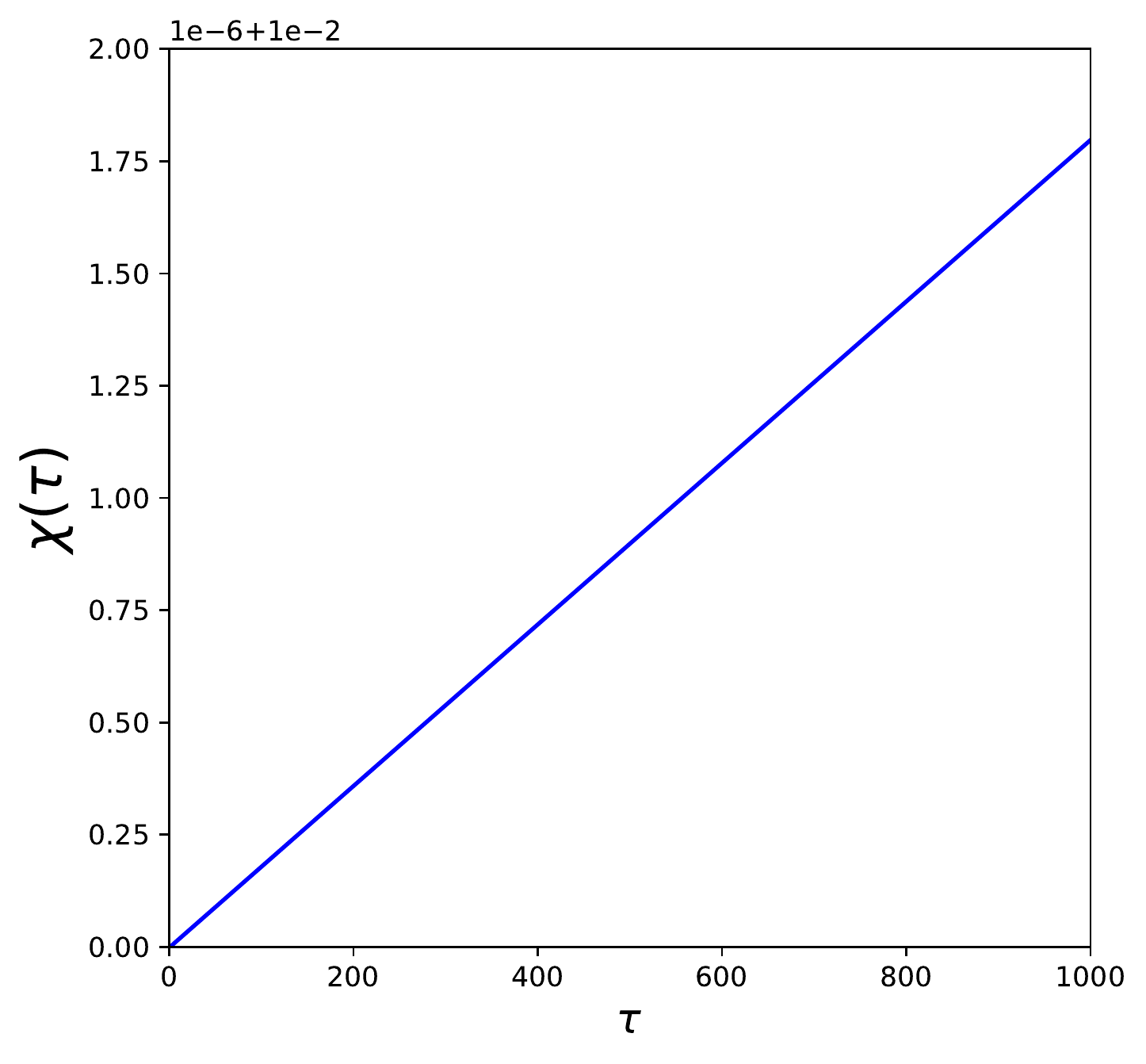} 
 \includegraphics[scale=.4]{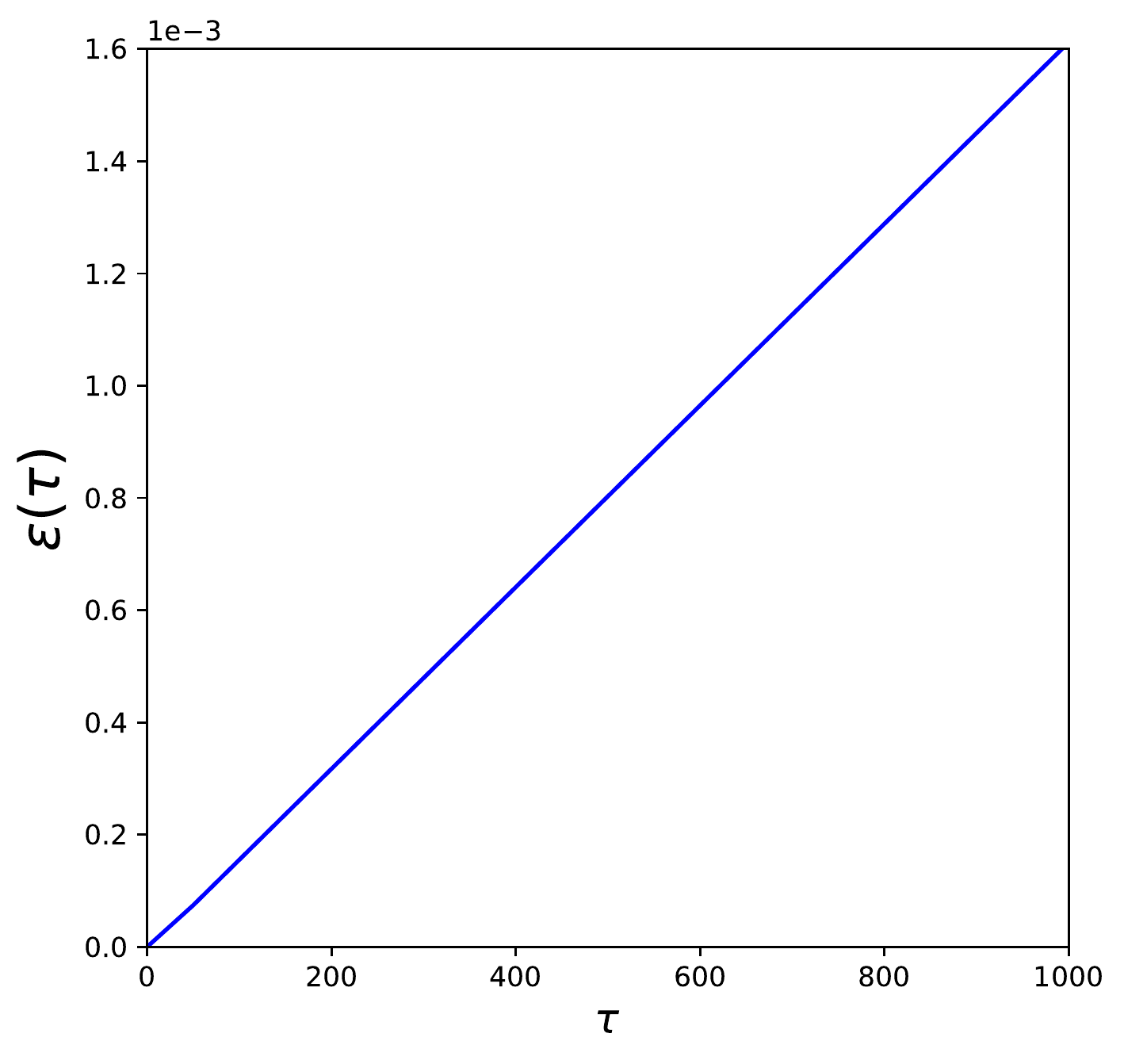}   
\caption[Plots for $\chi(\tau)$  and $ \epsilon(\tau)$ for case 5]{Plots for Geometric quantities  $\chi(\tau)$  and $ \epsilon(\tau)$ for $\chi_{in} = \frac{1}{100}$ and $ f(X) = X $. }\label{plotchiepsilonlatem5} 
\end{figure}
\begin{figure}[!ht]
  \centering 
 \includegraphics[scale=.4]{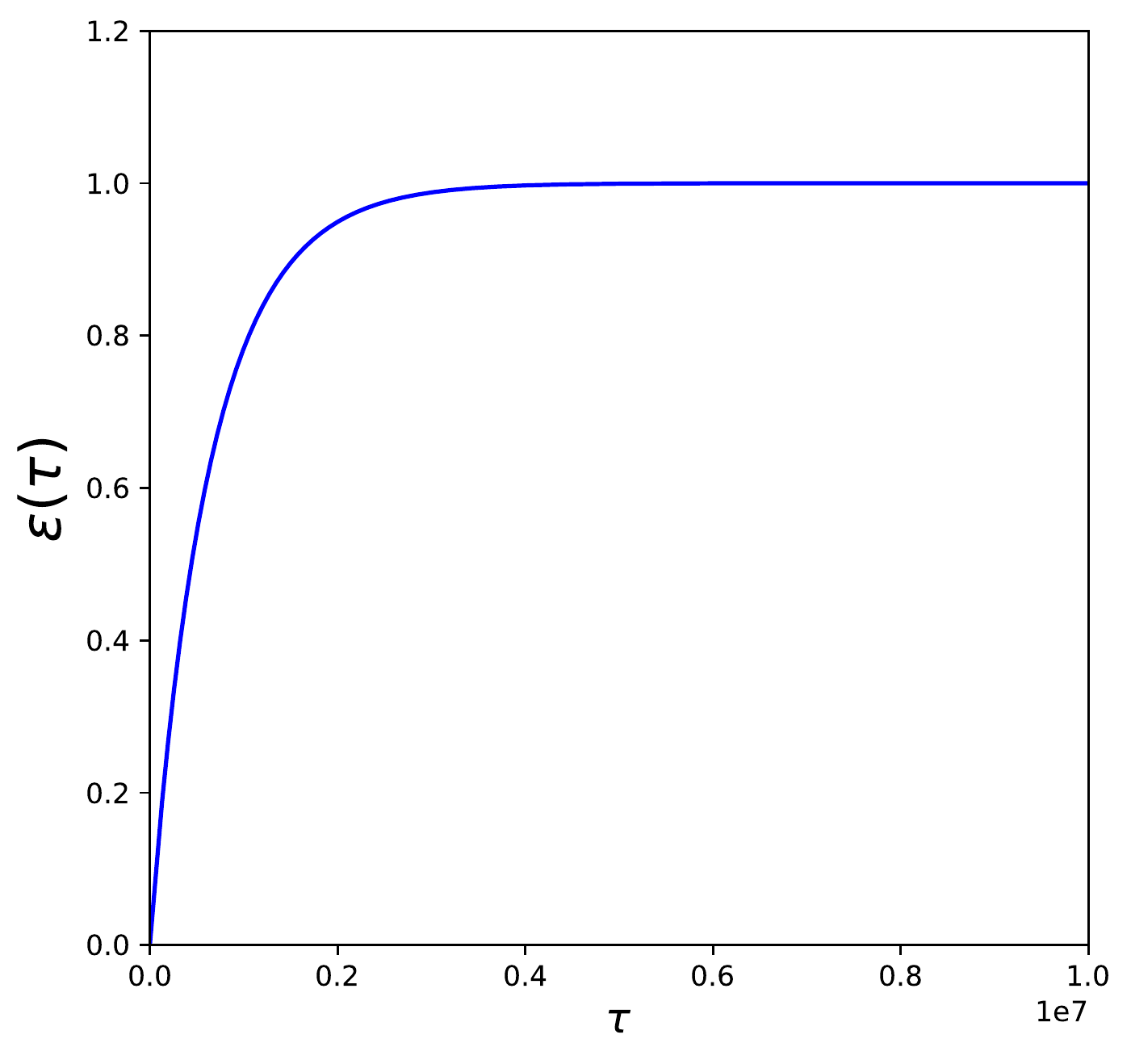} 
  \includegraphics[scale=.4]{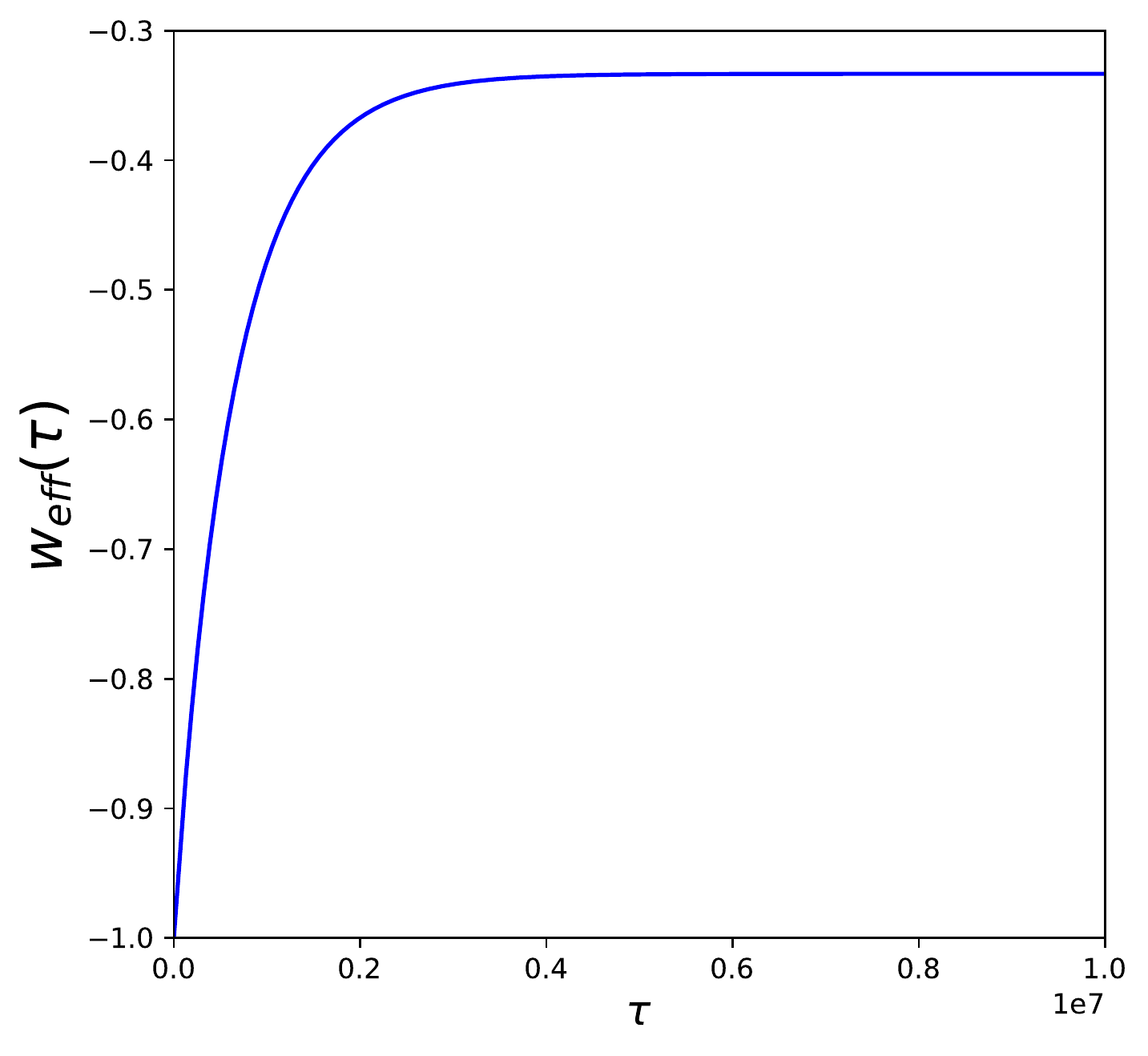}   
\caption[Plots for  $\epsilon(\tau)$  and $ w_{eff}(\tau)$ for case 5]{Plots for Geometric quantities  $\epsilon(\tau)$  and $ w_{eff}(\tau)$ for $\chi_{in} = \frac{1}{100}$ and $ f(X) = X $. }\label{plotwefflatem5} 
\end{figure}

Figure \ref{plotchiepsilonlatem5} shows numerical simulation of the $\chi(\tau)$ and $ \epsilon(\tau) $ . It is clear from the Figure \ref{plotchiepsilonlatem5} that both quantities varies linearly. We extend time from $ \tau = 100 $ to $ \tau =  10^{7}$. Figure \ref{plotwefflatem5} shows the late time evolution of $ \epsilon(\tau)$ and $ w_{eff}$. It is evident from the Figure \ref{plotwefflatem5} that   $ \epsilon(\tau)$ and $ w_{eff}$ sets to approximately $ \simeq 1 $ and $ \simeq -0.33 $ at late-times.

\section{Summary} \label{sec6} 
 Here we have studied background cosmology in five different incarnations of the Model I. Cosmology of Model I has been studied earlier in ref \cite{Tsamis:2016boj}. For consistency here, we reproduce all their result and note also the problems arise in Model I. In Model I, exit from inflation to radiation dominated era is possible. Whereas in Model $\rom{2} $, the transition from inflation era to RD era is not possible. In this model universe inflates for a long time and eventually transit over to $w_{eff} = -\frac{1}{3}$ and then evolve with the same equation of state forever. Analogous to Model $\rom{2}$, Model $ \rom{3}$ also shows similar cosmology. Here inflation does not last long in comparison to Model $ \rom{2} $. Similarly Model $\rom{4}$ gives rise to inflation for very short time and transit over to $ w_{eff}= -\frac{1}{3} $ and remain unchanged for the entire evolution. However, the transition from inflation to other era is not possible. Model $\rom{5}$ shows similar behavior as model $\rom{4}$. Here inflation also lasts for very short time and transit to RD era does not exist in this model. To summarize except Model $\rom{1}$ all these models are not suitable for having unification of inflation with RD era. Nevertheless all the Nonlocal models discussed here have inflationary solutions. 
 
  Overall the analysis done in this work could shed light on constructing an unified cosmological model using non-local terms in action.
  
 \section{Acknowledgements} This work was partially funded by DST grant no. SERB/PHY/2017041 .We are thankful to Richard Woodard for useful discussions. U.K. is indebited to arun rana for  carefully reading the draft.

\end{document}